\documentclass[iop]{emulateapj}

\bibpunct[; ]{(}{)}{,}{a}{}{,}
\defcitealias{kempner2004}{KD04}

\usepackage{epsfig}
\usepackage{subfigure}
\usepackage{graphics}

\newcommand{\chan}{{\it Chandra}}

\newcommand{\kms}{$\,\rm{km\,s^{-1}}$}

\newcommand{\rproj}{$R_{\rm proj}$}
\shorttitle{MINOR MERGERS IN A2142 AND RXJ1720}
\shortauthors{OWERS ET AL.}
\begin{document}
\title{Minor merger-induced cold fronts in Abell~2142 and 
RXJ1720.1+2638\altaffilmark{*}}
\author{Matt S. Owers\altaffilmark{1}, Paul E.J. Nulsen\altaffilmark{2}, 
Warrick J. Couch\altaffilmark{1}}
\altaffiltext{1}{Center for Astrophysics and Supercomputing, Swinburne 
University of Technology, Hawthorn, VIC 3122, Australia; 
mowers@astro.swin.edu.au}
\altaffiltext{2}{Harvard Smithsonian Center for Astrophysics, 60 Garden Street, 
Cambridge, MA 02138, USA}
\altaffiltext{*}{Some of the data presented herein were obtained at the W.M. 
Keck Observatory, which is operated as a scientific partnership among the 
California Institute of Technology, the University of California and the 
National Aeronautics and Space Administration. The Observatory was made possible
by the generous financial support of the W.M. Keck Foundation.}

\begin{abstract}
We present evidence for the existence of substructure in the 
``relaxed appearing'' cold front clusters Abell~2142 and RXJ1720.1+2638. The 
detection of these substructures was made possible by comprehensive 
multi-object optical spectroscopy obtained with the Hectospec and
DEep Imaging Multi-Object Spectrograph instruments on the 6.5m MMT and 10m 
Keck~II telescope, respectively. These observations produced 956 and 400 
spectroscopically confirmed cluster members within a projected radius of
3\,Mpc from the centers of Abell~2142 and RXJ1720.1+2638, respectively. The 
substructure manifests itself as local peaks in the spatial distribution 
of member galaxies and also as regions of localized velocity substructure. For 
both Abell~2142 and RXJ1720.1+2638, we identify group-scale substructures which,
when considering the morphology of the cold fronts and the time since 
pericentric passage of a perturber estimated from the cold front radii, could 
plausibly have perturbed the cluster cores and generated the cold fronts
observed in \chan\ images. The results presented here are consistent with cold
fronts being the result of merger activity and with cold fronts in relaxed 
appearing clusters being due to minor merger activity.

\end{abstract}


\section{Introduction}

One of the triumphs of modern cosmology is the success the hierarchical 
formation model has had in explaining the observed large scale structure of 
galaxies. In this model, small-scale perturbations in the matter density 
distribution at early epochs intensify due to the effects of gravity, collapse 
and merge together to form the galaxies, groups, and clusters observed at later 
epochs. Since clusters are the structures to form last, the process of 
hierarchical formation is expected to reveal itself dramatically in such 
systems. This is borne out by observations, where many clusters in the local
universe are found to have recently undergone significant growth via either a 
major merger with another cluster or through minor mergers with smaller clusters
and groups of galaxies. This offers a unique opportunity to study the effects of
large scale structure formation on the constituent galaxies and intra cluster 
medium (ICM). However, both the detection and distinction of clusters undergoing
major or minor mergers can be an observationally expensive exercise in the 
absence of easily detectable signatures of each process \citep{owers2009c}.

In this vein, the ``cold fronts'' first detected and characterized by the
\chan\ X-ray satellite \citep{markevitch2000, vikhlinin2001b}, thanks to its 
excellent sensitivity and unrivalled spatial resolution, may play an important 
role. Cold fronts are contact discontinuities in the ICM which form at abrupt
interfaces between low and high entropy plasma 
\citep[for a review see][]{markevitch2007}. In the scenario put forth by 
\citet{markevitch2000}, commonly referred to as the ``remnant core'' scenario, 
the edges delineate the low entropy gas of two cool cores, which have just 
survived a pericentric passage, and the surrounding hot shocked atmospheres 
which were stripped from the cores during the merger. This remnant core
scenario is observed in a number of clusters, e.g., 1ES0657-558 
\citep[otherwise known as the Bullet cluster;][]{markevitch2002},  
Abell~1758 \citep{david2004}, Abell~115 \citep{gutierrez2005}, Abell~2744 
\citep{kempner2004,owers2011}, Abell~2146 \citep{russell2010} and probably 
Abell~2069 \citep{owers2009c}. As the number of clusters observed by \chan\ 
increased, two things emerged (i)-The remnant core scenario cannot explain the 
majority of the observed cold fronts, and (ii)-There is a clear dichotomy when 
considering the X-ray morphology of clusters hosting cold fronts 
\citep{markevitch2007, owers2009c, ghizzardi2010}. One subset of cold front 
clusters exhibit clearly disturbed X-ray morphologies, while the second subset 
have cool cores and appear relaxed at large radii. The disturbed morphology 
present in the former subset is clearly related to major merger activity which 
manifests itself at other wavelengths. The cold fronts due to the remnant core 
scenario are generally found in these clusters, but other merger-related 
processes also produce cold fronts \citep[e.g., the ``ram pressure slingshot''
feature in Abell~168 reported by][see also \citealt{poole2006} for different cold fronts produced in 
merger simulations]{hallman2004}. In these cases, cold fronts are excellent 
signatures of major merger activity \citep{owers2009b, owers2009a}. For the 
subset with relaxed large scale X-ray morphologies, the best evidence for merger
activity comes from the observed cold fronts---the evidence for merger activity 
at other wavelengths in these clusters is generally ambiguous or non-existent.

The leading hypothesis for the formation of cold fronts in clusters with
a regular X-ray morphology is the relative motion
of the low entropy core gas---termed ``sloshing''\footnote{For clarity, 
throughout the paper we will refer to clusters which have regular/relaxed X-ray 
morphologies with cold fronts in their cores as sloshing cold front clusters.} 
by \citet{markevitch2001}---with respect to the quasi-static cluster potential. 
The simulations of \citet{tittley2005,ascasibar2006} and \citet{roediger2011} 
showed that a 
gravitational perturbation in the form of an infalling subcluster without a 
gaseous atmosphere can induce long-lasting oscillations of the the dark matter 
and gas cores. While the precise details differ (e.g., in the 
\citeauthor{tittley2005} scenario, the dark matter oscillates, dragging the gas 
with it, while in \citet{ascasibar2006} the dark matter is nearly static while 
the gas oscillates) in all cases the relative motion of low entropy cool core 
gas with respect to the cluster on larger scales
produces cold fronts where it meets the higher entropy gas at larger radii. 
Hydrodynamical perturbations in the form of minor merger-related weak shocks 
\citep{churazov2003}, shocks from fast moving galaxies \citep{roediger2011}, or 
acoustic-gravity waves related to bulk motions outside of the cluster cores 
\citep{fujita2004a} have also been investigated by simulations. In these 
scenarios, the core gas is displaced from the static dark matter core and 
oscillates around the potential minimum as the system relaxes back to its 
equilibrium configuration. Several authors have suggested that outbursts from 
AGN may provide a sufficient perturbation to induce sloshing 
\citep{markevitch2001,Hlavacek2011} although this possibility is yet to be 
fully explored. The emerging consensus is that the cold fronts seen in relaxed 
appearing cluster are due to minor merger activity. This leads to the 
interesting implication that cold fronts can be used as both a signpost of 
merger activity and, in combination with X-ray morphology, a tool to 
differentiate clusters undergoing major and minor mergers.

Much of the focus on sloshing cold front clusters has been at X-ray wavelengths 
or has involved simulations designed to explore mechanisms which are capable of 
reproducing the cold fronts along with a relaxed morphology at larger radii.
To date, no studies have focused on identifying possible perturbers by 
utilizing the spatial and kinematic information provided by multiobject 
spectroscopy (MOS)---an issue critical to the minor merger scenario where a 
recently merged perturber is not apparent at X-ray wavelengths, but may 
manifest itself as a dynamical substructure. Here, we set out to address this 
problem using the powerful combination of spatial and redshift information 
provided by multiobject spectroscopy in an attempt to detect merger related 
substructure and characterize its dynamical properties. Such observations are 
a crucial test for constraining simulations and for showing that cold fronts 
are excellent probes of ongoing merger activity. In this paper we 
present and analyze comprehensive MOS observations of the sloshing 
cold front clusters Abell~2142 and RXJ1720.1+2638 (hereafter A2142 and RXJ1720, respectively) which were selected from the 
subsample of relaxed appearing cold front clusters in \citet{owers2009c}. 
Previous MOS follow up of cold front clusters taken from 
the disturbed subset in \citet{owers2009c} has demonstrated that MOS 
observations are an essential ingredient required for understanding the merging 
history of a cluster \citep{owers2009a,owers2009b}.

A2142 was one of the first clusters targeted with the \chan\ X-ray 
telescope and, along with the observations of Abell~3667 \citep{vikhlinin2001b},
characterized for the first time the cold front phenomenon 
\citep{markevitch2000}. The {\it Einstein} X-ray images of A2142 
presented in \citet{oegerle1995}, revealed an asymmetric X-ray morphology, while
their analysis of 103 spectroscopically confirmed members of this cluster
yielded a mean redshift of $z=0.0905\pm0.0004$ and a velocity dispersion 
$\sigma=1280^{+94}_{-76}$\kms. 
\citet{oegerle1995} performed a Dressler-Schectman test \citep{dressler1988}
and found marginal evidence for dynamical substructure, while their data also  
revealed a high peculiar velocity for the second ranked member of 
$\sim 1600$\kms.  X-ray observations using ROSAT
revealed a highly elliptical X-ray morphology with a luminous compact core, the 
centroid of which is offset from that of the larger scale emission 
\citep{buote1996,henry1996}. The temperature map of \citet{henry1996} showed 
that this compact core was cooler than the surrounding gas and that the 
temperature distribution was asymmetric on large scales. The high ellipticity, 
offset core, asymmetric temperature distribution and evidence for dynamical 
structure were interpreted by both \citeauthor{henry1996} and 
\citeauthor{buote1996} as evidence for a later stage merger (i.e., viewed 
$\sim 1$\,Gyr after the merger) in A2142. Hints of 
a substructure to the northwest of the core in A2142 are seen in the weak 
lensing and SZ maps presented in \citet{umetsu2009}, and also the lensing data 
of \citet{okabe2008}.

RXJ1720 was categorized as an extended X-ray source in the {\it Einstein} Slew 
Survey \citep{elvis1992}. The \chan\ observations of RXJ1720 revealed two 
cold fronts within 100\arcsec\, of the cluster center, while the X-ray emission
outside of this radius showed a regular morphology \citep{mazzotta2001}. These
\chan\ observations of RXJ1720 provided the first clear indication that cold 
fronts can also be found in relaxed appearing cool core clusters. 
\citet{mazzotta2001} interpreted the cold fronts as evidence for either a late 
stage major merger viewed after several pericentric passages or the asynchronous
collapse of two nearly cospatial density perturbations with different linear 
scales. Using a deeper \chan\ observation, \citet{mazzotta2008} found that the 
spiral features revealed in the temperature and residual maps are more readily
explained by the sloshing scenario.  \citeauthor{mazzotta2008} also reported an 
interesting correlation between the spiral structure and radio emission from a 
mini-halo and suggested that the origin of this diffuse radio emission may be 
related to the sloshing process. At optical wavelengths, RXJ1720 was detected 
with the C4 cluster finding algorithm using the Sloan Digital Sky Survey (SDSS) 
\citep{miller2005}. From 26 spectroscopically confirmed cluster members
\citet{miller2005} measured a mean redshift of $z=0.1603$ and a velocity 
dispersion $\sigma=878$\kms\, while the Dressler-Schectman test revealed no 
evidence for dynamical substructure. The projected mass maps from the weak 
lensing analysis of \citet{okabe2010} showed that RXJ1720 harbors a second mass 
condensation just north of the main cluster component which is indicative of
merger activity.

For both clusters there are tantalizing hints for merger activity. The goal of
this paper is to seek conclusive evidence of dynamical substructure related to
merger activity by using the extensive spectroscopy collected at the 6.5m 
MMT and 10m Keck II Telescope using the Hectospec and DEep Imaging 
Multi-Object Spectrograph instruments, respectively. We describe our target selection and 
MOS observations in Section~\ref{obs} and our methods for cluster member selection 
and substructure detection in Section~\ref{anal}. We interpret our results
in Section~\ref{disc} and present a summary and concluding remarks in 
Section~\ref{sum}. Throughout this paper we assume a standard $\Lambda$CDM 
cosmology where $H_0=70$\kms$\,{\rm Mpc}^{-1}$, $\Omega_m=0.3$ and 
$\Omega_V=0.7$. Assuming this cosmology, at the systemic redshifts of A2142
and RXJ1720, 1 arcsecond corresponds to a physical scale of $1.68$\,kpc and
$2.76$\,kpc, respectively.

\section{Observations and Data Reduction}\label{obs}

\subsection{Photometric Catalogs}

The spectroscopic target catalogs for both A2142 and RXJ1720 were 
drawn from the SDSS and included objects within
a projected distance from the cluster center of $R_{\rm proj}=3$\,Mpc, or $\sim 30\arcmin$ and $\sim 18\arcmin$ for 
A2142 and RXJ1720, respectively. The catalogs were filtered to 
contain only objects identified as galaxies by the SDSS pipeline morphological 
classification scheme \citep{stoughton2002}. Objects near to bright stars
and misclassified as galaxies by the automated SDSS pipeline were identified by 
eye and rejected from the catalog. Objects with existing SDSS spectroscopy and
having redshifts which placed them well outside the redshift range expected for 
the cluster of interest were also removed from the sample. We transformed the 
SDSS $g$ and $r-$band magnitudes onto the Johnson-Cousins $B$ and $R$ system 
using Equations A5 and A7 of \citet{cross2004}. 

\subsection{MMT/Hectospec Observations}\label{MMT}

The majority of the spectra used in this paper were collected at
the 6.5m MMT using the Hectospec multi-object spectrograph 
\citep{fabricant2005}.
Hectospec is a bench-mounted spectrograph which has the capability
to simultaneously observe 300 objects through 1\arcsec.5 diameter fibers
spread over a 1 degree field of view. The observations were performed
with the 270 groove mm$^{-1}$ grating resulting in spectra with $\sim 6$\,\AA\
resolution covering the wavelength range $3500 - 10000$\,\AA. Approximately 30 
fibers were allocated to blank sky regions to facilitate sky subtraction during 
the data reduction. Due to the large number of targets compared to the number
of available fibers and fiber packing limitations, 6 and 8 configurations
were required to obtain adequate spectroscopic completeness for RXJ1720
and A2142, respectively. The exposure time, magnitude limit and seeing 
conditions for each MMT/Hectospec configuration are summarized in 
Table~\ref{MMT_obs}.

\begin{figure*}
{\includegraphics[angle=90,width=0.48\textwidth]{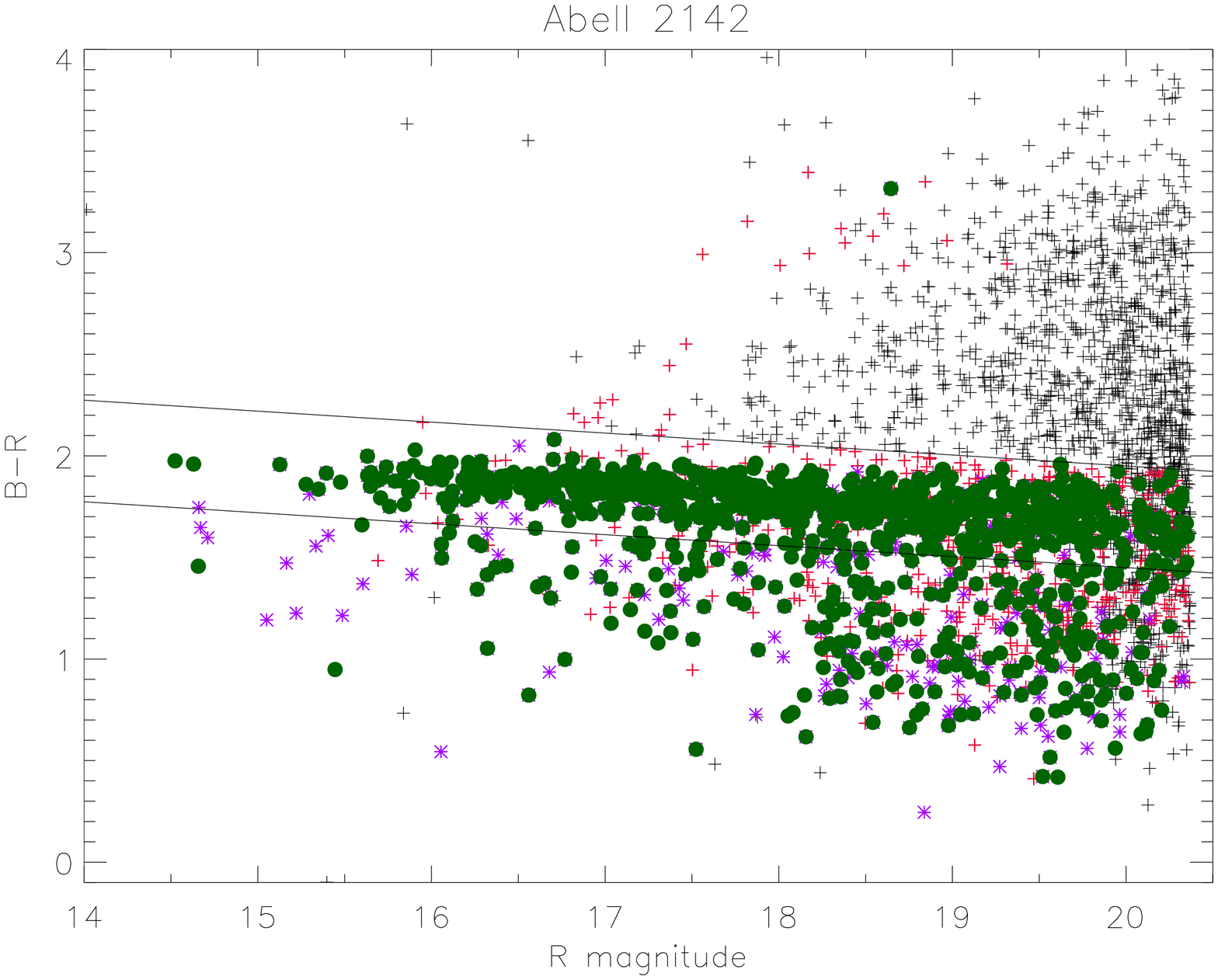}}
{\includegraphics[angle=90,width=0.48\textwidth]{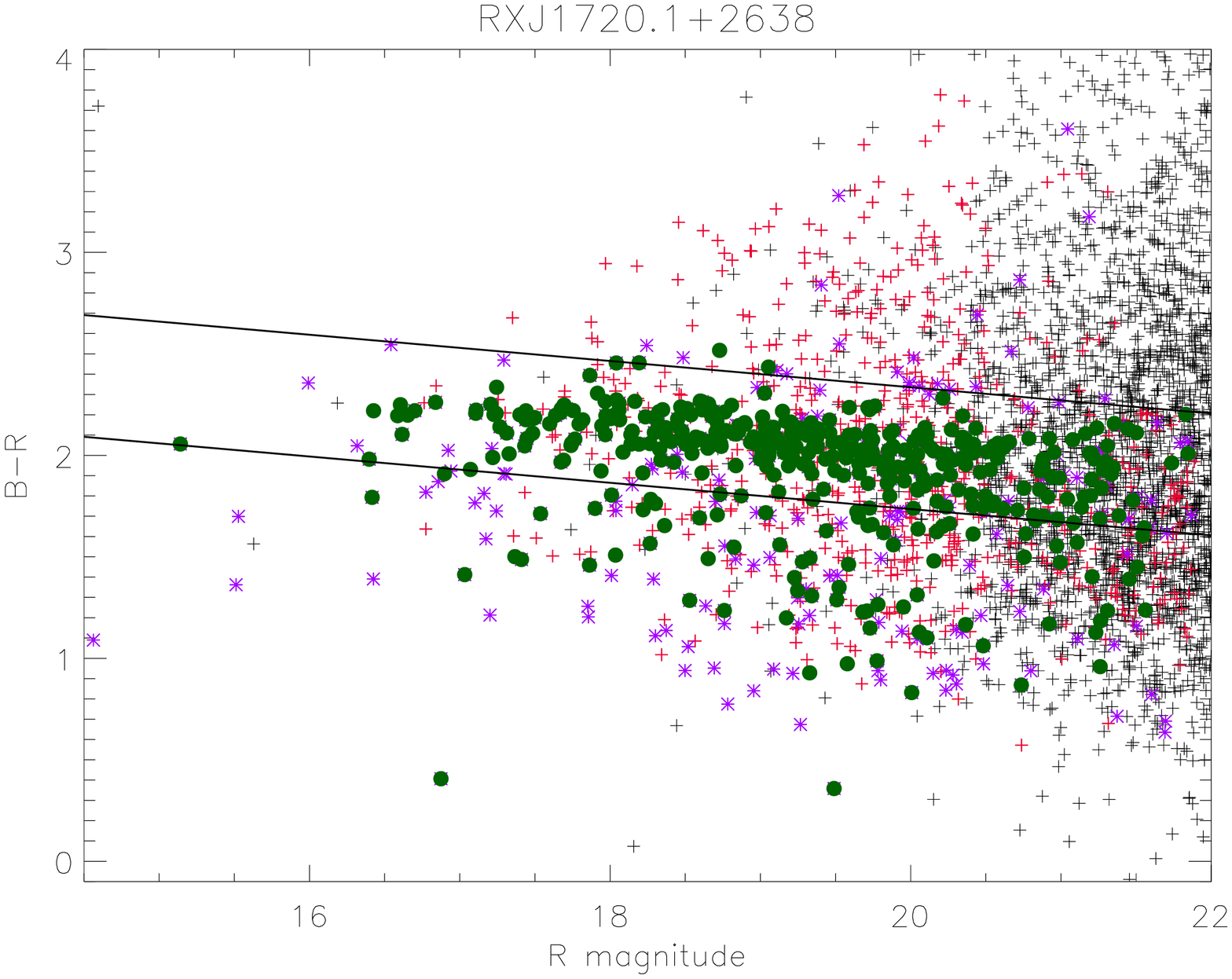}}
\caption{Color magnitude diagrams for the objects within 3\,Mpc of A2142 
({\it left panel}) and RXJ1720 ({\it right panel}). Filled green circles
are spectroscopically confirmed cluster members, blue stars are foreground 
objects, red crosses are background objects and black crosses are objects 
without a spectrum. The black lines show the estimates used for the red and 
blue limits defining the cluster red envelope which was used for the cluster 
membership allocation procedure described in Section~\ref{memsel}. The red limit
was used in the ranking process described in Sections~\ref{MMT} and \ref{keck}.}
\label{redseq}
\end{figure*}

The fiber configurations were generated with the 
XFITFIBS\footnote{http://www.cfa.harvard.edu/mmti/hectospec.html} software where
the ranking capability was utilized to ensure that the most likely cluster members
had a better chance of being allocated to a configuration. For RXJ1720, objects
were ranked based on their position on the color-magnitude relation 
(Figure~\ref{redseq}) and also their \rproj\ in the 
sense that objects lying redward of the cluster red-sequence, i.e., above the 
bold black line in Figure~\ref{redseq}, and with $2.5 < R_{\rm proj}<3$\,Mpc
were ranked lowest while those galaxies which lie on or blueward of the 
red-sequence and with $R_{\rm proj} < 0.5$\,Mpc were ranked highest. The 
slope of the line delineating those galaxies 
redward of the red sequence is calculated from the best fit to the red sequence 
slope versus redshift diagram in Figure 3. of \citet{lopezcruz2004}. The ranking
procedure was similar for the A2142 configurations, except that objects 
lying redward of the cluster red-sequence were simply not included for 
observation (Figure~\ref{redseq}), while a number of objects identified as being
part of local projected overdensities (aside from the central overdensity) were 
given higher rankings and a number of fainter ($20 < R \leq 20.5$) objects were 
included with low rankings as filler objects to be allocated to spare fibers.

The spectra were reduced at the Smithsonian Astrophysical Observatory 
Telescope Data Center\footnote{http://tdc-www.harvard.edu} (TDC) using the 
SPECROAD pipeline\footnote{http://tdc-www.harvard.edu/instruments/hectospec/specroad.html} \citep{mink2007}. 
Redshifts were also determined at the TDC using the IRAF cross-correlation 
XCSAO software \citep{kurtz1992} and each spectrum was assigned a redshift 
quality of ``Q'' for a reliable redshift, ``?'' for questionable and ``X'' for 
a bad redshift measurement. These observations provided 1771 and 815 quality 
``Q'' redshift measurements for extragalactic objects in the A2142 and 
RXJ1720 fields, respectively.

\begin{deluxetable*}{ccccccc}
\tabletypesize{\scriptsize}
\tablecolumns{3}
\tablewidth{0pc}
\tablecaption{Summary of the MMT/Hectospec observations.\label{MMT_obs}}
\tablehead {\colhead{Cluster} & \colhead{Date} &\colhead{Magnitude limit}&\colhead{Exposure Time (s)} & ${\rm N}_{\rm spec} $& ${\rm N}_{\rm z}$&\colhead{Seeing}} 
\startdata
RXJ1720 & 2006 May 4&$R<20.5$ &$3\times 900\rm s$& 208 & 189 & - \\
 & 2006 May 4&$R<20.5$ &$3\times 900\rm s$ & 186 & 176& - \\
 & 2007 Mar 15&$R<20.5$ &$3\times 900\rm s$ & 189& 171 & 0\arcsec.97 \\
 & 2007 Mar 16&$R<20.5$ &$3\times 1200\rm s$ & 177 & 92& 0\arcsec.63 \\
 & 2007 April 16&$R<20.5$ &$3\times 1200\rm s$ & 161 & 122 &1\arcsec.5 \\
 & 2007 April 20&$R<20.5$ &$5\times 600\rm s$ & 166 & 121 &1\arcsec.3 \\
A2142 & 2009 February 5&$R<20.5$ &$3\times 1200\rm s$ & 266 & 255& 1\arcsec.08 \\
& 2009 February 21&$R<20.5$ &$3\times 1200\rm s$ & 261 & 174&1\arcsec.30 \\
& 2009 February 24&$R<20.5$ &$1\times 1200\rm s$ & 263 & 159&0\arcsec.56 \\
& 2009 March 1&$R<20.5$ &$3\times 1200\rm s$ & 263 & 249 & 0\arcsec.84 \\
& 2009 April 22&$R<20.5$ &$3\times 960\rm s$ & 260 & 255 &0\arcsec.60 \\
& 2009 April 23&$R<20.5$ &$3\times 1200\rm s$ & 259 & 244& 1\arcsec.21 \\
& 2009 April 23&$R<20.5$ &$3\times 1200\rm s$ & 255 & 240 &0\arcsec.61 \\
& 2009 April 24&$R<20.5$ &$3\times 1200\rm s$ & 242 & 229 &0\arcsec.86 \\
\enddata
\end{deluxetable*}

\subsection{Keck/DEIMOS Observations}\label{keck}

The MMT/Hectospec data for RXJ1720 were supplemented with spectra 
obtained at the Keck II telescope using the Deep Imaging Multi-Object 
Spectrograph (DEIMOS). In MOS mode, DEIMOS utilizes a slitmask to collect
spectra for $\sim 100$ target objects positioned within a 
$\sim 16.7$\arcmin$\times5$\arcmin\ field of view. The observations 
presented here made use of the 600 line mm$^{-1}$ grating which, in combination 
with a 1\arcsec.5 slitwidth, gives a spectral resolution of $\sim 7.0 $\,\AA. The 
GG400 order-blocking filter was used and the grating was tilted to a central 
wavelength of 5500\,\AA\ which yields a useable spectral range of 
$\sim 4000 - 9000$\,\AA.

Slit masks were generated with the DSIMULATOR task within the DEIMOS IRAF 
package. The input target catalog was limited to contain objects with $R <22$ 
and within \rproj$=2$\,Mpc ($\sim12$\arcmin). The catalog
was culled of those objects found to be non-cluster members from the previous
MMT/Hectospec observations, while those
spectroscopically confirmed cluster members with high 
quality Hectospec spectra (i.e. S/N$\gtrsim 8$) were also removed. The 
DSIMULATOR software allows a two tier ranking of targets: First, the catalog is 
split into subsamples with objects in the first subsample given first priority 
in slit allocation. Second, the objects within subsamples are given weights 
such that the highest weighted objects are given a higher priority during slit 
allocation. We divide the catalog into 5 subsamples: the first and second 
contain objects lying on or blueward of the red sequence with $R \leq 21.5$ and 
$21.5 < R \leq 22$, respectively, the third contains objects
already observed with Hectospec, the fourth and fifth contain objects lying
redward of the red sequence with $R \leq 21.5$ and $21.5 < R \leq 22$, 
respectively. The object weights were assigned according to \rproj. This subsampling and weighting scheme means that objects within the
central 500\,kpc lying on or blueward of the red sequence and with $R \leq 21.5$
are most likely to be allocated a slit, while objects with $21.5 < R \leq 22$
lying redward of the red sequence and 2\,Mpc from the cluster center are least 
likely to be allocated a slit on the mask. The position angle of each slit
may be tilted by $\pm 30\deg$ with respect to the mask position angle (which is 
aligned with the spatial direction on the CCD). Where the position angle of a
galaxy's major axis (taken from the SDSS shape parameters) does not exceed this
limitation, we tilt the slit to align with the major axis. In practice, this 
was only possible in a few cases and the majority of slits were therefore 
assigned a position angle of $5\deg$ with respect to the mask. Using this setup,
we generated 6 slitmasks which were observed on 2009 May 24. 
Table~\ref{keck_obs} summarizes these observations.

\begin{deluxetable*}{ccccccc}
\tabletypesize{\scriptsize}
\tablecolumns{7}
\tablewidth{0pc}
\tablecaption{Summary of the Keck/DEIMOS observations of RXJ1720.\label{keck_obs}}
\tablehead {\colhead{$\alpha$ (J2000)} &\colhead{$\delta$ (J2000)} &\colhead{Mask PA} &\colhead{Exposure Time (s)} & \colhead{${\rm N}_{\rm spec}$} & \colhead{${\rm N}_{\rm z}$} & \colhead{Seeing}}
\startdata 
17:20:05.02 &  26:39:41.5 & 180.0&$2\times 1100 + 1340\rm s$& 70 & 61 & 0\arcsec.57 \\
17:20:14.28 &  26:35:38.2 & 0.0  &$2\times 1100 + 1340\rm s$ & 67 & 67& 0\arcsec.60 \\
17:20:03.70 &  26:37:31.9 & 0.0  &$2\times 1100 + 1200\rm s$ & 62& 57 & 0\arcsec.50 \\
17:20:16.73 &  26:37:25.3 & 180.0 &$2\times 1100 + 1200\rm s$ & 60 & 57& 0\arcsec.63 \\
17:19:46.02 &  26:37:31.9 & 0.0  &$2\times 1100\rm +1500s$ & 68 & 68 &0\arcsec.54 \\
17:20:34.44 &  26:37:31.9 & 180.0  &$2\times 1100 + 800\rm s$ & 62 & 50 &0\arcsec.61 \\
\enddata
\end{deluxetable*}

The data were reduced with the {\sf spec2d} 
software\footnote{http://deep.berkeley.edu/spec2d/} developed for the DEEP2 
galaxy redshift survey \citep{davis2003}. Briefly\footnote{a more thorough 
explanation can be found at http://deep.berkeley.edu/spec2d/primer.html}, the 
software uses the internal Quartz flats taken on the afternoon before the 
observations to find and rectify the slits on the CCD, to determine the 
throughput of each slit and to correct for fringing. We utilize blue and 
red specific arc lamps to determine the 2D wavelength solution which is initially
estimated using the DEIMOS optical model. The science exposures are combined 
using an inverse variance weighed mean after cosmic ray correction and sky 
subtraction. The 1D spectra are then extracted using optimal extraction 
\citep{horne1986} and rebinned to linear wavelength bins using the IRAF task 
RSPECTEXT.

The redshifts are again measured with the XCSAO cross-correlation package in 
IRAF \citep{kurtz1992} and each spectrum was visually assessed (by MSO) and 
assigned a redshift quality of ``Q'' for a reliable redshift, ``?'' for 
questionable and ``X'' for a bad redshift measurement. These observations
produced 333 quality ``Q'' redshift measurements for extragalactic objects in the 
RXJ1720 field.

\subsection{Redshift precision}

The redshift precision is determined here using repeat observations taken during
the observating runs. For the MMT/Hectospec observations of A2142, there 
were 234 objects having two quality ``Q'' redshift determinations which were 
reobserved due to their lower signal to noise spectra (none were taken for 
RXJ1720). Taking the difference of the redshifts of these repeat observations 
(in the sense that we subtract the subsequent higher S/N redshift measurement 
from the lower S/N initial redshift) we use biweight estimators to find a mean 
redshift difference of $<\Delta cz> = 2.3\pm2.3$\kms\ and a standard deviation 
$\sigma(\Delta cz) = 34.2\pm2.3$\kms\ 
indicating a single measurement uncertainty of 
$\sigma(<\Delta cz>)/\sqrt 2 = 24.2\pm1.6$\kms.
This single measurement uncertainty is consistent with the median of the 
redshift uncertainties determined by the XCSAO software of $24.7$\kms. 

We can test the veracity of this internal precision measurement by comparing the
repeat observations taken with MMT/Hectospec and from the SDSS. There are 301 objects
which have redshift measurements from both our A2142 MMT/Hectospec 
observations and the SDSS. Taking the difference between the MMT/Hectospec
and SDSS redshifts, we find $<\Delta cz> = -4.8\pm1.8$\kms\ and 
$\sigma(\Delta cz) = 31.3\pm1.5$\kms. This standard deviation is well 
encompassed by the median of the individual redshift uncertainty measurements 
for the SDSS ($\overline{cz_{\rm err}}=47.5$\kms) and the MMT/Hectospec 
($\overline{cz_{\rm err}}=21.9$\kms) observations.
Similarly, while there were no repeat observations within the MMT/Hectospec or
Keck/DEIMOS redshift catalogues for RXJ1720, there were 52 common measurements 
between the SDSS and MMT/Hectospec. The mean difference in the redshift measurements
common to the SDSS and Hectospec catalogues was $<\Delta cz> = -31.6\pm7.4$\kms\
and $\sigma(\Delta cz) = 49.8\pm6.3$\kms. The quadrature sum of the median 
value for the individual uncertainties on the SDSS 
($\overline{cz_{\rm err}}=45.8$\kms) and the MMT/Hectospec 
($\overline{cz_{\rm err}}=28.9$\kms) repeat observations  well encompasses the 
scatter in  $<\Delta cz> $ implying that the individual uncertainties provide a 
good measure of the real uncertainties associated with these measurements. 

Given the robustness of the MMT/Hectospec redshift measurements and associated 
uncertainties, we now use the 33 MMT/Hectospec objects which were reobserved 
with Keck/DEIMOS to determine the real uncertainties in the DEIMOS measurements,
finding $<\Delta cz> = -69.5\pm19.0$\kms\ with scatter 
$\sigma(\Delta cz) = 111.6\pm18.5$\kms. Comparing the median values of the
individual uncertainty measurements (for MMT/Hectospec: 
$\overline{cz_{\rm err}}=38.5$\kms\ and for Keck/DEIMOS: 
$\overline{cz_{\rm err}}=52.3$\kms) it can be seen that the scatter is not well 
accounted for by the quadrature sum of the uncertainties. We conclude that the 
real redshift uncertainties on the DEIMOS measurements are $\sim 100$\kms.

Since there are significant offsets measured in the repeat observations for RXJ1720,
when combining the MMT/Hectospec, Keck/DEIMOS and SDSS catalogs we correct
each Keck/DEIMOS and SDSS redshift measurement by the $<\Delta cz>$ value
measured above such that they are consistent with the MMT/Hectospec 
measurements. Where there are repeat measurements, for both the A2142 and 
RXJ1720 catalogs we use the measurement with the lowest measured uncertainty. 
Where a Keck/DEIMOS measurement is involved, we double the redshift uncertainty,
consistent with the results outlined above. The final catalogs contain 1121 and 
1635 single redshift measurements for extragalactic objects in the field 
surrounding RXJ1720 and A2142, respectively. We tabulate the position,
redshift, redshift uncertainty and redshift source of objects in the final 
catalogs in Tables~\ref{a2142_cztab} and \ref{rxj_cztab} for
A2142 and RXJ1720, respectively.

\begin{deluxetable*}{ccccc}
\tabletypesize{\scriptsize}
\tablecolumns{3}
\tablewidth{0pc}
\tablecaption{Combined redshift catalog for A2142.\label{a2142_cztab}}
\tablehead{\colhead{R.A. (J2000)} & \colhead{decl. (J2000)} & \colhead{($cz$)} & 
\colhead{$cz$ uncertainty} & \colhead{$cz$ source}\\
    deg.      &     deg.       &   \kms     &    \kms  &            }
\startdata
239.036453  & 27.235901 &  57055.78 &  18.59     &      MMT/Hectospec \\
239.042587  & 27.177689 &  28939.35 &  30.64     &      MMT/Hectospec \\
239.045349  & 27.174765 &  28811.72 &  56.53     &      MMT/Hectospec \\
239.048401  & 27.247974 &  96294.95 &  58.26     &      MMT/Hectospec \\
239.049500  & 27.187420 &  24531.00 &  48.49     &      MMT/Hectospec \\
239.053802  & 27.157728 &  86498.82 &  35.92     &      MMT/Hectospec \\
239.055679  & 27.257549 &  33359.37 &  35.87     &      MMT/Hectospec \\
239.062576  & 27.132755 &  97387.20 &  21.20     &      MMT/Hectospec \\
239.078674  & 27.185207 &  67723.70 &  27.55     &      MMT/Hectospec \\
239.079071  & 27.307705 &  26771.70 &  19.48     &      MMT/Hectospec \\
\enddata
\tablecomments{This table is available in its entirety in a machine-readable form in the online
journal. A portion is shown here for guidance regarding its form and content.}
\end{deluxetable*}

\begin{deluxetable*}{ccccc}
\tabletypesize{\scriptsize}
\tablecolumns{3}
\tablewidth{0pc}
\tablecaption{Combined redshift catalog for RXJ1720.\label{rxj_cztab}}
\tablehead{\colhead{R.A. (J2000)} & \colhead{decl. (J2000)} & \colhead{($cz$)} & 
\colhead{$cz$ uncertainty} & \colhead{$cz$ source}\\
    deg.      &     deg.       &   \kms     &    \kms  &            }
\startdata
259.713308  &  26.620569 &  119987.97 &  102.64   &         MMT/Hectospec \\
259.718571  &  26.644934 &   47540.42 &  144.19   &         MMT/Hectospec \\
259.719717  &  26.702864 &  111920.71 &  101.03   &         MMT/Hectospec \\
259.726667  &  26.565548 &   72713.53 &   28.23   &         MMT/Hectospec \\
259.727325  &  26.544941 &   97930.56 &   28.17   &         MMT/Hectospec \\
259.728700  &  26.560825 &   72656.78 &   80.44   &         MMT/Hectospec \\
259.731187  &  26.554123 &   98086.70 &   55.13   &         MMT/Hectospec \\
259.732817  &  26.646864 &   36395.72 &   35.47   &         MMT/Hectospec \\
259.738025  &  26.516831 &   30848.81 &   21.77   &         MMT/Hectospec \\
259.739112  &  26.738709 &  117214.30 &   35.72   &         MMT/Hectospec \\
\enddata
\tablecomments{This table is available in its entirety in a machine-readable form in the online
journal. A portion is shown here for guidance regarding its form and content.}
\end{deluxetable*}

\subsection{Spectroscopic completeness}

\begin{figure*}
{\includegraphics[angle=90,width=0.48\textwidth]{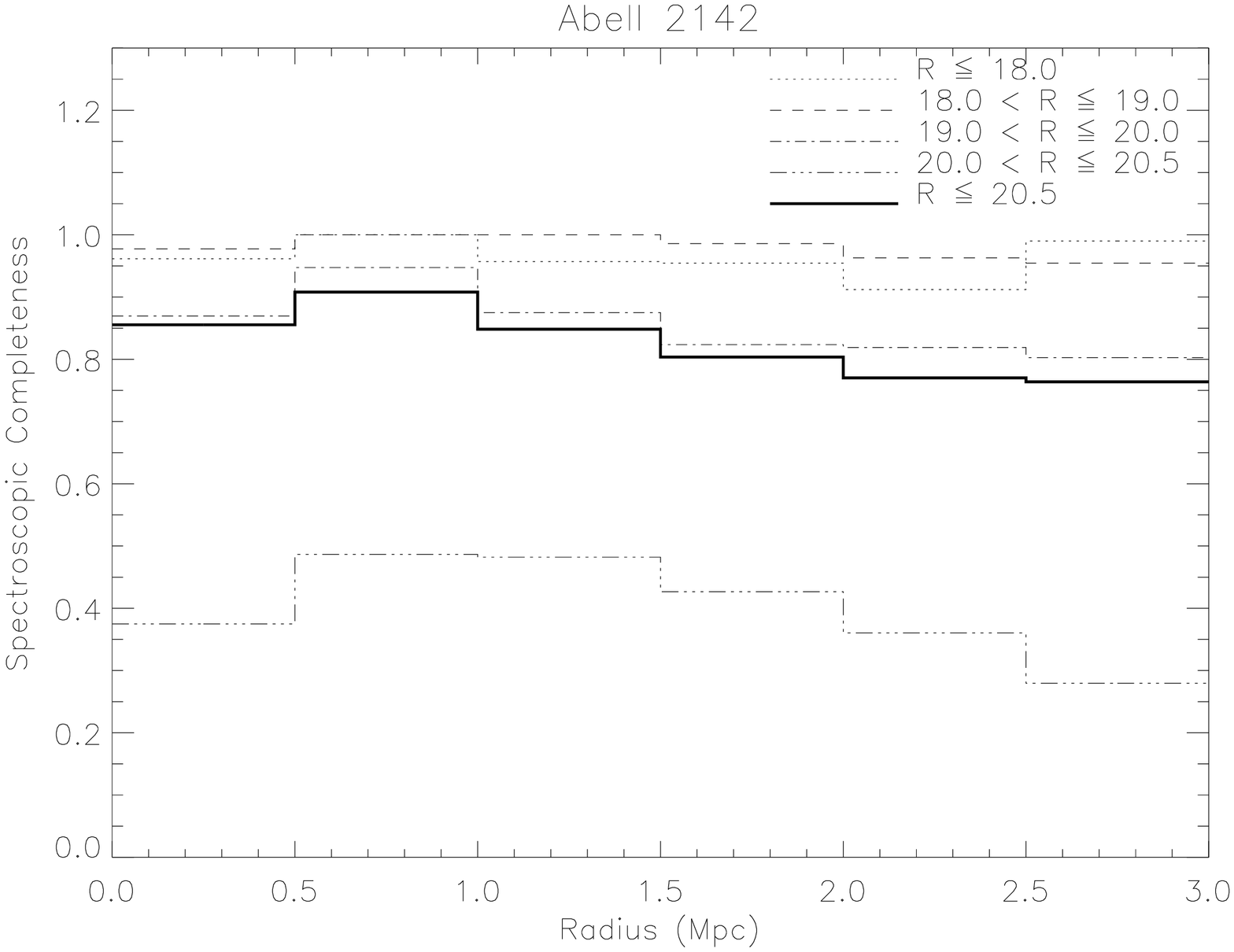}}
{\includegraphics[angle=90,width=0.48\textwidth]{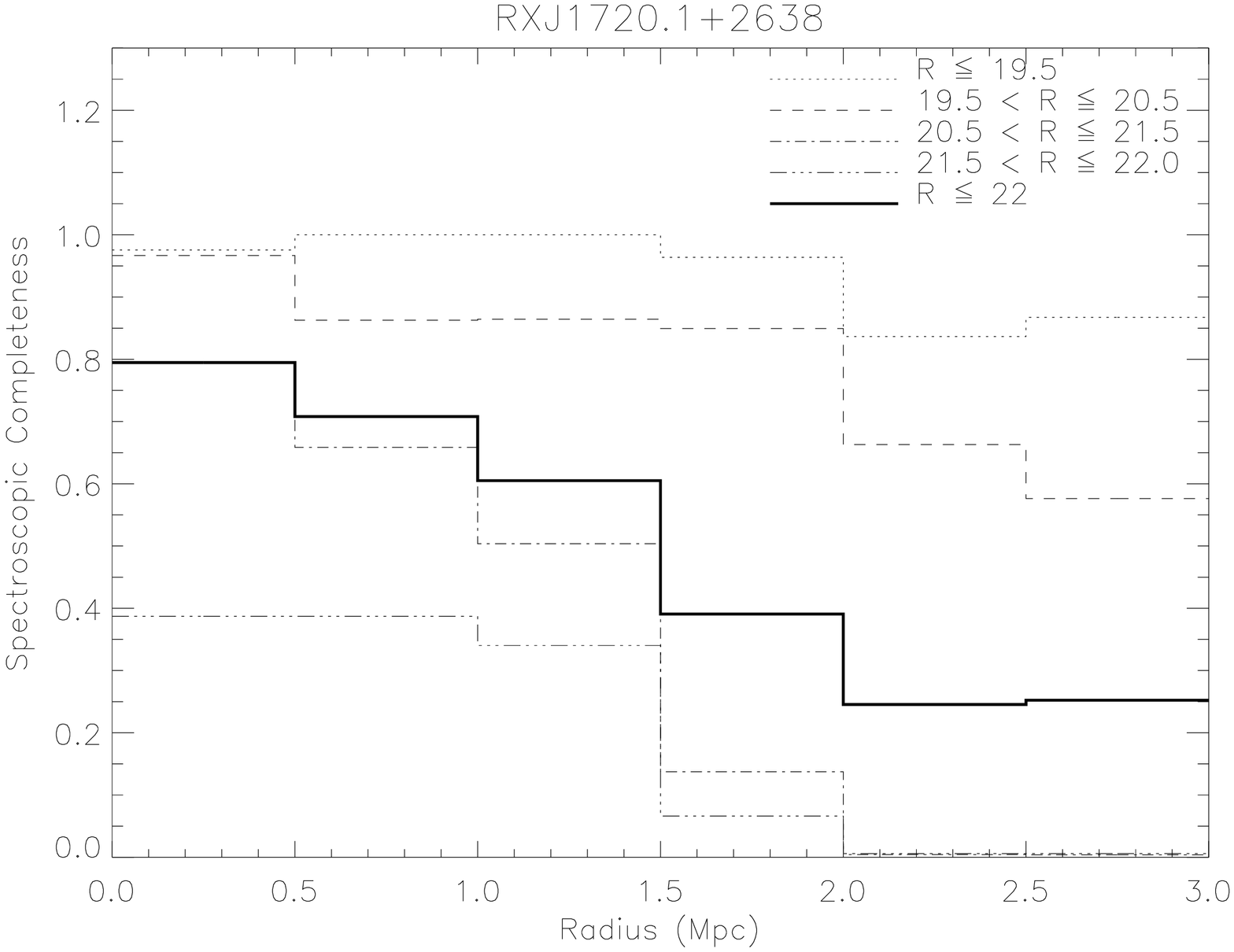}}
\caption{Spectroscopic completeness as a function of radius in different 
magnitude bins (listed in the top right corner) for A2142 
({\it left panel}) and RXJ1720 ({\it right panel}).}
\label{compl_radius}
\end{figure*}

An understanding the spectroscopic completeness of our redshift catalogs
is imperative when interpreting the results of tests for substructure, in 
particular for those that rely on finding local overdensities in the spatial
distribution of the cluster members. More specifically, we need to understand
what fraction of potential cluster members are in the photometric catalog
but have not been assigned a redshift. Therefore, we define the spectroscopic
completeness as the ratio of the number of objects with quality ``Q'' redshift 
measurements to the number of objects in the photometric catalog. As seen in
the right panel of Figure~\ref{redseq}, objects lying redward of the cluster 
red sequence are unlikely to be cluster members. Therefore, given our goal of
understanding how redshift incompleteness affects the cluster members, 
for RXJ1720 we only measure the spectroscopic completeness for the subsample of 
objects lying below the upper black bold line defining the red limit of the 
cluster red sequence in the right panel of Figure~\ref{redseq}. This 
also allows a more meaningful comparison to the A2142 data. 
Figure~\ref{compl_radius} shows the spectroscopic completeness as a function of 
\rproj\ for both clusters for different magnitude bins. For 
A2142, it can be seen in the left panel of Figure~\ref{compl_radius} that
for the high priority targets (i.e. those with $R \leq 20.$) the spectroscopic 
completeness is always higher than $80\%$, while it drops to $30 - 40\%$ for
the low priority faint objects. We therefore achieve a high level of 
spectroscopic completeness to $\sim 3$ magnitudes fainter than the 
characteristic magnitude of the cluster luminosity function, $M^*$. 

\begin{figure*}
{\includegraphics[angle=0,width=0.3\textwidth]{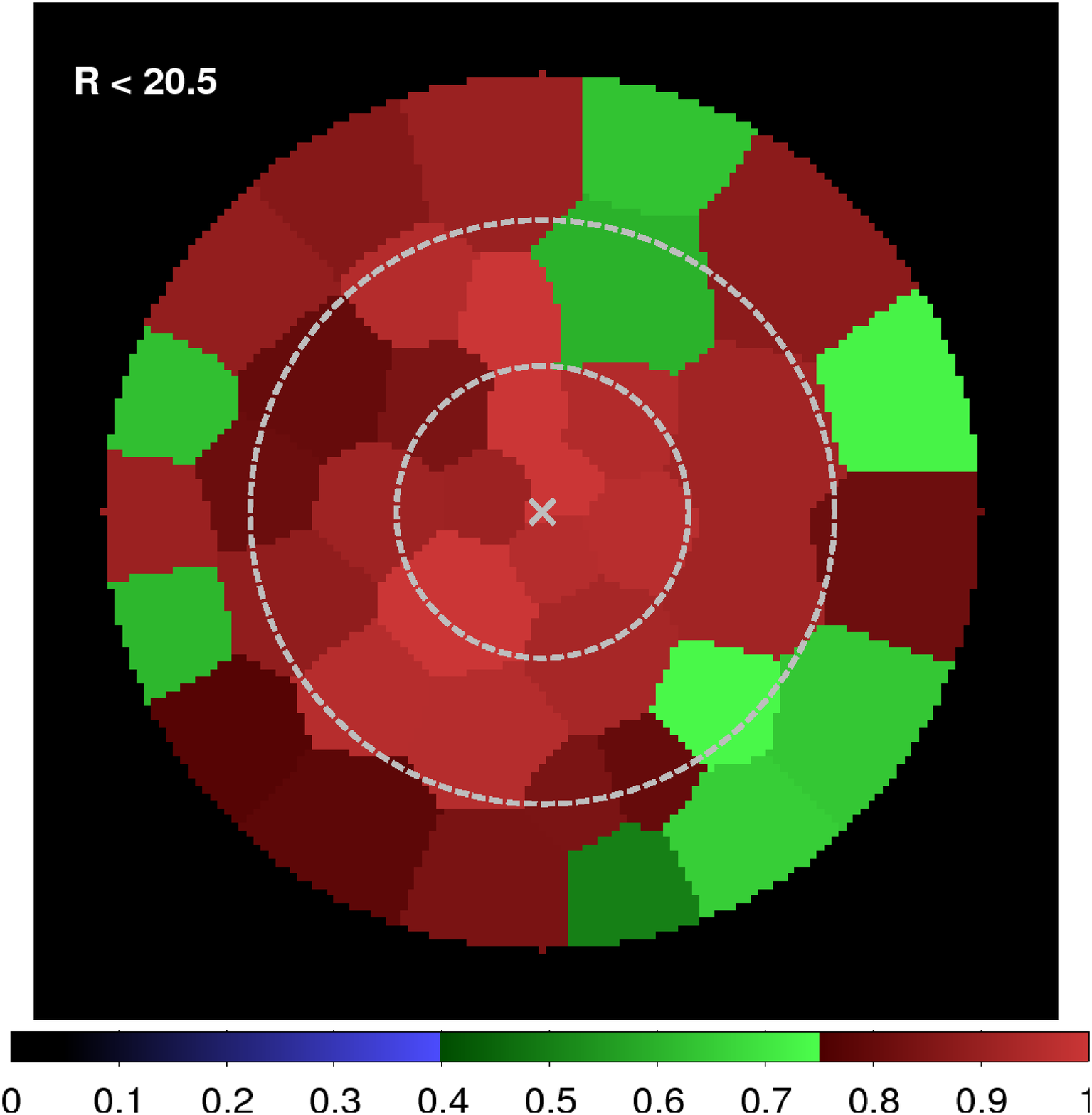}}
{\includegraphics[angle=0,width=0.3\textwidth]{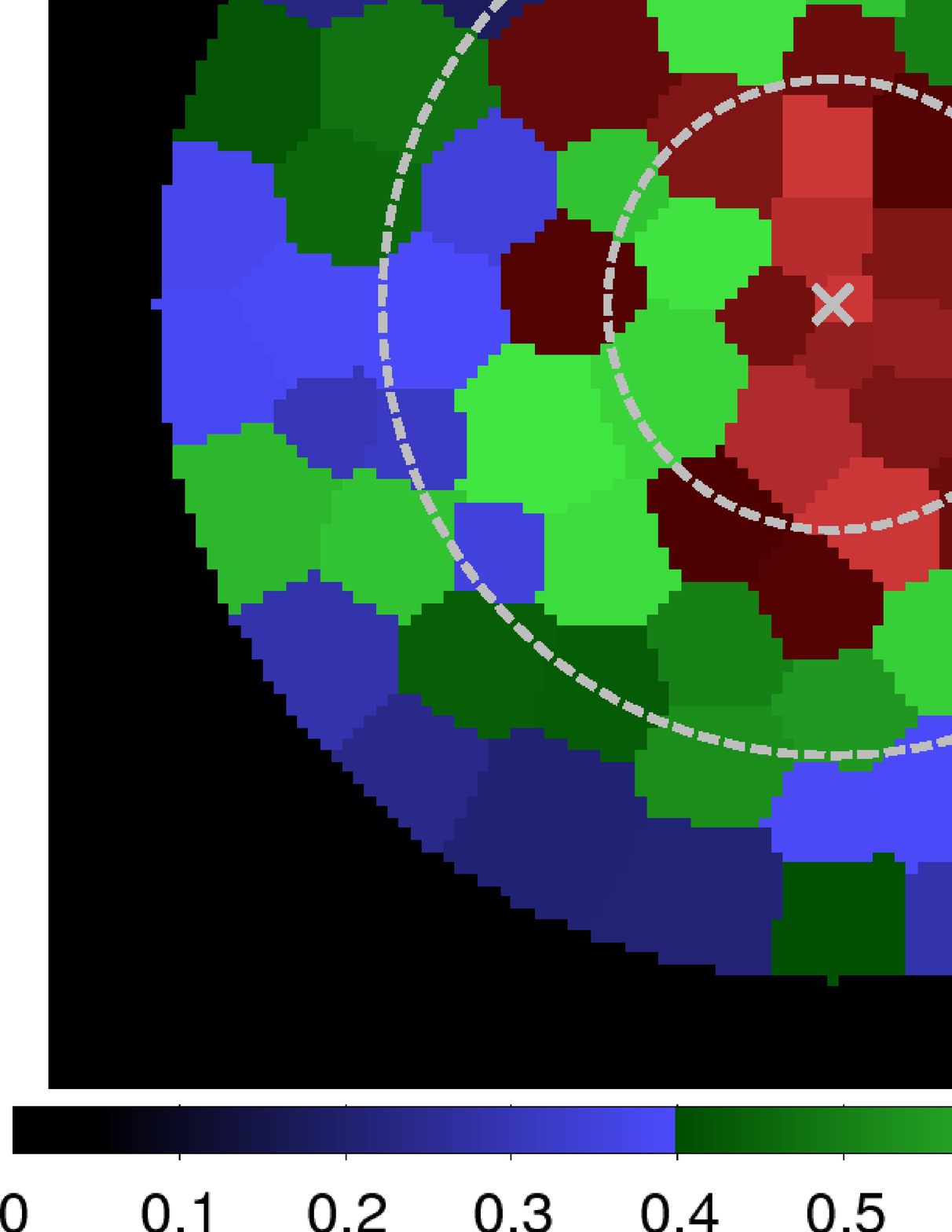}}
{\includegraphics[angle=0,width=0.3\textwidth]{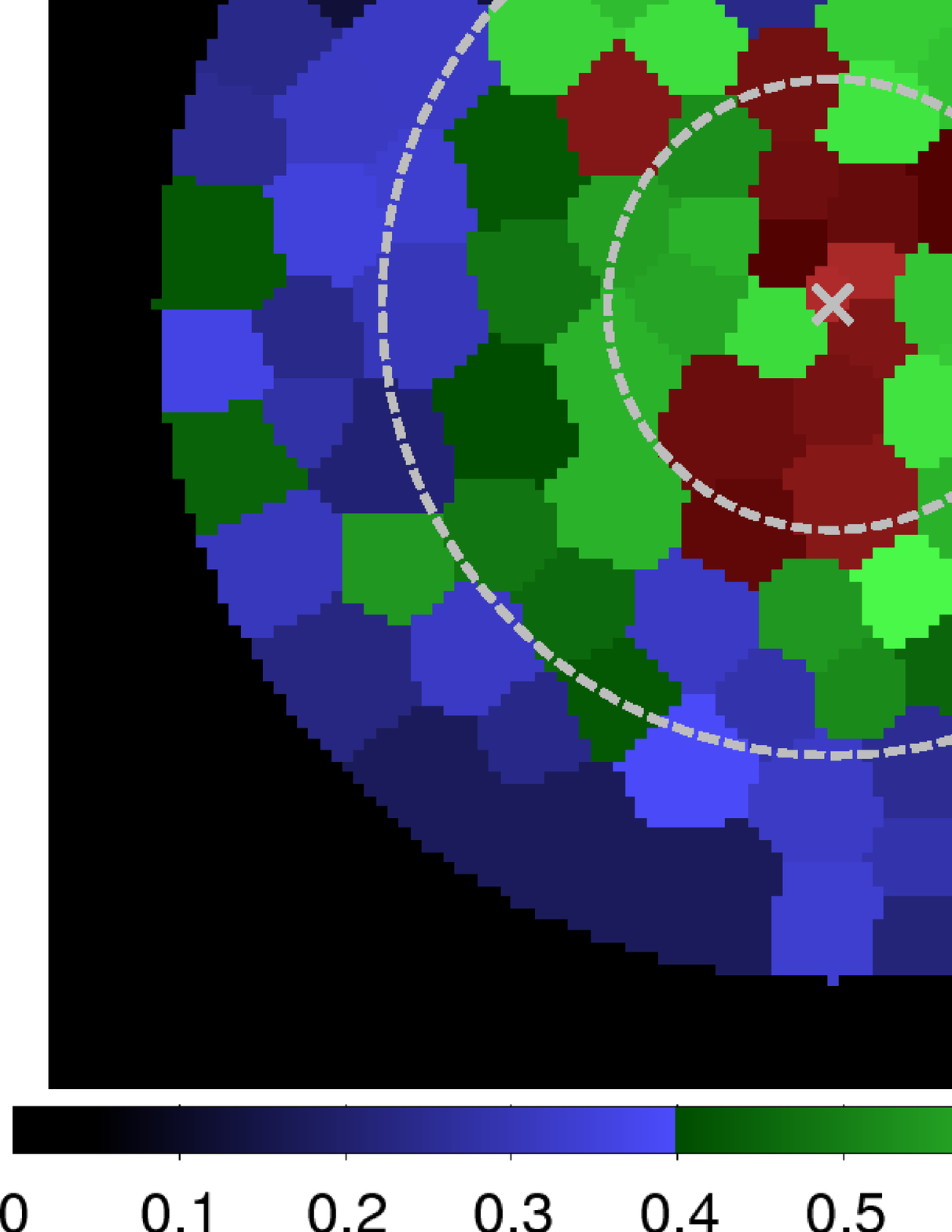}}
\caption{Spectroscopic completeness maps for RXJ1720 for objects with magnitudes
$R \le 20.5$ ({\it left panel}), $R \le 21.5$ ({\it middle panel}) and 
$R \le 22$ ({\it right panel}). The dashed circles have radii of 1 and 2\,Mpc 
and the cross marks the position of the central BCG.}
\label{compl_map}
\end{figure*}

With regard to RXJ1720, the right panel in Figure~\ref{compl_radius} shows that
we achieve spectroscopic completeness above $\sim 85\%$ for magnitudes brighter 
than $R = 20.5$ (apart from magnitudes $19.5 < R \leq 20.5$ at radii outside 
2\,Mpc). At fainter magnitudes than $R =20.5$ our data are limited to the 
central $\sim 2$\,Mpc region covered by the Keck/DEIMOS spectroscopy. For these
observations, the smaller FOV and more limited number of objects observed per 
pointing (compared to the Hectospec observations) meant that it was difficult 
to obtain multiple pointings of the same regions over the entire central area 
within 2\,Mpc of the cluster center. Therefore, the spectroscopic completeness 
at these faint magnitudes is limited, but still respectable, at $> 50\%$ for 
radii $< 1.5$\,Mpc and $20.5 < R \leq 21.5$. As expected, the low priority 
$21.5 < R \leq 22.0$ objects have low spectroscopic completeness ($\sim 40\%$ 
for radii $< 1.5$\,Mpc). Given the more heterogeneous coverage for RXJ1720, we 
further explore the spatial distribution of the spectroscopic completeness. 
Briefly, we use the WVT binning algorithm by 
\citet{diehl2006}, which is a generalization of 
\citeauthor{Cappellari2003}'s \citeyearpar{Cappellari2003} 
Voronoi binning algorithm, to produce an adaptively binned image of the spatial 
distribution of objects in the photometric subsample such that each bin 
contains at least 20 galaxies. This binning is then applied to the spatial
distribution of objects in the spectroscopic sample. The resulting image is
divided by the binned image of the spatial distribution of objects in the 
photometric sample to produce the completeness map. These maps are produced at
three limiting magnitudes of interest: $R \le 20.5,\, R \le 21.5\, 
{\rm and}\, R \le 22.0$ and are shown in Figure~\ref{compl_map}. The leftmost
map in Figure~\ref{compl_map} shows the completeness for the magnitude limit of
the MMT/Hectospec observations and confirms that the spectroscopic completeness 
is excellent and generally above $80\%$. The middle and right panels reveal 
a more patchy spectroscopic completeness which is high in the central 1\,Mpc and
falls off towards the cluster outskirts.

\section{Analysis}\label{anal}

\subsection{Allocation of cluster members}\label{memsel}

\begin{figure*}
{\includegraphics[angle=90,width=0.48\textwidth]{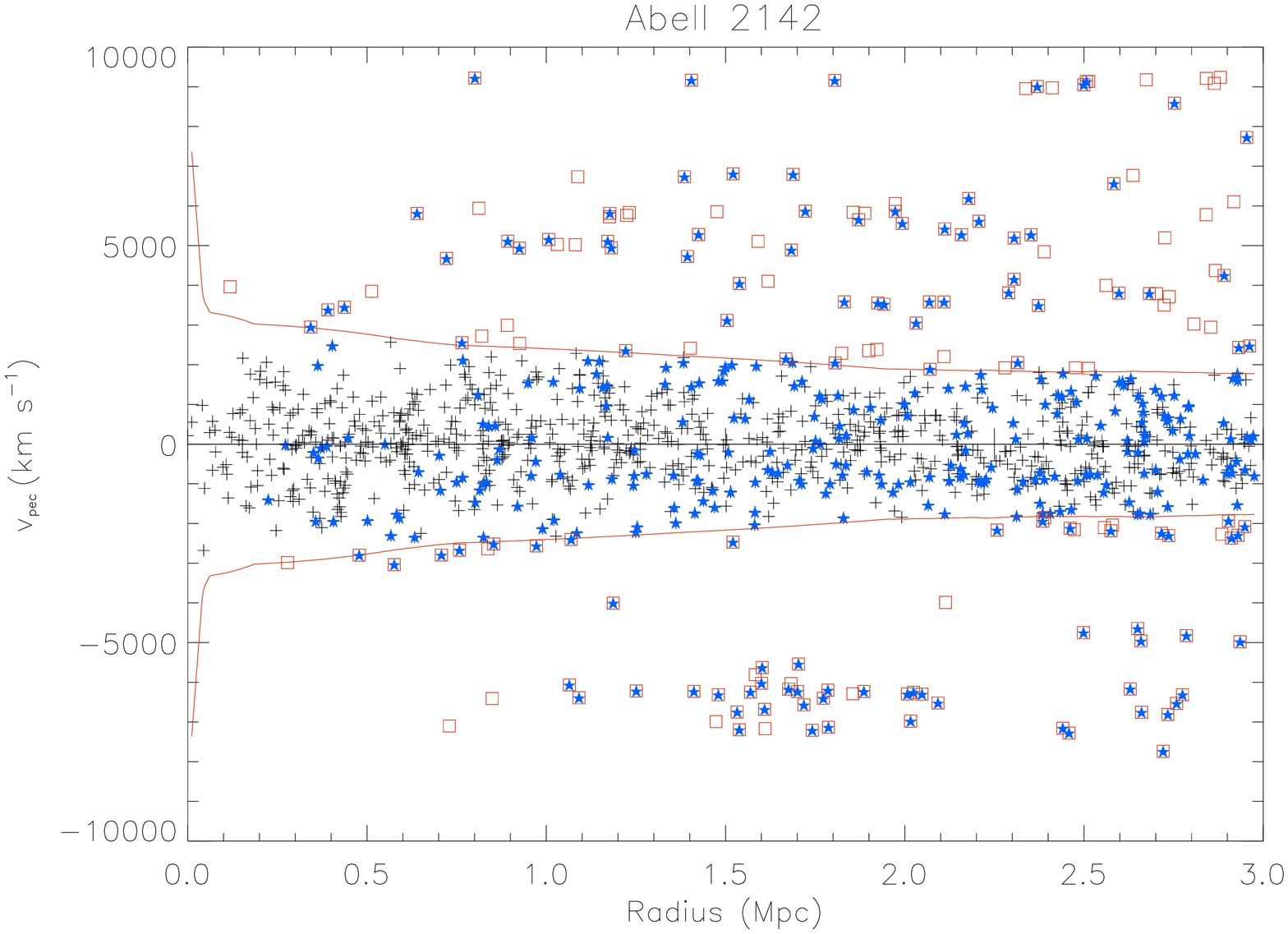}}
{\includegraphics[angle=90,width=0.48\textwidth]{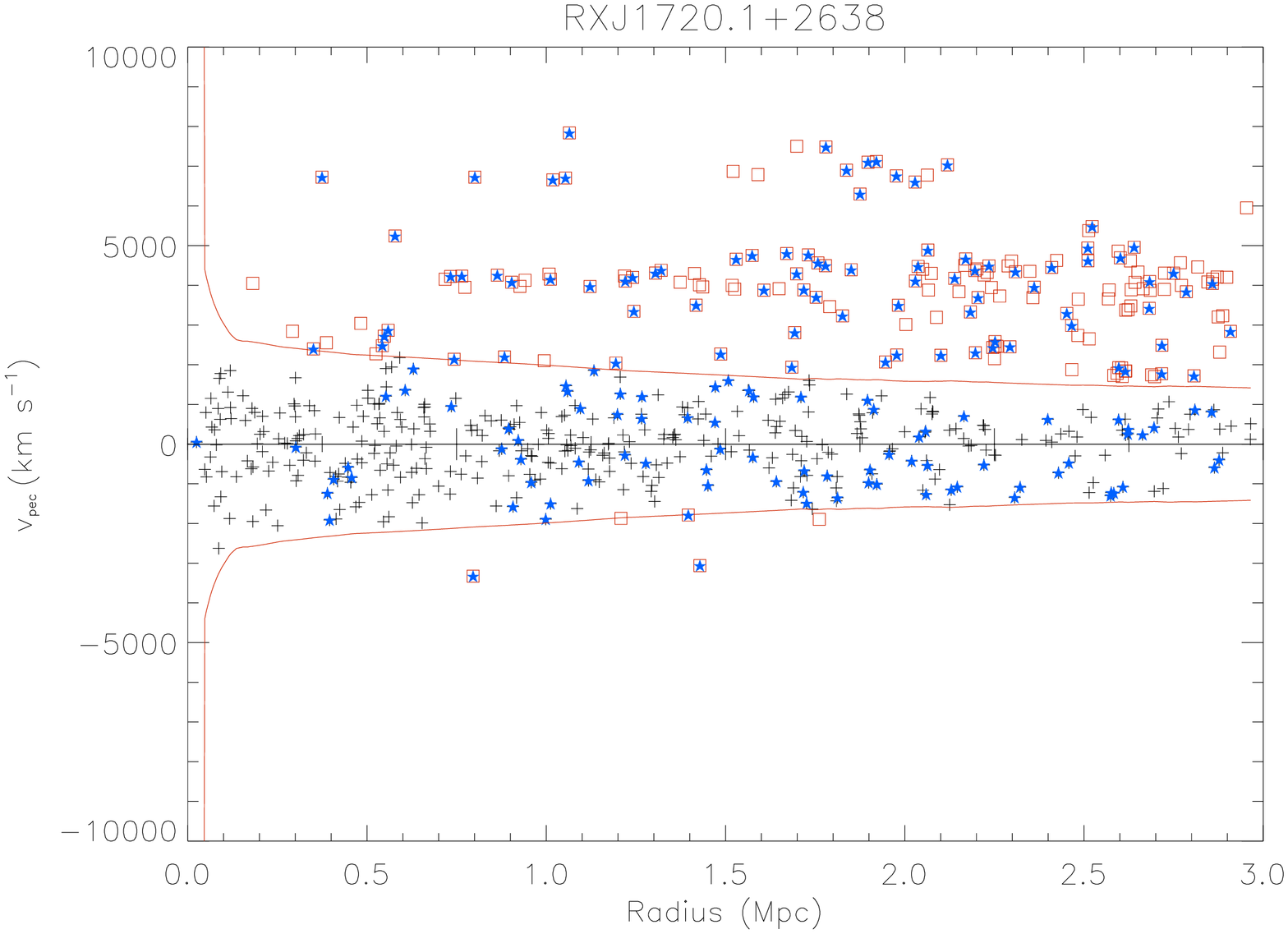}}
\caption{Phase space diagrams used for defining cluster membership for 
A2142 ({\it left panel}) and RXJ1720 ({\it right panel}). The red curves 
show the cluster limits in phase-space and are determined from the cluster mass 
profile (see the text for a description). Black crosses show cluster members 
which lie on the red sequence, while blue stars within the phase space limits 
show those cluster members which lie blueward of the red sequence. Red boxes 
show fore- and background galaxies. Red boxes filled with blue stars show 
non-members which are bluer than the cluster red sequence.}
\label{vpec_rad}
\end{figure*}

We allocate cluster membership using a two-step procedure. In the first step, 
crude cluster membership is achieved by selecting galaxies within 3\,Mpc of the 
brightest cluster galaxy (BCG) and with a peculiar velocity of $\pm 10000$\kms\ 
with respect to the BCG redshift. The shifting gapper method used previously in 
\citet{owers2009b,owers2009a} does not adequately remove the remaining 
interlopers. This is because the depth and completeness of the A2142
redshift catalog means gaps in velocity phase space are more likely to 
be ``filled'', while the proximity of a nearby structure in velocity space for
RXJ1720 causes the shifting gapper method to fail (Figure~\ref{vpec_rad}). 
Therefore, we employ the parametric technique introduced by 
\citet{denhartog1996} where the maximum possible velocity at the projected 
radius of each galaxy is calculated given the mass profile of the cluster, which
is derived from the virial theorem 
\citep[for a detailed discussion see][]{wojtak2007a,wojtak2007b}. Determination
of the virial mass relies critically on the assumption that the particles 
tracing the underlying gravitational potential are virialized and that their 
distribution is spherically symmetric. The most likely galaxies for which these
assumptions apply are elliptical and S0 type galaxies, while spiral galaxies 
are more likely to be either line of sight interlopers, or an infalling 
population on radial orbits and, therefore, not virialized 
\citep{biviano1997, biviano2004}. For these reasons, 
we use only galaxies on the red sequence shown in Figure~\ref{redseq}, which 
we assume is primarily composed of elliptical and S0 galaxies, to determine the 
mass profiles used in 
defining the velocity limits for cluster membership. Furthermore, to reduce
any biases in the measurements of the mean projected separations used for the 
virial mass estimator, we restrict the red sequence galaxies to a magnitude 
range with high spectroscopic completeness (i.e., $R < 19.5$ for RXJ1720 and 
$R < 19.0$ for A2142). The results of the membership allocation are presented 
in Figure~\ref{vpec_rad} where it can be seen that this method of interloper 
removal does an excellent job. For reference, we highlight the galaxies lying
on the blue cloud using blue stars in Figure~\ref{vpec_rad}. We define blue 
cloud galaxies as those lying blueward of the lower limit defining the red 
sequence in Figure~\ref{redseq}. Inspection of the distribution of blue cloud 
galaxies confirms that they have a very different phase-space distribution when 
compared to the red sequence galaxies, thus justifying our decision not to use 
them when estimating the virial mass. We note
that our procedure rejects a number of galaxies which lie close to the 
boundaries defined by the maximum velocity profile but also appear to be well
separated from the more clear-cut non-members. These consist primarily of
blue cloud galaxies and, given their large peculiar velocities, are likely
to be galaxies on radial orbits on their first passage onto the cluster. We
prefer the more conservative cut here, since these galaxies appear to be 
distributed evenly in phase space and are thus unlikely to be part of the
substructures we seek to detect, while their inclusion in the sample may affect
the tests we plan to use for detecting substructure in the later sections.

The final cluster member catalogs contain 956 and 400 members for A2142
and RXJ1720, respectively. From the ensemble of member redshifts we
use biweight estimators \citep{beers1990} to measure cluster redshifts of
$z_{\rm clus}=0.09005\pm0.00012$ and $z_{\rm clus}=0.16010\pm0.00018$ and peculiar 
velocity dispersions $\sigma_{\rm vpec}=995\pm21$\kms\ and 
$\sigma_{\rm vpec}=882\pm29$\kms\ for A2142 and RXJ1720, respectively, where
the uncertainties are $1\sigma$ and are estimated using the jack-knife 
resampling approach \citep{beers1990}.

\subsection{Substructure Detection}

Ultimately, our goal in this study is to test the hypothesis that cold fronts
in clusters with otherwise regular X-ray morphologies can be caused by the
gravitational perturbation of an infalling substructure during a minor merger
\citep{markevitch2001,ascasibar2006,markevitch2007,owers2009c,roediger2011}. 
To that end, we use our samples of spectroscopically confirmed cluster members 
to search for merger-related substructures in the 1D velocity, 2D spatial and
3D velocity-plus-spatial distributions.

\subsubsection{Testing the shape of the peculiar velocity distribution}\label{1D}

\begin{figure*}
{\includegraphics[angle=90,width=0.45\textwidth]{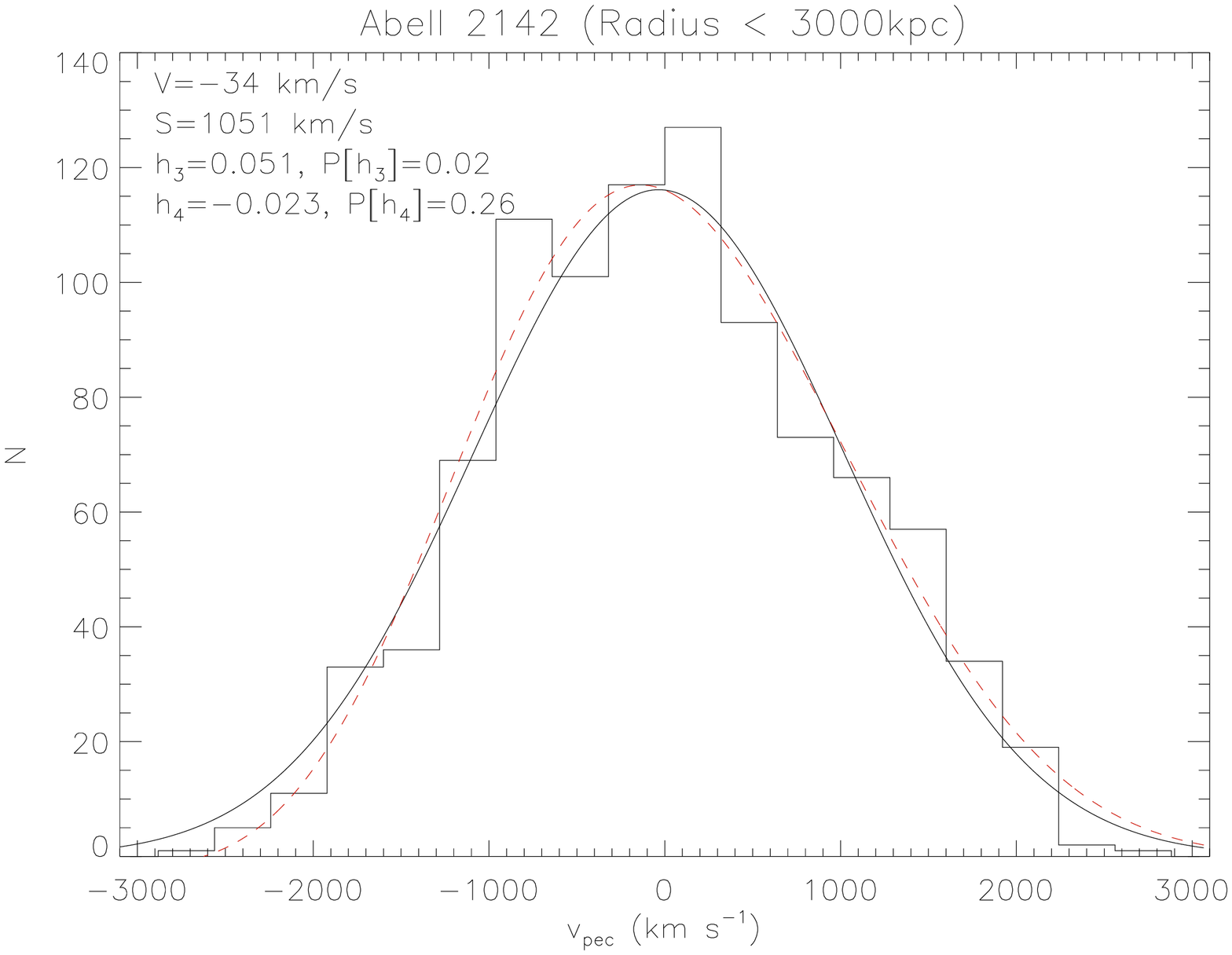}}
{\includegraphics[angle=90,width=0.45\textwidth]{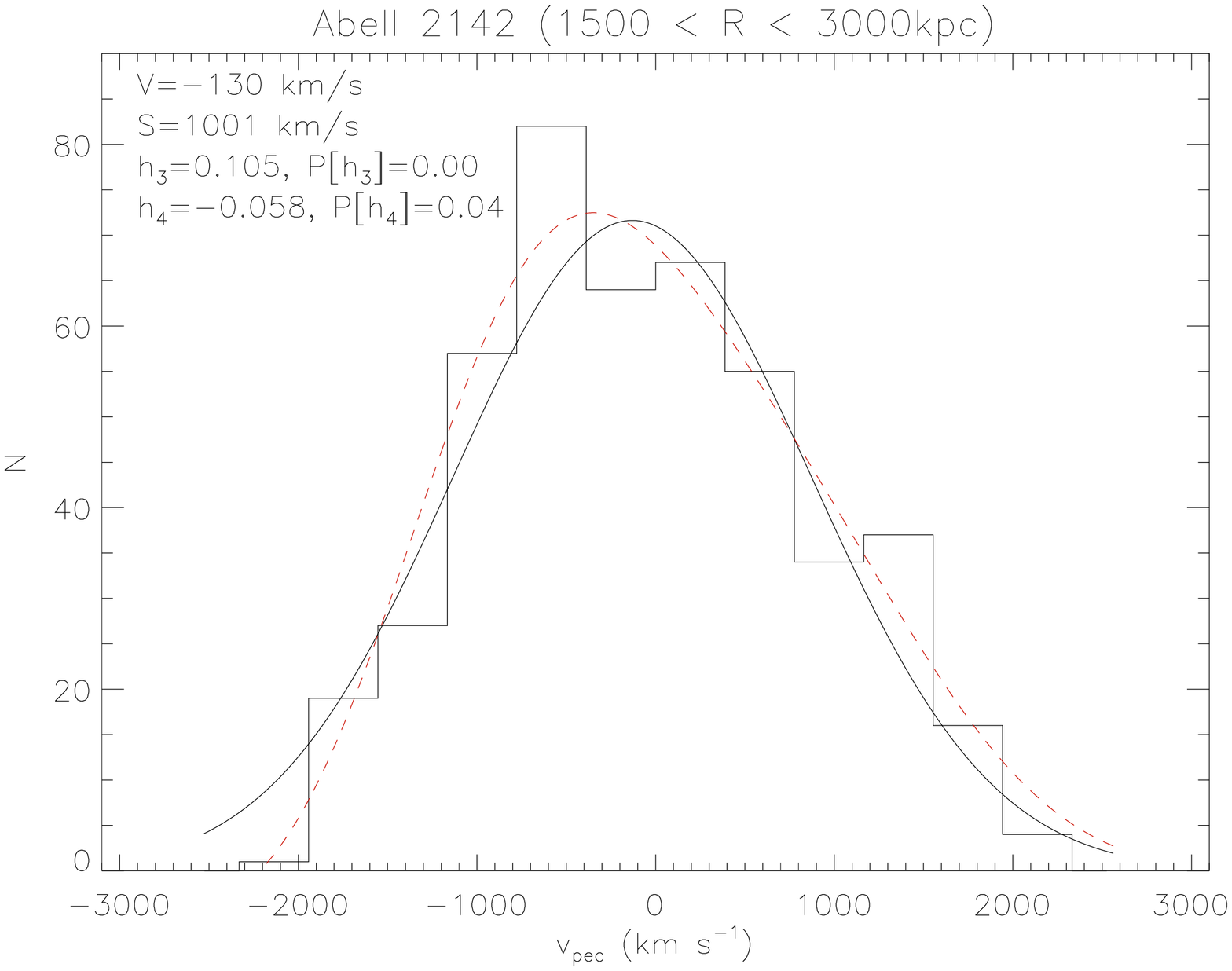}}\\
{\includegraphics[angle=90,width=0.45\textwidth]{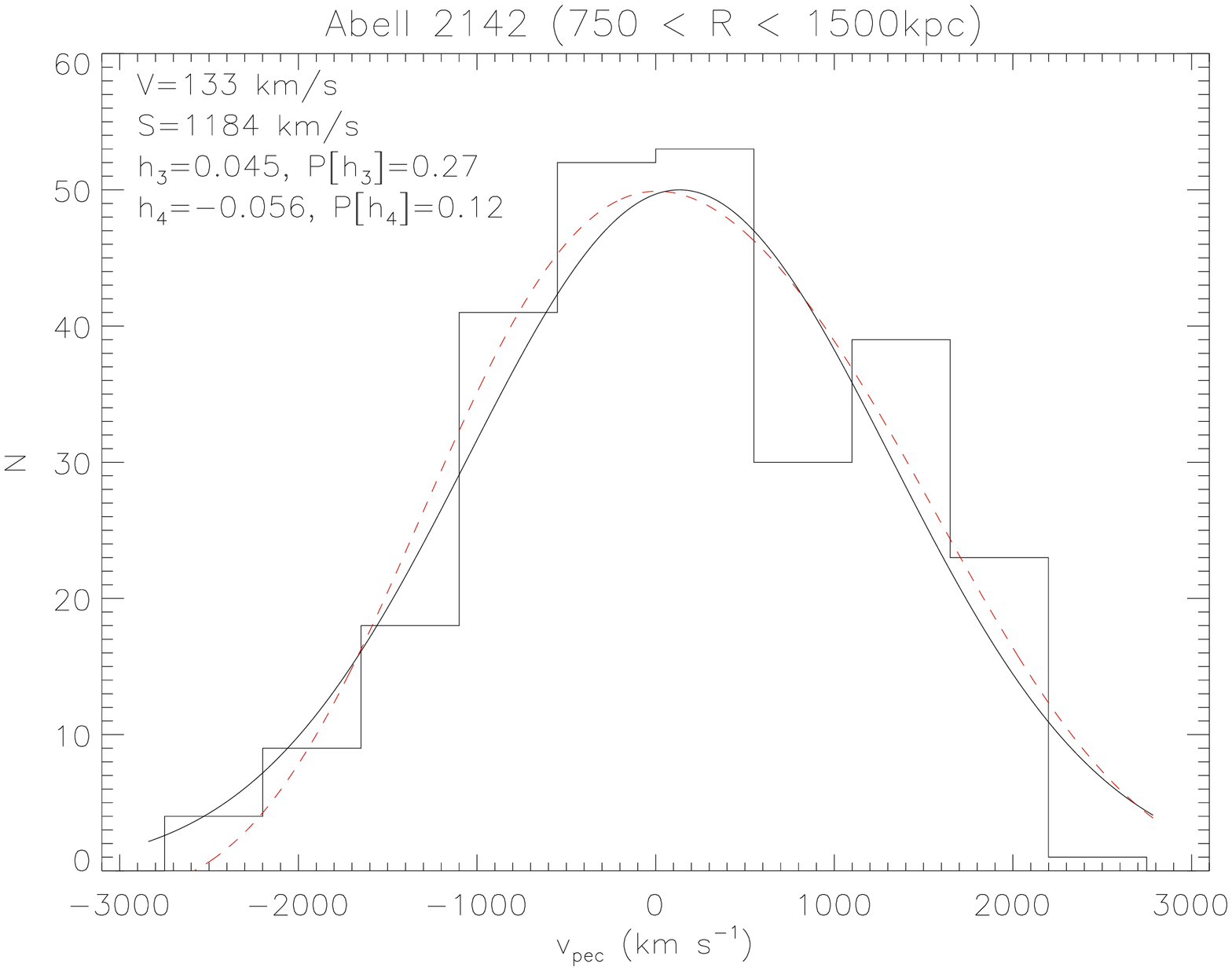}}
{\includegraphics[angle=90,width=0.45\textwidth]{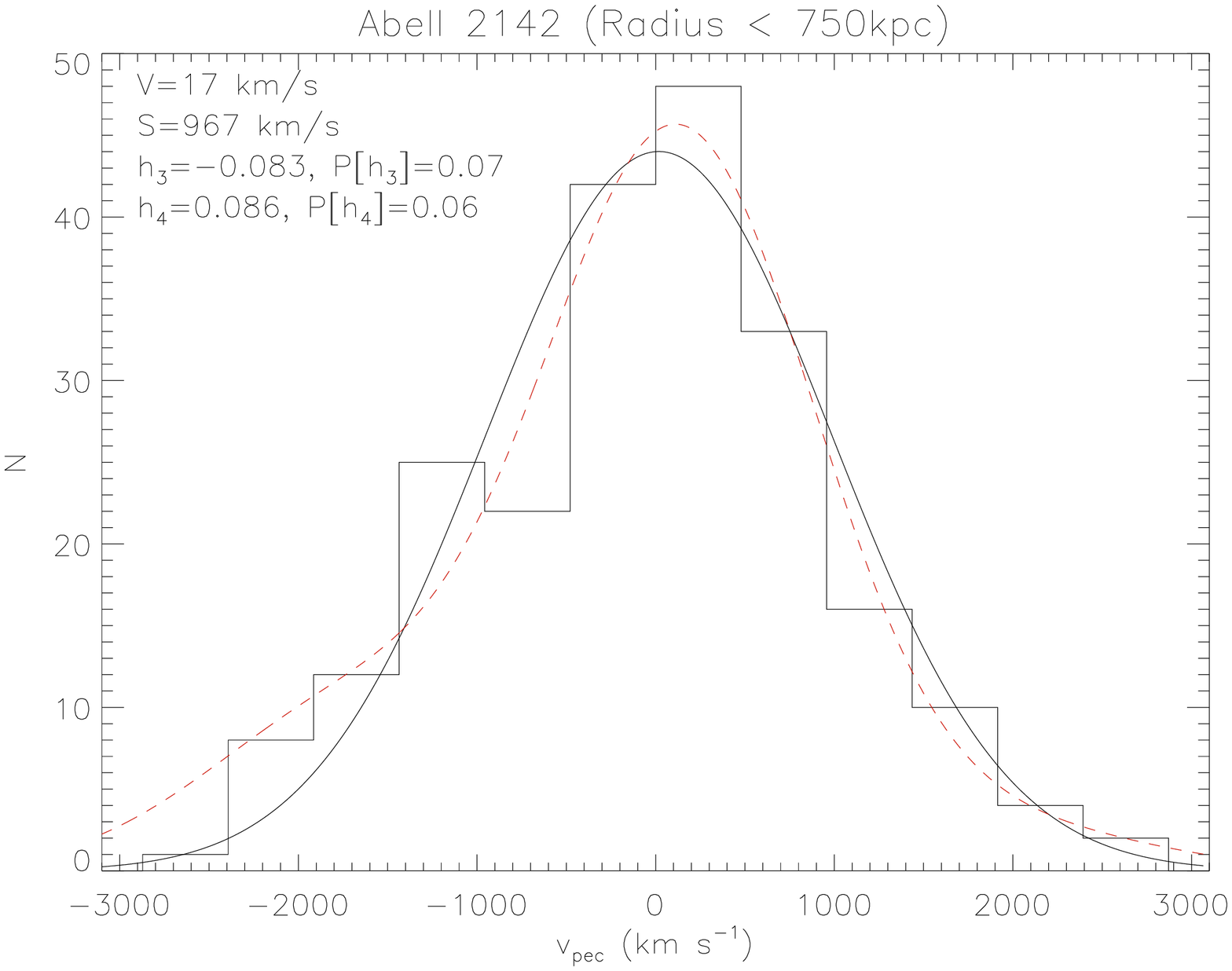}}
\caption{The velocity distribution for all 956 cluster members 
({\it top left panel}), 463 members with $1500 <$\rproj$ < 3000$\,kpc 
({\it top right}), 270 members with  $750 < $\rproj$ < 1500$\,kpc ({\it bottom left}) 
and 223 members with \rproj$ < 750$\,kpc ({\it bottom right}). In each panel, the 
best fitting Gaussian parameters, the Gauss-Hermite terms $h_3$ and $h_4$, 
representing the asymmetric and symmetric deviations from a Gaussian shape, and 
their associated level of significance (see text) are given in the upper right 
corner. The histograms show the observed velocity distribution and the bin size
is set to $3S/(N_{\rm gal})^{(1./3.)}$, where $N_{\rm gal}$ is the number of galaxies 
listed above. The solid black line shows the best fitting Gaussian, while the 
dashed red line shows the Gauss-Hermite reconstriction of the velocity 
distribution.}
\label{a2142_GH}
\end{figure*}

\begin{figure*}
{\includegraphics[angle=90,width=0.45\textwidth]{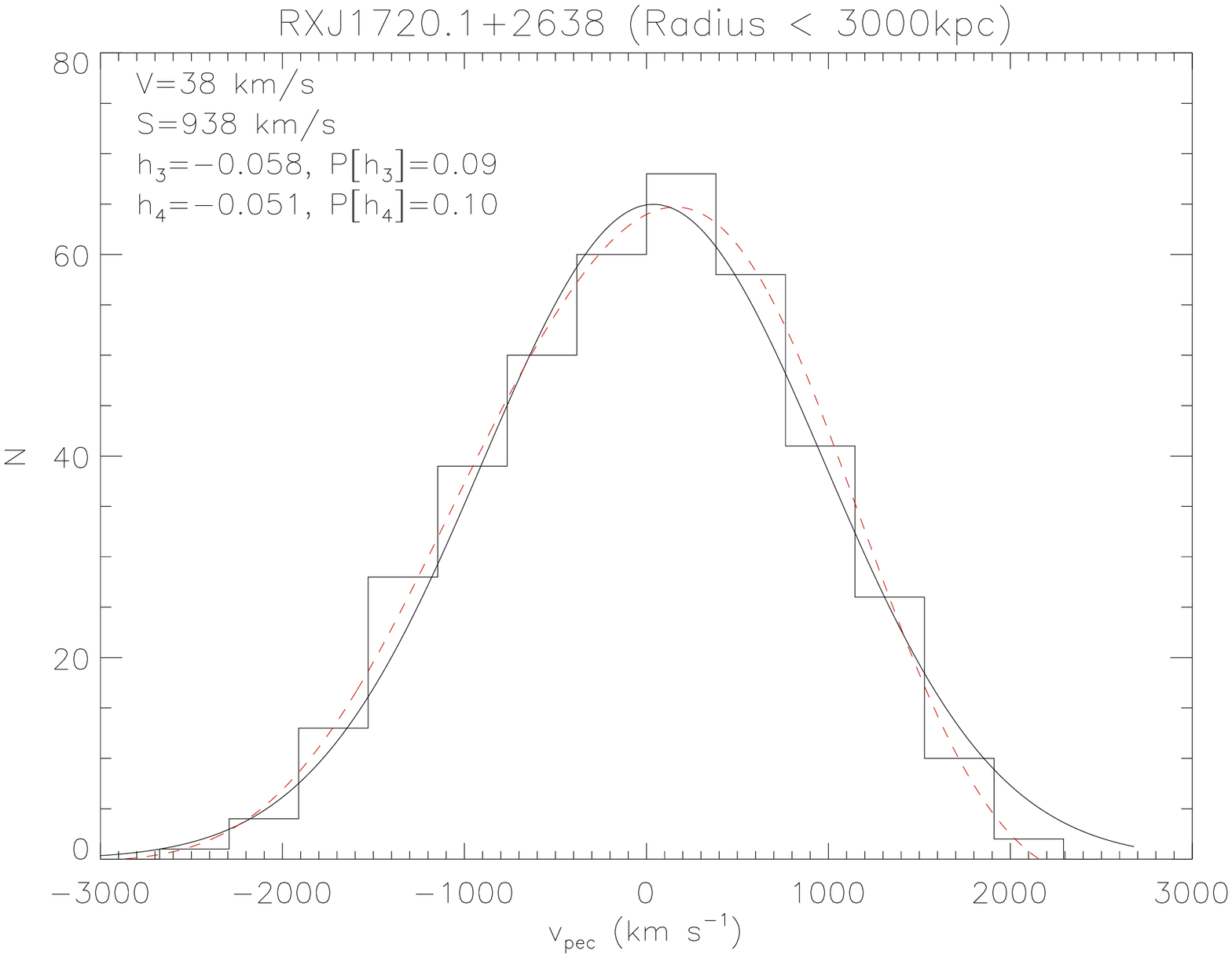}}
{\includegraphics[angle=90,width=0.45\textwidth]{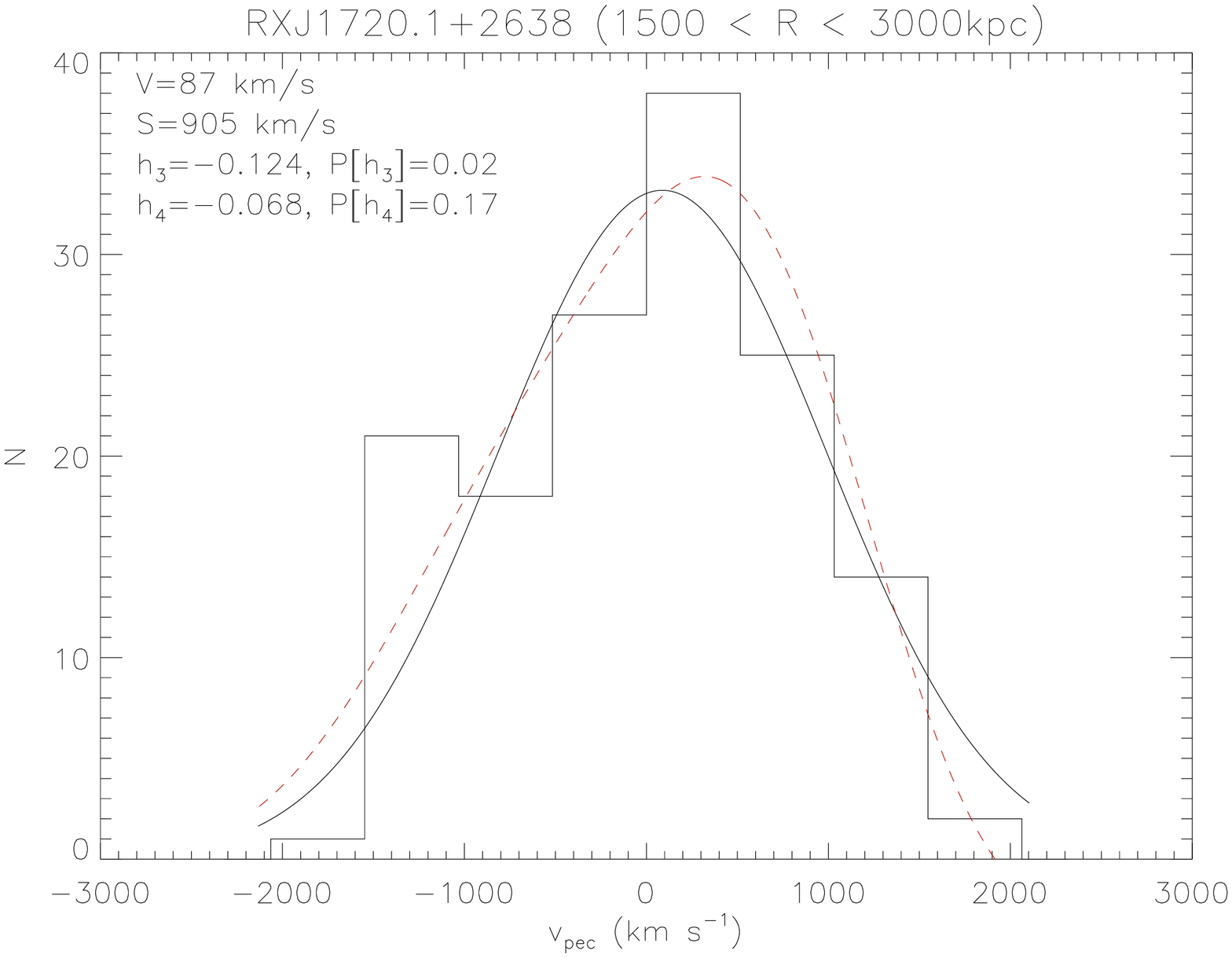}}\\
{\includegraphics[angle=90,width=0.45\textwidth]{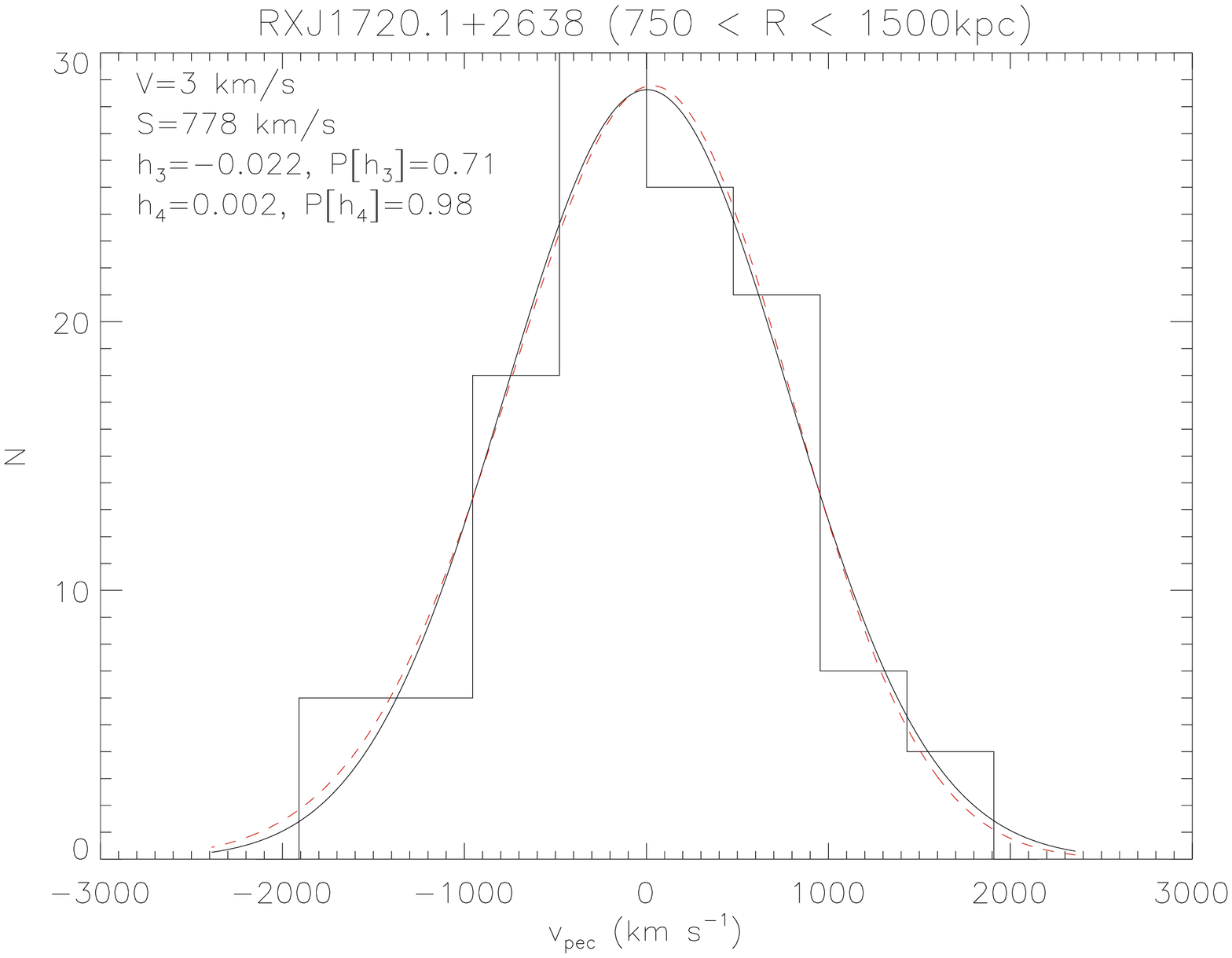}}
{\includegraphics[angle=90,width=0.45\textwidth]{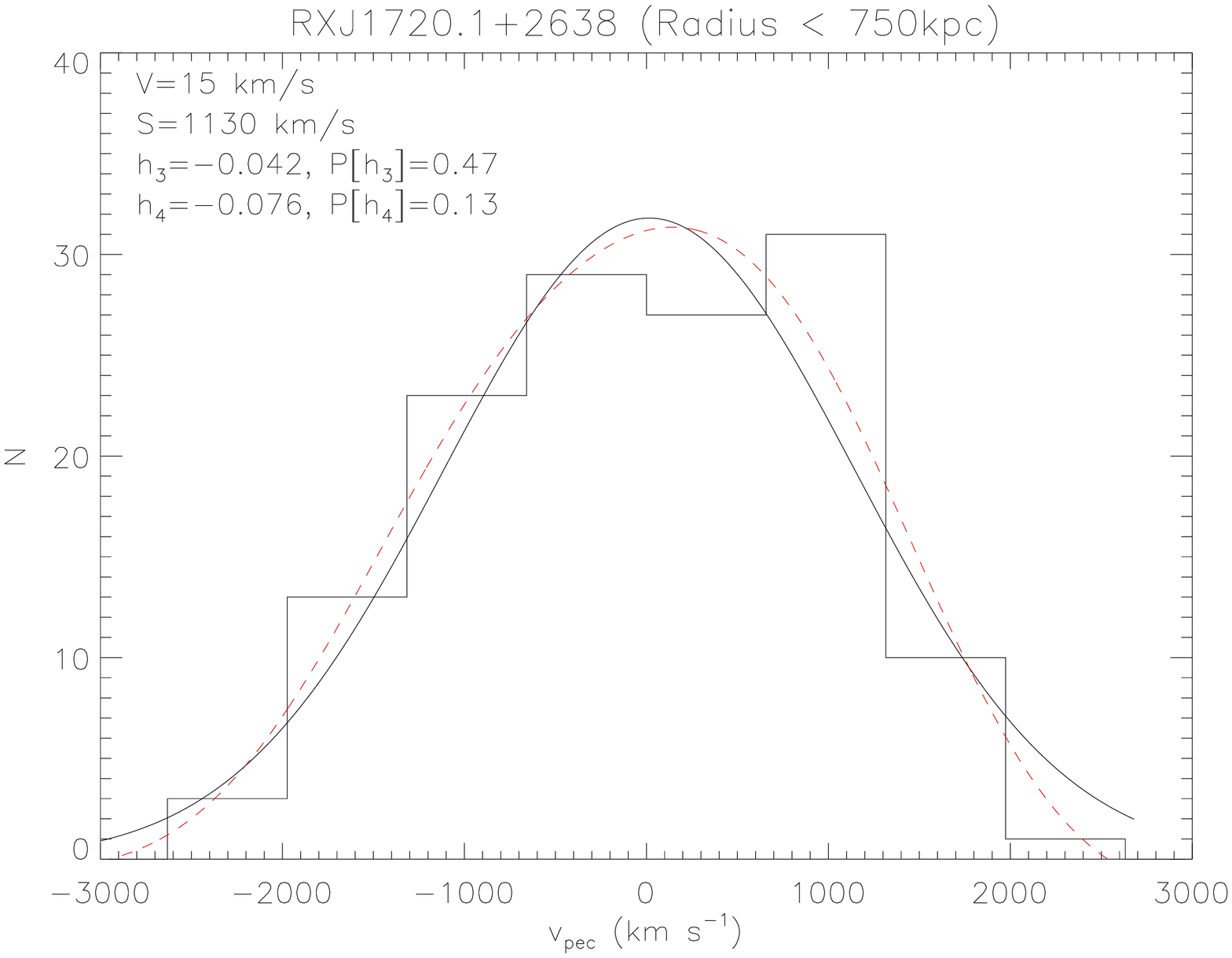}}
\caption{The same as Figure~\ref{a2142_GH} but for RXJ1720. The number
of galaxies in each subsample shown in the {\it top left, top right, bottom left
and bottom right} are 400, 146, 117 and 137 galaxies, respectively.}
\label{rxj1720_GH}
\end{figure*}

A first order test of the dynamical regularity of a cluster comes from the 
shape of the peculiar velocity distribution which, in a dynamically relaxed
cluster, is expected to take an approximately Gaussian shape. Gross departures 
from a Gaussian shape, particularly in the form of a skewness or bimodality, 
often indicate the existence of merger activity 
\citep[e.g.,][]{zabludoff1993,owers2011}. We choose to use the Gauss-Hermite
reconstruction method \citep[see also \citealt{owers2009a}]{zabludoff1993} 
to test for departures from a Gaussian shape. 
Briefly, the velocity distribution is described by a series of Gauss-Hermite 
functions with the Gauss-Hermite moments $h_0 \simeq 1$ multiplying the zeroth 
order term, which is the best fitting Gaussian with mean $V$ and standard 
deviation $S$, while the $h_3$ and $h_4$ terms approximate asymmetric and 
symmetric deviations from a Gaussian. Radial changes in the velocity 
distribution profile, e.g., the decrease in the velocity dispersion with radius 
observed in many clusters \citep{denhartog1996}, can cause significant $h_4$ 
terms which are not related to merger activity. Our aim is to detect
merger related substructure rather than non-Gaussian shapes caused by a 
changing velocity dispersion profile so, in addition to measuring the 
Gauss-Hermite distribution for the entire cluster member sample, we also repeat 
the Gauss-Hermite reconstruction for members with 
$1500 < $\rproj$ < 3000,\, 750 < $\rproj$ < 1500\, {\rm and}\,$\rproj$ < 750$\,kpc. To 
determine the significance of the $h_{3,4}$ terms, we generate 10,000 Gaussian 
random distributions with the same number of data points, mean and standard 
deviation as the observed velocity distribution and measure the $h_{3,4}$ value 
for each random velocity distribution. We define $P[h_{3,4}]=2\,{\rm min} 
\{P[h_{3,4}({\rm sim}) < h_{3,4}({\rm obs}),\, P[h_{3,4}({\rm sim}) > 
h_{3,4}({\rm obs})]\}$ where $P[h_{3,4}({\rm sim}) < h_{3,4}({\rm obs})]$ and
$P[h_{3,4}({\rm sim}) > h_{3,4}({\rm obs})]$ are determined from the number of 
times a $h_{3,4}$ value smaller and larger than the observed $h_{3,4}$ term is
measured in the 10,000 random distributions. The results of this analysis are 
shown in Figure~\ref{a2142_GH} for A2142 and Figure~\ref{rxj1720_GH} for 
RXJ1720. 

The A2142 velocity distributions for the radial ranges \rproj
$ < 3000, 1500 <$\rproj$ < 3000$\,kpc have significant positive $h_3$ terms. 
This indicates that 
the peak of the distribution is more negative and that the distribution has more
power in the positive peculiar velocity tails than expected from the 
best-fitting Gaussian. The $1500 <$\rproj$ \leq 3000$\,kpc velocity distribution also
has a significant negative $h_4$ term, which indicates the velocity 
distribution has a broader, more boxy shape than the best fitting Gaussian.
The velocity distribution in the range $750 < $\rproj$ \leq 1500$\,kpc does not appear
to have a shape which differs significantly from a Gaussian one, but we note
that the measured velocity dispersion, $S = 1184$\kms\ is larger than that 
measured at other radii. The peculiar velocity distribution in the central 
750\,kpc returns marginally significant negative $h_3$ and positive $h_4$ values
indicating that there are symmetric and asymmetric distortions present, although
the marginal significances do not allow strong conclusions to be drawn.

The velocity distribution for the entire RXJ1720 member sample (top left panel 
in Figure~\ref{rxj1720_GH}) is well approximated by a Gaussian---small 
deviations are present at the $5-6\%$ level in both the $h_3$ and $h_4$ terms 
but they are only significant at the $10\%$ level. The velocity distribution 
for the cluster members with $1500 <$\rproj$ < 3000$\,kpc, (top right panel in 
Figure~\ref{rxj1720_GH}) has a negative $h_3$ term which is significant at the
$2\%$\, confidence level. This indicates that the velocity distribution in this
radial bin has a peak which is shifted towards positive peculiar velocities and 
a tail with slightly more power at negative peculiar velocities. The remaining
two velocity distributions (lower panels in Figure~\ref{rxj1720_GH}) do not
harbor any significant deviations from a Gaussian shape.

\subsubsection{Substructure in the spatial distribution of galaxies}\label{2d}

Previous studies have shown that clusters in the process of merging may contain
substructures which have little to no significant detectable impact on the 
shape of the peculiar velocity distributions, particularly in cases where the 
merger plane is perpendicular to our line of sight 
\citep{pinkney1996,owers2009a,owers2009b}. In these cases, the substructure may 
be conspicuous as local peaks in the projected surface density of the galaxies 
which indicate the existence of compact, bound subclusters. Here, we utilize our
comprehensive spectroscopically confirmed cluster member samples to search for 
spatially compact substructures.

\begin{figure*}
{\includegraphics[angle=90,width=0.45\textwidth]{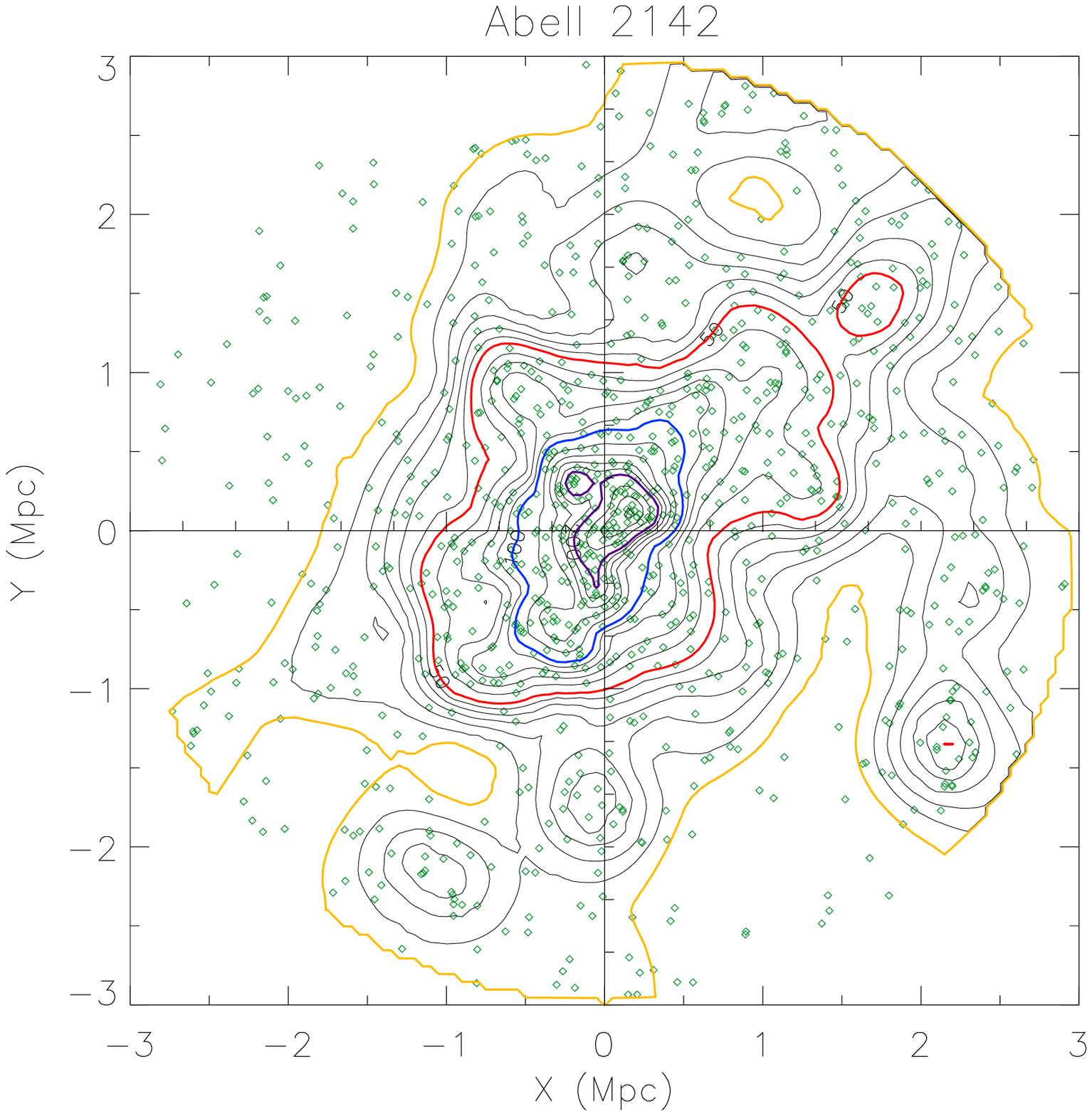}}
{\includegraphics[angle=90,width=0.45\textwidth]{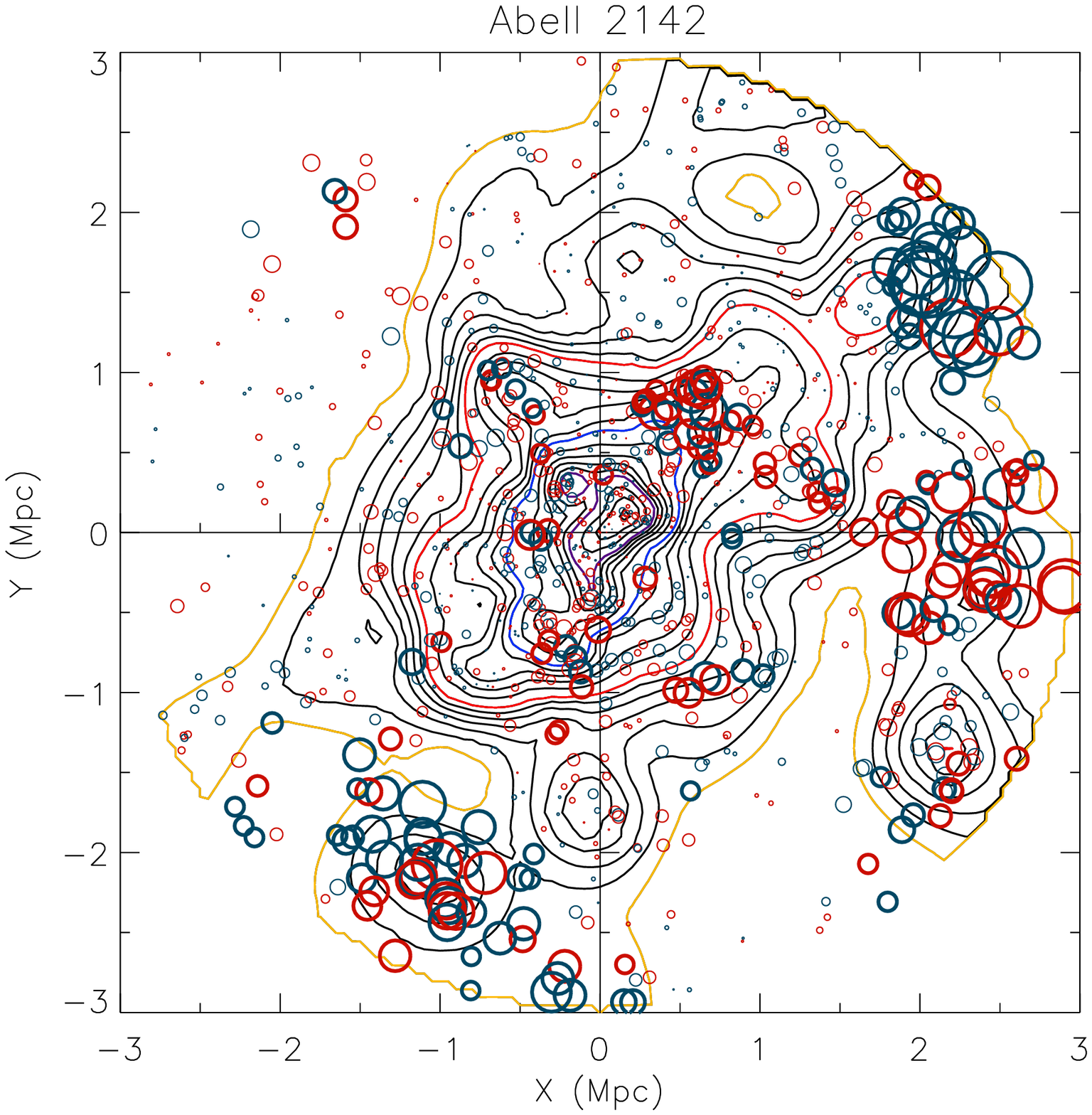}}\\
{\includegraphics[angle=0,width=0.45\textwidth]{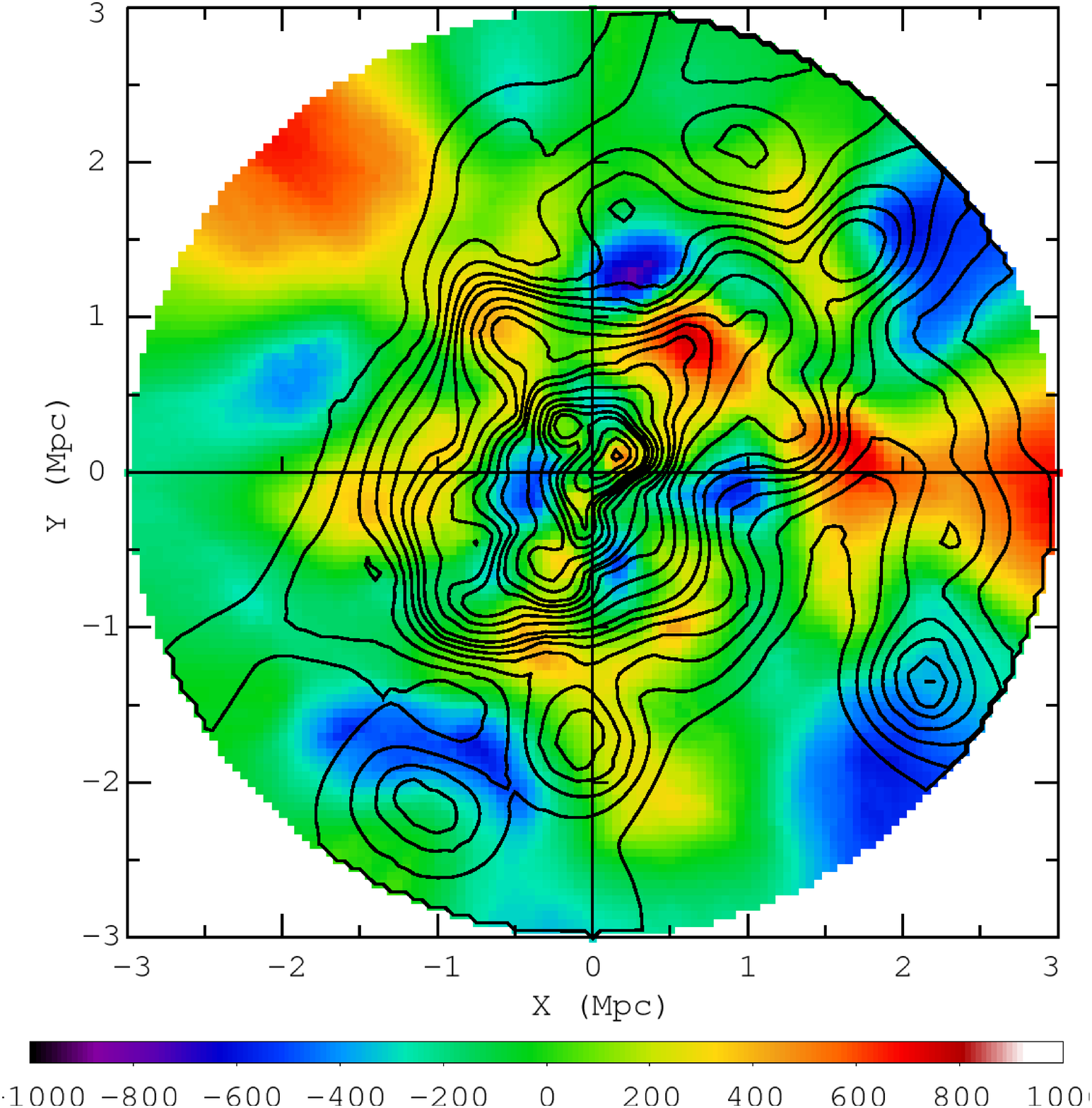}}
{\includegraphics[angle=0,width=0.45\textwidth]{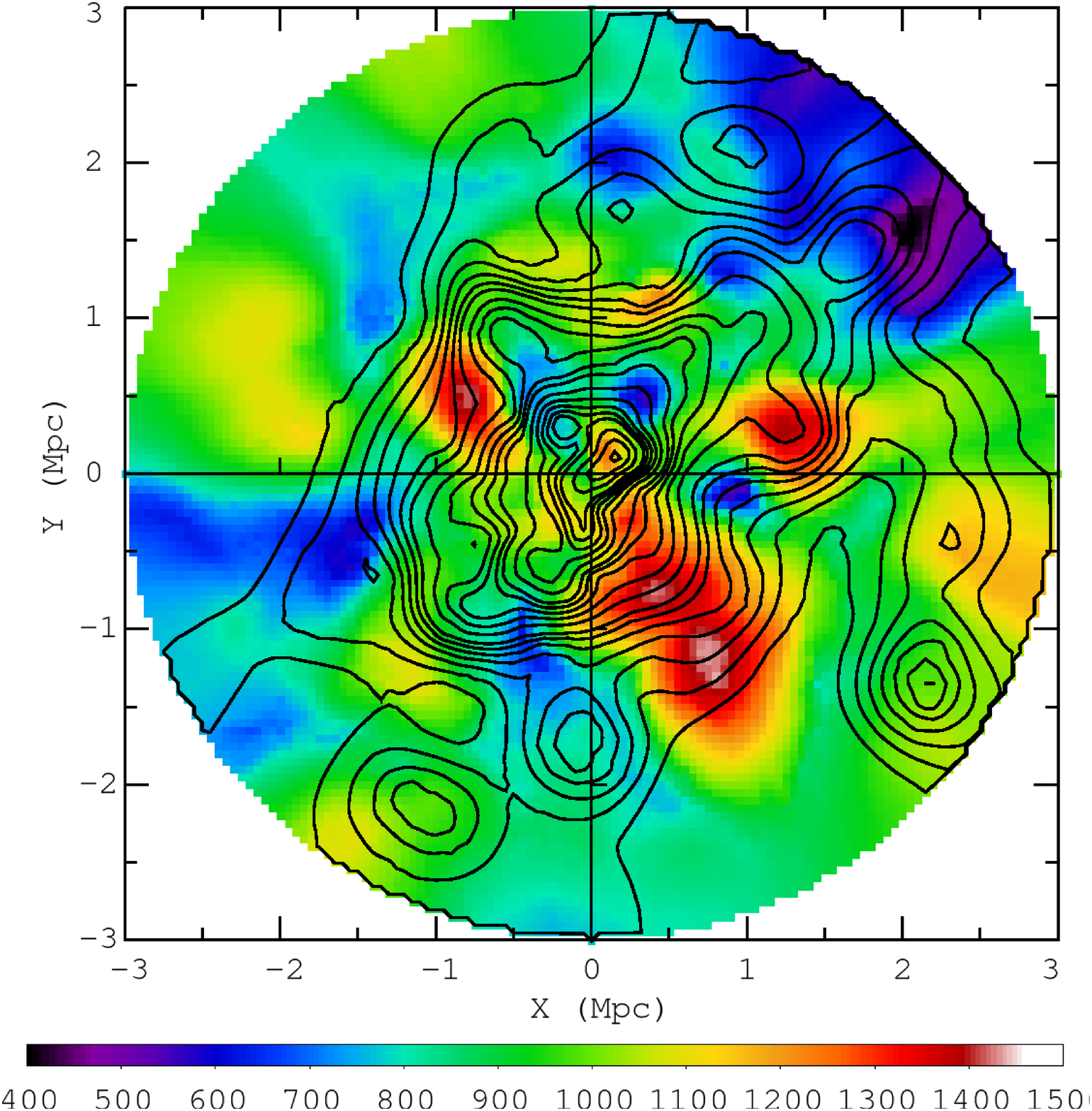}}
\caption{{\it Top left panel:} Isopleths showing the galaxy surface density 
distribution. The levels are not spaced equally, but are spaced at intervals
of 5, 10, 20 and 50 galaxies/Mpc$^2$ between the levels 20-50, 50-100, 100-200 
and above 200 galaxies/Mpc$^2$, respectively. The 20, 50, 100 and 200 
galaxies/Mpc$^2$ levels are colored yellow, red, light blue and dark blue, 
respectively. Green diamonds show the member positions. {\it Top right panel:}
Results of the $\kappa$-test where the circle size gives an indication of the
difference of local velocity distribution compared to the global cluster 
velocity distribution. Clusterings of large emboldened circles indicate 
significant departures. Blue and red circles are centered on galaxies which have
negative and positive peculiar velocities, respectively. {\it Bottom panels:} 
Maps of the mean velocity ({\it left}) and velocity dispersion ({\it right})
fields generated using Equations~\ref{weighted_mean} and \ref{weighted_sigma}. 
The units of the values on the colorbars are \kms.
Contours in the top right and bottom panels are the same as those shown in the 
top right panel.
}
\label{a2142_substructure_maps}
\end{figure*}

\begin{figure*}
{\includegraphics[angle=90,width=0.45\textwidth]{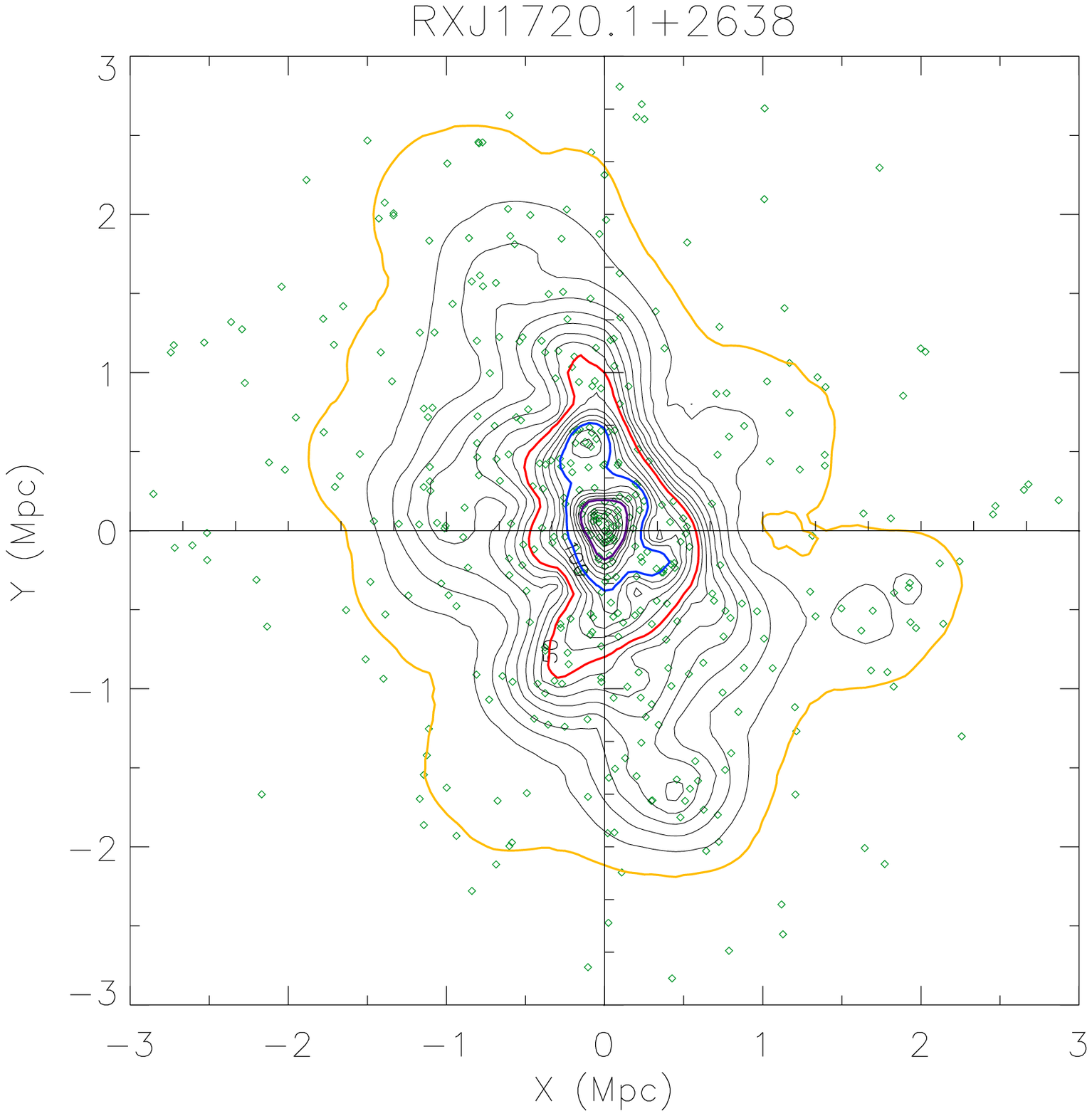}}
{\includegraphics[angle=90,width=0.45\textwidth]{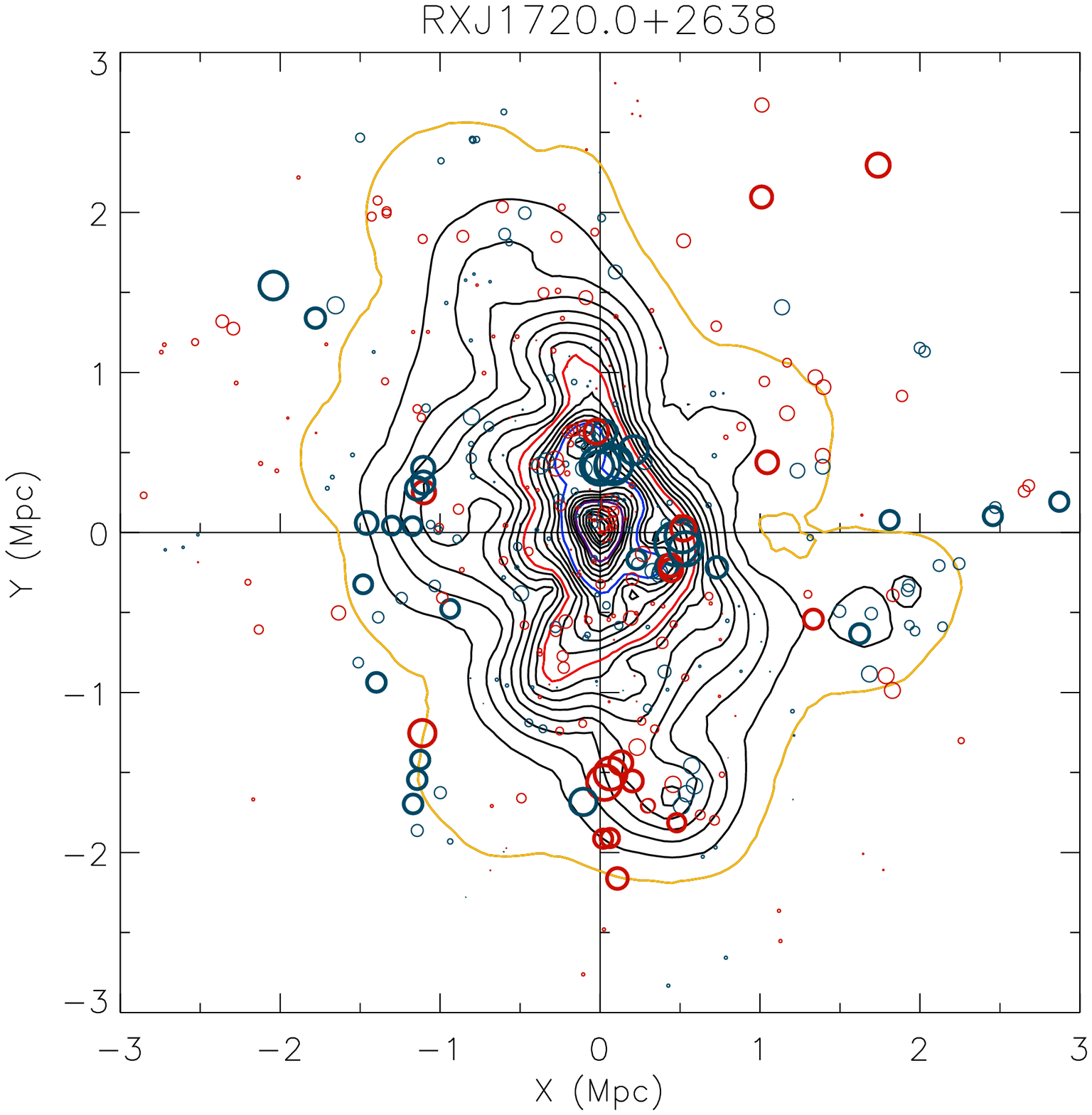}}\\
{\includegraphics[angle=0,width=0.45\textwidth]{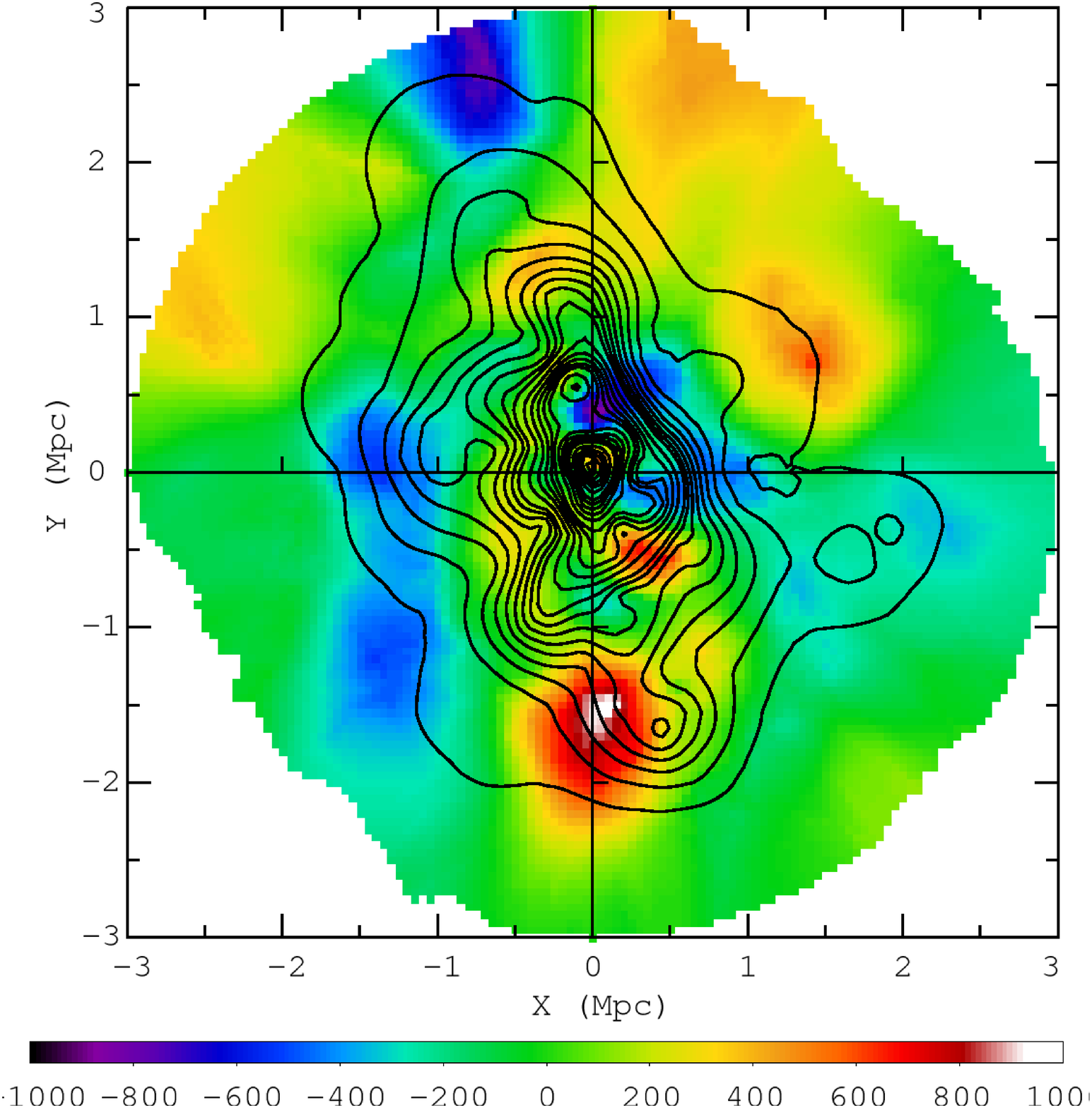}}
{\includegraphics[angle=0,width=0.45\textwidth]{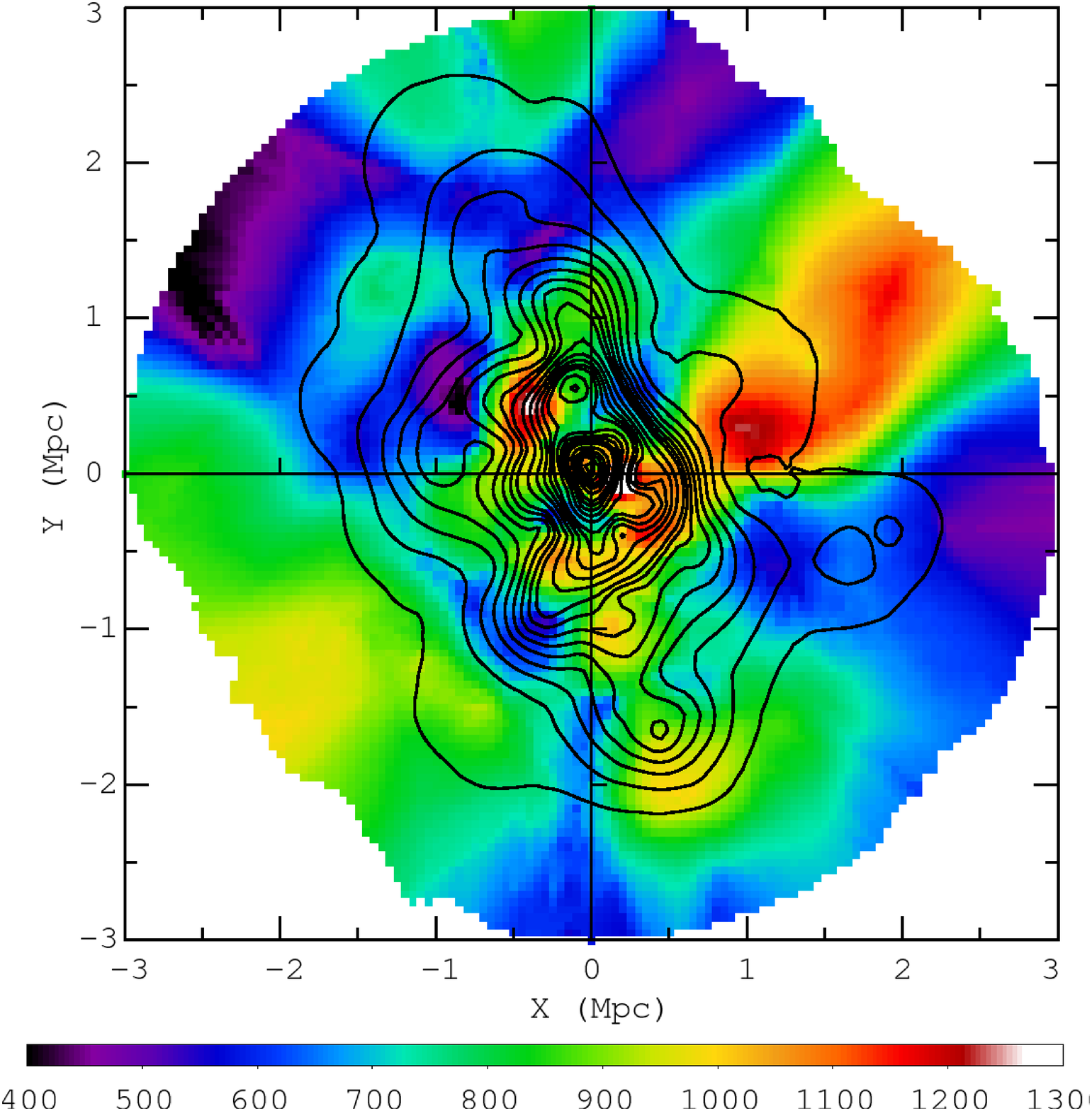}}
\caption{The same as Figure~\ref{a2142_substructure_maps} but for RXJ1720. Here, the 
lowest isopleth is 10 galaxies/Mpc$^2$.}
\label{rxj1720_substructure_maps}
\end{figure*}

Clusters exhibit a large dynamic range in their projected galaxy surface 
density distributions. Therefore, in order to both resolve spatially close 
substructures and minimize the effects of noise in low density regions, an 
adaptive smoothing kernel is required. We use a Gaussian kernel with FWHM equal 
to the radius of the $N$th nearest neighbor in projection, where 
$N=\sqrt{(n_{\rm mem})}$ and $n_{\rm mem}$ is the number of members in 
the sample, to smooth the projected galaxy distribution. The results of the 
smoothing are shown in the top left panels of 
Figures~\ref{a2142_substructure_maps} and \ref{rxj1720_substructure_maps} for 
A2142 and RXJ1720, respectively. Both clusters exhibit a dense core of 
galaxies surrounding the central BCG, elongated morphologies on larger scales
and a number of local overdensities. The central regions of A2142 appear to be 
irregular, with a number of local overdensities within 1\,Mpc of the cluster 
center, the most significant of which occurs $\sim 180$\,kpc to the NW, roughly 
coincident with the second BCG. There are also four large conglomerations of 
galaxies in the outskirts of A2142; one $\sim 2.4$\,Mpc to the SE, one 
$\sim 1.7$\,Mpc due south, one $\sim 2.5$\,Mpc to the SW and one $2.2$\,Mpc to 
the NW. Most notably for RXJ1720, there is a second peak in the projected
galaxy density slightly east of due north and $\sim 550$\,kpc from the cluster
center, while there is a less significant overdensity $\sim 380$\,kpc to the 
SW. There is also an overdensity $\sim 1.7$\,Mpc just west of south in the 
outskirts of RXJ1720.


\subsubsection{Substructure in the local kinematics}\label{3D}

Having identified evidence for substructure in both the velocity and spatial 
distributions, we now search for correlations between the local kinematics and 
the projected local overdensities in the galaxy distribution. As a first attempt
at this, we utilize the $\kappa$-test
which identifies kinematic substructures by comparing the local velocity
distribution of the $N=\sqrt{n_{\rm mem}}$ nearest neighbors around each galaxy to 
the global velocity distribution \citep{colless1996}. The comparison is 
quantified using the Kolmogorov-Smirnov (KS) test, which returns the likelihood,
$P_{\rm KS}$, that the local and global velocity distributions are drawn from the 
same parent distribution. The overall level of substructure within the cluster 
can be quantified by summing the individual $-log[P_{\rm KS}]$ to obtain the 
$\kappa$-statistic. The significance of the $\kappa$-statistic is obtained by 
comparing the observed value to the distribution of 10,000 remeasurements of 
the $\kappa$-statistic which have been obtained from samples where the member 
velocities are randomly redistributed amongst the positions, which are held 
fixed, thus erasing coherent dynamical substructure. For A2142 and RXJ1720
we measure $\kappa$-statistics of 836 and 256 at a significance of $5.6\sigma$ 
and $2.5\sigma$, respectively, indicating that A2142 harbors significant 
dynamical substructure, while RXJ1720 shows evidence for dynamical substructure,
although at a lower significance level.

The results of the $\kappa$-test for the full field of view for A2142
and RXJ1720 are shown in the top right panels of 
Figures~\ref{a2142_substructure_maps} and \ref{rxj1720_substructure_maps}, 
respectitvely. At the position of each member galaxy, we plot a circle 
with radius $R \propto -log[P_{\rm KS}]$. Each circle is color coded so that members
with positive $v_{\rm pec}$ are red while members with negative $v_{\rm pec}$ are blue. 
Furthermore, those members having $-log[P_{\rm KS}]$ values occurring 
in only $5\%$ of the random realizations are plotted with thick lines. 
Clusterings of these large, emboldened circles indicate that there
exists localized velocity substructure. For comparison to the substructures 
revealed in Section~\ref{2d}, we also overplot the corresponding surface
density contours. 

\begin{figure*}
{\includegraphics[angle=90,width=0.45\textwidth]{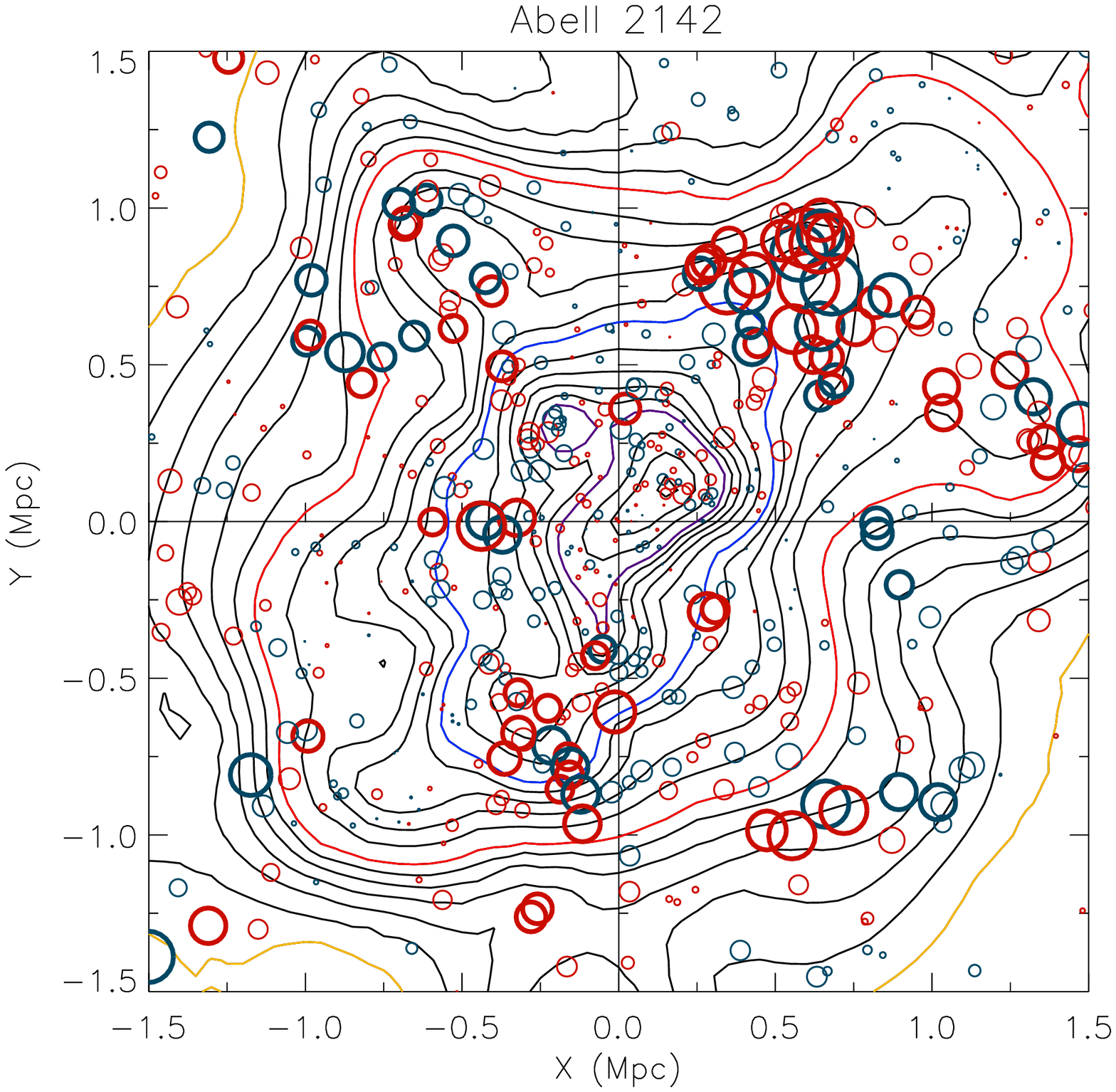}}
{\includegraphics[angle=90,width=0.45\textwidth]{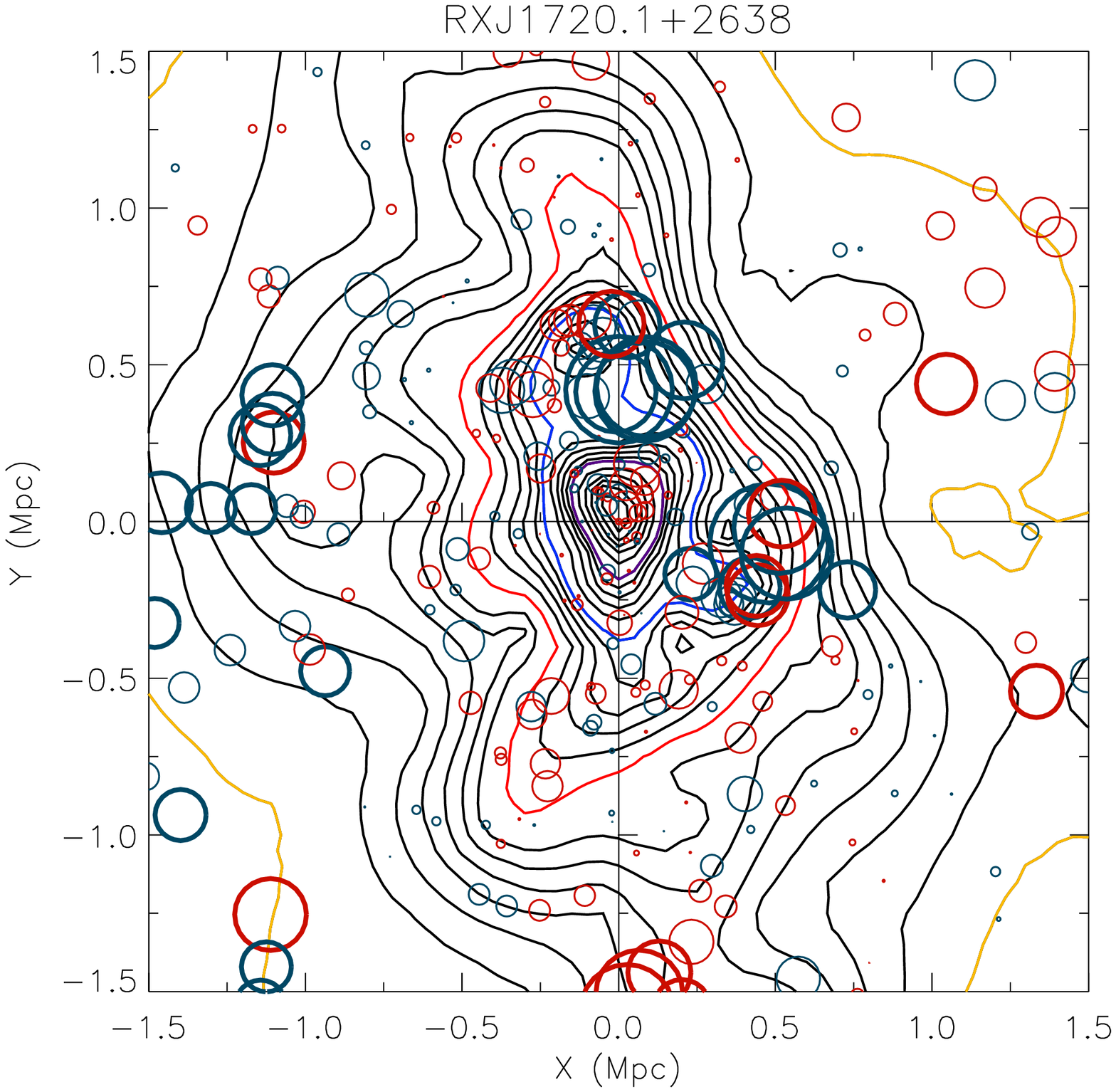}}
\caption{Zoomed version of the {\it top right} panels in 
Figures~\ref{a2142_substructure_maps} and \ref{rxj1720_substructure_maps} 
showing the central 1.5\,Mpc regions.
}
\label{zoomed_kstat}
\end{figure*}

Within a 1\,Mpc radius of the center of A2142 
(left panel of Figure~\ref{zoomed_kstat}), there is one region
of large clustered circles $\sim 1$\,Mpc to the NW which is coincident with an 
extension in the galaxy surface density contours. There is no significant 
clustering of large circles coincident with the local galaxy overdensity 
associated with the second BCG $\sim 180$\,kpc to the northwest. We note, 
however, that this does not rule out the existence of a compact dynamical 
substructure here, particularly given 
the enhanced projection effects due to the close proximity to the main cluster 
core which serve to dilute differences in the velocity distribution of a real
substructure and the main cluster. In the cluster outskirts (top right panel of 
Figure~\ref{a2142_substructure_maps}), 
there are three concentrations of large circles, two of which 
coincide approximately with local overdensities in the spatial distribution
of galaxies. The most significant of these is located $\sim 2.6$\,Mpc to the
northwest and contains a large number of members with negative $v_{\rm pec}$. The
second lies $\sim 2.4$\,Mpc to the SE, while the third, which is not associated 
with a local overdensity in the galaxy distribution, lies $\sim 2.3$\,Mpc
due west. There is a fourth region $\sim 2.4$\,Mpc to the SW and coinciding 
with a local overdensity in the galaxy distribution which contains a more mild
concentration of large circles. Further investigation of this
region is required to confirm the existence of dynamical substructure here.

Considering the central $1$\,Mpc region of RXJ1720 (right panel in 
Figure~\ref{zoomed_kstat}) it can be seen that there is a 
clustering of large circles associated with the prominent 
secondary galaxy surface density substructure located $\sim 550$\,kpc 
to the north of the cluster center. There is also a smaller clustering of large
circles $\sim 380$\,kpc to the SW which is coincident with one of the more minor
substructures in the galaxy surface density distribution, indicating there is 
dynamical substructure there. At larger radii (top right panel of 
Figure~\ref{rxj1720_substructure_maps}), we note that there are a number 
of larger circles associated with the galaxy overdensity $\sim 1.7$\,Mpc just
west of south, although only a fraction of those are deemed to be significantly
larger than expected, indicating that further investigation is required before
concluding that there is local dynamical substructure here.

The $\kappa$-test is an ideal tool for detection of substructure but provides 
little information on kinematic properties which are key to understanding
and characterizing substructure. For the purpose of characterizing these 
substructures, we produce maps tracing the mean velocity and velocity dispersion
fields which are presented in the lower panels of 
Figures~\ref{a2142_substructure_maps} and \ref{rxj1720_substructure_maps} for
A2142 and RXJ1720, respectively. Briefly, these maps were produced by 
generating a grid of $50\times 50$\,kpc pixels and determining at each pixel the
weighted mean, $\overline{v_w}$, 
\begin{equation}\label{weighted_mean}
\overline{v_w} = {\sum\limits_{i=1}^N w_i v_{{\rm pec}, i} \over 
\sum \limits_{i=1}^N w_i}
\end{equation}
and dispersion, $\sigma(v_w)$, 
\begin{equation}\label{weighted_sigma}
\sigma(v_w)^2 = {\sum\limits_{i=1}^N w_i (v_{{\rm pec},i} - 
\overline{v_w})^2 \over \sum \limits_{i=1}^N w_i}
\end{equation}
of the $N=\sqrt{n_{\rm mem}}$ nearest neighbors, where $v_{{\rm pec},i}$ is the peculiar velocity 
of the $i$th near neighbor, $w_i=1-(r_i/R)$, $r_i$ is the radial distance of the
$i$th nearest member from the pixel center and $R=r_{i=N}$. 

Examining the results for A2142 in the bottom two panels of 
Figure~\ref{a2142_substructure_maps} we note that there is a 
general coincidence of regions which contain clustered large circles seen in 
the $\kappa$-test bubble plot and regions where either one or both of the 
$\overline{v_w}$ and $\sigma(v_w)$ fluctuate away from the overall cluster 
values. The region of clustered large circles $\sim 1$\,Mpc to the NW coincides
with a region where $\overline{v_w} \sim 750$\kms\, while the most significant 
clustering of large bubbles located $\sim 2.6$\,Mpc to the NW coincides with a 
region where $\overline{v_w} \sim -600$\kms\ and $\sigma(v_w) \sim 400$\kms. The
substructure $\sim 2.4$\,Mpc to the SE has mainly negative $\overline{v_w}$ values 
which peak at $\overline{v_w} \sim -550$\kms\ just to the north of the local 
peak in the galaxy density. Similarly, the velocity field surrounding the local 
galaxy overdensity $\sim 2.5$\, Mpc to the southwest has values of 
$\overline{v_w} \sim -450$\kms, while the substructure $\sim 2.5$\,Mpc due west has
$\overline{v_w} \sim 500$\kms. These maps also reveal that the kinematics 
immediately surrounding the second BCG $\sim 180$\,kpc NW of the cluster center 
are different, with $\overline{v_w} \sim 500$\kms\ and 
$\sigma(v_w) \sim 1200$\kms.

For RXJ1720 (Figure~\ref{rxj1720_substructure_maps}) we find a 
general agreement between the regions which have fluctuations in either one or 
both of $\overline{v_w}$ and $\sigma(v_w)$ and regions containing the most 
significant local departures from the global velocity distribution according to 
the $\kappa$-test. The region surrounding the local projected 
galaxy overdensity located $\sim 550$\,kpc north of the cluster center shows
very complex velocity structure. The regions just SW and NE of this substructure
have $\overline{v_w} \simeq -800$\kms\, and $\simeq 250$\kms, 
respectively, while the the region immediately to the SE of this substructure 
has $\sigma(v_w) \simeq 1200$\kms. The velocity field coincident with the 
substructure $\sim 400$\,kpc to the SW has $\overline{v_w} \simeq 450$\kms, while 
the velocity field surrounding the local galaxy overdensity $\sim $1.7\,Mpc to 
the south has $\overline{v_w} \sim 400$\kms, and shows a peak of 
$v_w \sim 900$\kms\, just to the east of the local peak in the galaxy density.

\begin{figure*}
{\includegraphics[angle=90,width=0.95\textwidth]{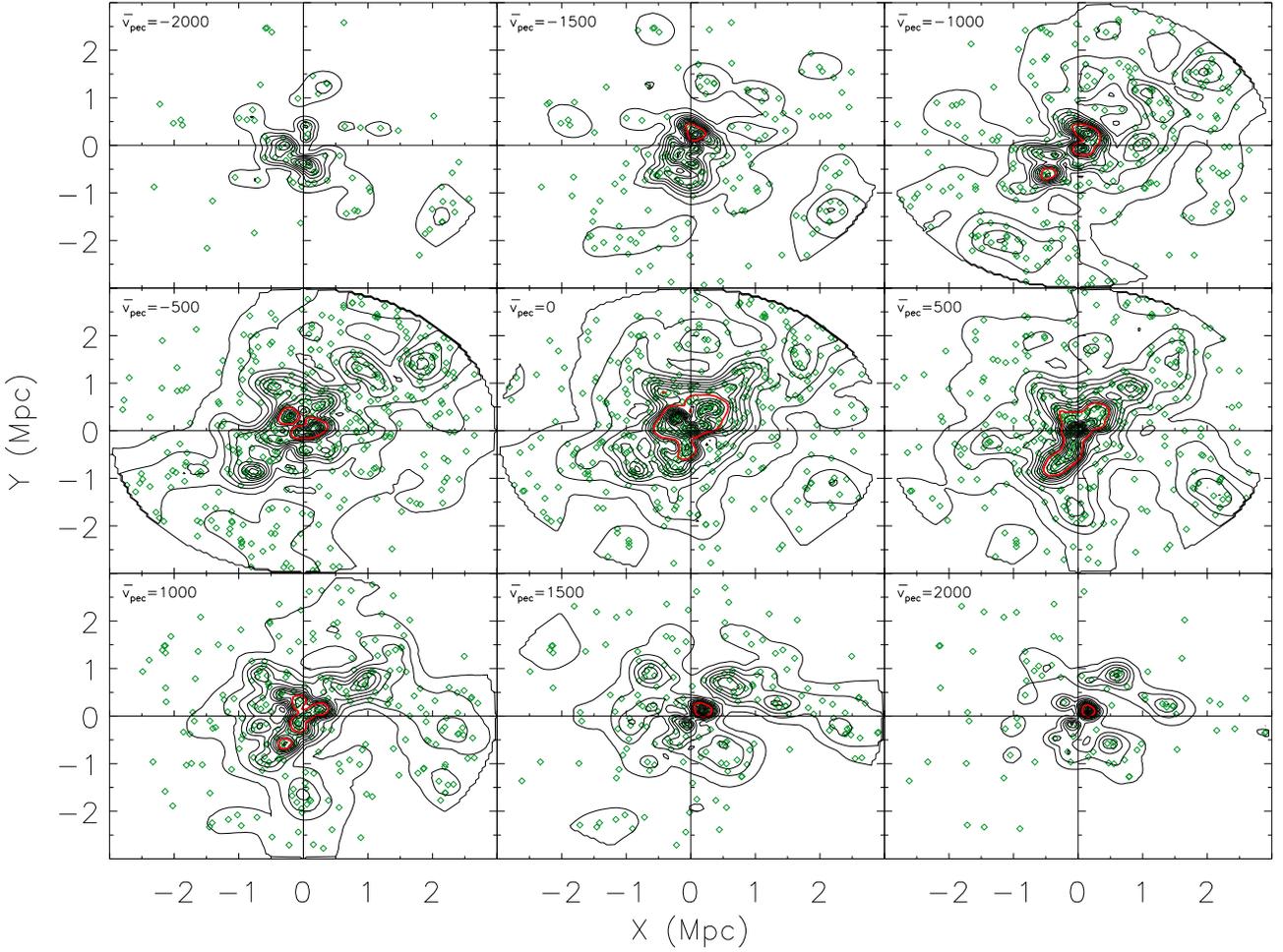}}
\caption{Tomograms for A2142 revealing overdensities in position-velocity space
centered at the velocities listed in the upper left of each panel. The spatial
smoothing uses the same adaptive kernel as that used in the top left panel of
Figure~\ref{a2142_substructure_maps}, while the velocity smoothing uses a fixed
sigma of 300\kms. The green diamonds show the positions of galaxies lying within
$\pm 600$\kms\ of the central velocity, ${\overline v}_{\rm pec}$. The contours show
the galaxy density and start at 5~galaxies/Mpc$^{2}$/(1000\kms) with an
interval of 5~galaxies/Mpc$^{2}$/(1000\kms) up to the 
50~galaxies/Mpc$^{2}$/(1000\kms) level (red contour). Above 
50~galaxies/Mpc$^{2}$/(1000\kms), for clarity, the interval is 
10~galaxies/Mpc$^{2}$/(1000\kms).}
\label{a2142_tomo}
\end{figure*}

\begin{figure*}
{\includegraphics[angle=90,width=0.95\textwidth]{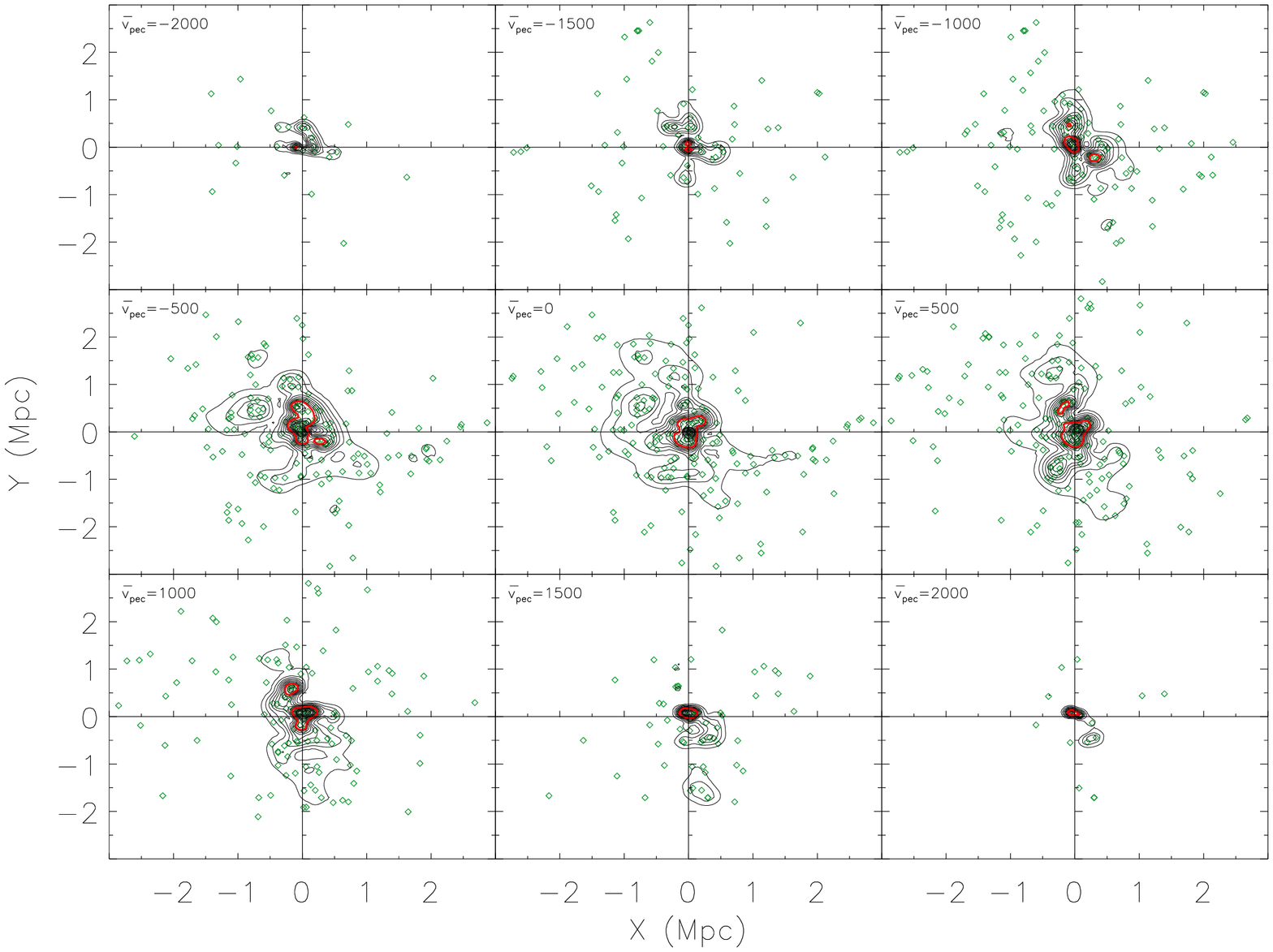}}
\caption{As for Figure~\ref{a2142_tomo}, but for RXJ1720. The spatial smoothing
uses the same adaptive kernel as that used in the top left panel of
Figure~\ref{rxj1720_substructure_maps}.}
\label{rxj1720_tomo}
\end{figure*}

The $\kappa$-test and the velocity field maps described above are ideal for 
identifying local kinematic substructure and go some way towards classifying
the dynamical properties of the substructure. However, the data at hand allow 
us to go one step further in the process of visualizing and characterizing 
substructure by producing ``tomograms'' of the galaxy density in peculiar
velocity slices. These tomograms are shown in Figures~\ref{a2142_tomo} and
\ref{rxj1720_tomo} for A2142 and RXJ1720, respectively. We produce
maps at nine central velocities where the $x-y-v_{\rm pec}$ distribution has
been smoothed with a 3D Gaussian Kernel with an adaptively varying 
$\sigma_{xy}$ (the same kernel described in Section~\ref{2d}) and
$\sigma_{v}=300$\kms. The central velocities used in the smoothing are listed 
in the top left corners of Figures~\ref{a2142_tomo} and \ref{rxj1720_tomo}
which show contours of the 3D galaxy density in units of galaxies per 
${\rm Mpc}^2$ per $1000$\kms. The central core is the dominant feature in
all of the tomograms for RXJ1720 (Figure~\ref{rxj1720_tomo}) which confirm the
complex nature of the dynamics surrounding the local galaxy overdensity located
$\sim 550$\,kpc to the N of the core---it appears most significant in the
${\overline v}_{\rm pec}=-1000, -500, 500\, {\rm and}\, 1000$\kms\ maps, but is relatively 
insignificant in the ${\overline v}_{\rm pec}=0$\kms\, map.  Unlike in RXJ1720,
the core does not dominate in all of the tomograms for A2142 
(Figure~\ref{a2142_tomo}) which are also more complicated than their 
counterparts in RXJ1720. Comparison of the tomograms with those maps presented 
in Figure~\ref{a2142_substructure_maps} reveals that those substructures seen in 
the cluster outskirts generally appear to be fairly well isolated structures
in velocity space, too. The tomograms show that the conglomeration of 
significant circles $\sim 1$\,Mpc to the NW of the core appears as a feature
in the ${\overline v}_{\rm pec} > 500$\kms\, maps, while the local overdensity associated with
the second BCG appears most prominently at ${\overline v}_{\rm pec} = 1500\, {\rm and}\, 2000$\kms.

\subsubsection{Characterizing the substructures with KMM}\label{kmm}

\begin{figure*}
{\includegraphics[angle=0,width=0.85\textwidth]{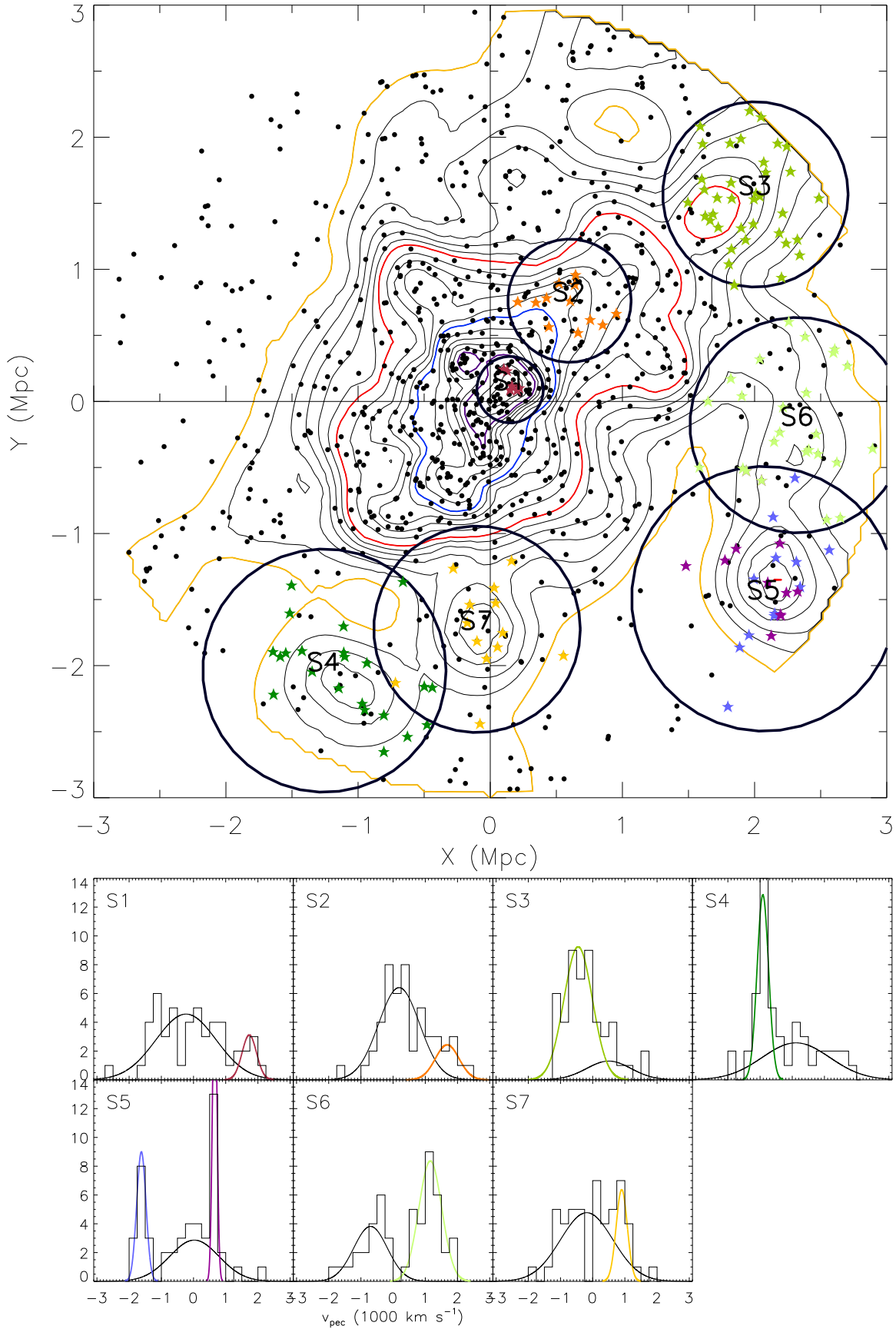}}
\caption{Top panel: The black circles show regions selected for the KMM 
analyses based on substructure revealed in Figure~\ref{a2142_substructure_maps} 
and \ref{a2142_tomo}. Black contours show galaxy surface density as in 
Figure~\ref{a2142_substructure_maps}. The stars are color coded to match the 
corresponding velocity components shown in the lower panels. Black points reveal
the positions of the remaining cluster members.}
\label{a2142_KMM}
\end{figure*}

\begin{figure*}
{\includegraphics[angle=0,width=0.85\textwidth]{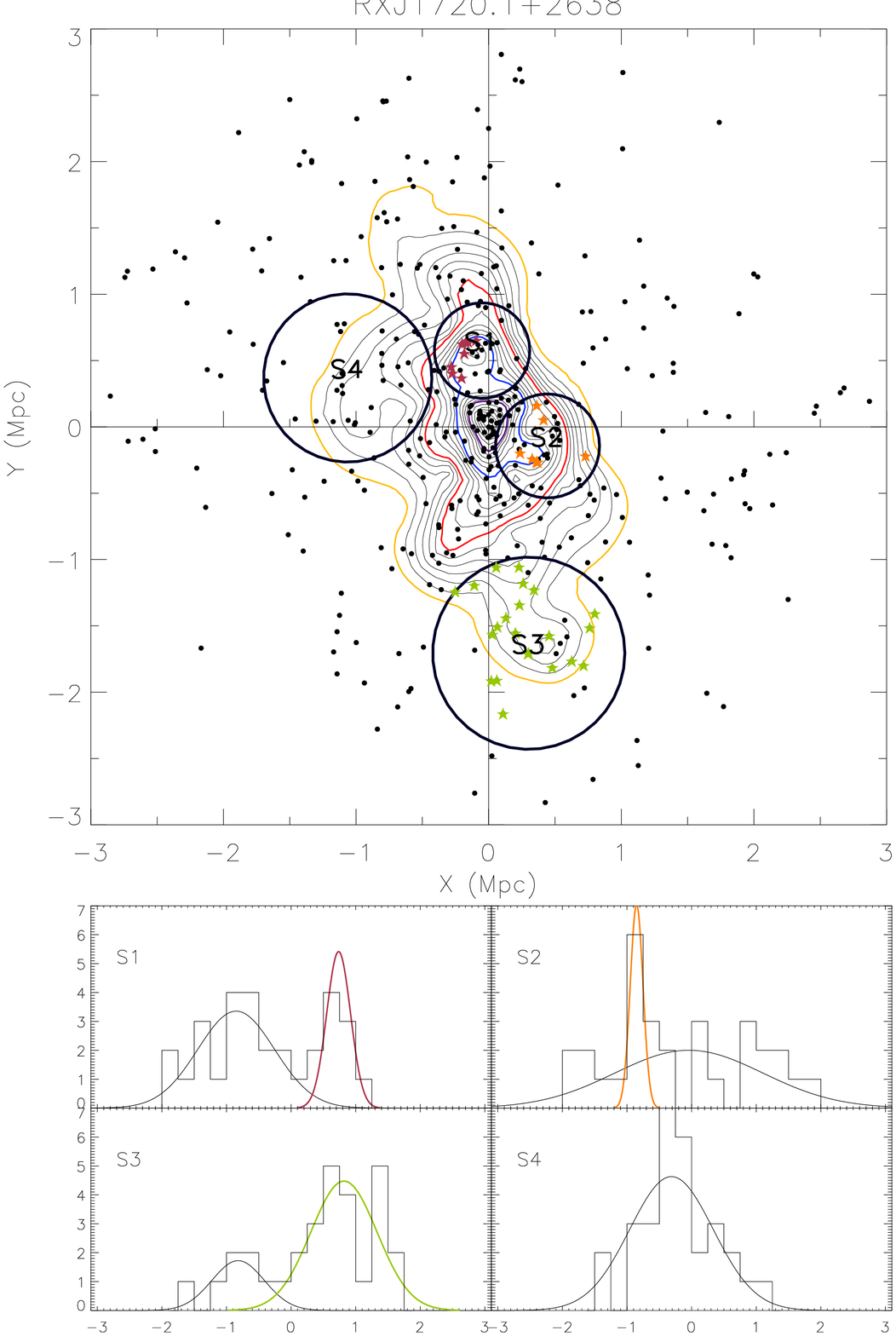}}
\caption{As for Figure~\ref{a2142_KMM} but for RXJ1720.}
\label{rxj1720_KMM}
\end{figure*}

Having identified substructure with a number of techniques in 
Sections~\ref{2d} and \ref{3D}, we are now in a position to characterize the 
dynamical properties of the substructures. To achieve this, we use the Kaye's 
Mixture Model algorithm of \citet[KMM;][]{ashman1994} to investigate the 
dynamics within the regions shown in Figures~\ref{a2142_KMM} and 
\ref{rxj1720_KMM}. The sizes of these regions were chosen to contain 
$\sim 1.5\sqrt{N_{\rm mem}}=50$ and 30 nearest neighbors for A2142 and RXJ1720, 
respectively. 
For each of these regions, the KMM algorithm was used to fit a 
minimum of two Gaussians to the velocity distribution. The first 
Gaussian component models the main cluster velocity distribution while the 
remaining Gaussian components model the superimposed substructure velocity 
distribution. The KMM algorithm requires initial estimates of the mean ($\mu$), 
variance ($\sigma^2)$ and the fraction of galaxies ($f$) belonging to each 
Gaussian velocity component. The initial estimate for the main cluster component
is set to $(\mu_{1,{\rm in}}, \sigma_{1,{\rm in}}, f_{1,{\rm in}})=(0\,{\rm km\,s}^{-1}, 1000\,{\rm 
km\,s}^{-1}, 0.8)$, while the initial estimates for the remaining Gaussian 
components are determined by inspection of the velocity distribution and are 
listed in Tables~\ref{a2142_KMM_fits} and \ref{rxj1720_KMM_fits}. We determine 
the significance of each bimodal fit by employing a parametric Bootstrap
method \citep[see][]{owers2011}. Briefly, we produce 1000 random resamplings of 
a Gaussian with 50 (for A2142) or 30 (for RXJ1720) data points and $\mu$ and 
$\sigma$ set to the values derived from the entire cluster sample. The KMM 
algorithm is then used to fit two Gaussians to the resampled distribution with 
the initial inputs set to the outputted KMM fits to the observed data. A 
measure of the improvement of the fit in going from a unimodal to bimodal 
Gaussian distribution is given by the Likelihood Ratio Test Statistic 
\citep[LRTS; see][]{ashman1994}. The distribution of LRTSs produced by the 
resampled distributions can be compared to the observed LRTS to determine if the
results of the fit to the observed distribution can be caused by random 
fluctuations in the data and this is listed as a $P$-value in 
Table~\ref{a2142_KMM_fits} and \ref{rxj1720_KMM_fits}. In a number of cases
(e.g., S5 and S6 for A2142 and S1 for RXJ1720) a three-mode partition was also
fitted to the data. For these cases, the $P$-value is evaluated by comparison
of the trimodal fit to the bimodal one.

\begin{deluxetable*}{cccccc}
\tabletypesize{\scriptsize}
\tablecolumns{3}
\tablewidth{0pc}
\tablecaption{KMM fits to the regions shown in Figure~\ref{a2142_KMM} for A2142.\label{a2142_KMM_fits}}
\tablehead {\colhead{Region} & 
\colhead{($X,\,Y$)}&           
\colhead{Substructure Input} & 
\colhead{Main output}&  
\colhead{Substructure output} & 
\colhead{P-value}\\            
 &
\colhead{(kpc, kpc)}&
\colhead{($\mu,\, \sigma,\, f$)} & 
\colhead{($\mu,\, \sigma,\, f$)}&
\colhead{($\mu,\, \sigma,\, f$)} & 
} 
\startdata
S1 &  (150, 90) & (1800, 300, 0.2) & (-232.8, 938.7, 0.86) & (1732.5, 224.2, 0.14) & 0.434\\
S2 &  (600, 763) & (1500, 300, 0.2) & (186.3, 638.3, 0.82) & (1681.0, 368.8, 0.18) & 0.007\\

S3 &  (2007, 1567) & (-600, 400, 0.2) & (500.1, 688.7, 0.18) & (-450.8, 442.5, 0.82) & $<0.001$ \\

S4 & (-1255, -2038) & (-900, 300, 0.2) &(114.7, 1047.5, 0.54)  & (-913.4, 178.5, 0.46)& $< 0.001$ \\

S5 &  (2072, -1494) & (-1500, 300, 0.2)& (272.0, 649.9, 0.72) & (-1608.3, 153.1, 0.28)& $< 0.001$\\
S5 &  (2072, -1494) & (-1500, 300, 0.2), (600, 200, 0.1)& (24.2, 765.6, 0.44) & (-1620.4, 144.1, 0.26), (658.8, 68.9, 0.2)& 0.007\\

S6 &  (2327, -180)&(1200, 300, 0.2) & (-701.1, 522.5, 0.40) & (1156.5, 357.8, 0.60)& $<0.001$\\

S6 &  (2327, -180)&(1200, 300, 0.2), ( -400, 200, 0.1)& (-1212.4, 423.8, 0.14) & (1147.8, 362.8, 0.60), (-433.7, 220.7, 0.26)& 0.32\\

S7 &  (-1725, -100)&(900, 300, 0.2) & (-185.0, 815.2, 0.78) & (902.6, 172.2, 0.22)& 0.11\\
\enddata
\end{deluxetable*}

\begin{deluxetable*}{cccccc}
\tabletypesize{\scriptsize}
\tablecolumns{3}
\tablewidth{0pc}
\tablecaption{KMM fits to the regions shown in Figure~\ref{rxj1720_KMM} for RXJ1720.\label{rxj1720_KMM_fits}}
\tablehead {\colhead{Region} & 
\colhead{($X,\,Y$)}&           
\colhead{Substructure Input} & 
\colhead{Main output}&  
\colhead{Substructure output} & 
\colhead{P-value}\\            
 &
\colhead{(kpc, kpc)}&
\colhead{($\mu,\, \sigma,\, f$)} & 
\colhead{($\mu,\, \sigma,\, f$)}&
\colhead{($\mu,\, \sigma,\, f$)} & 
} 
\startdata

S1 &  (-50, 576) & (700, 300, 0.2) & (-849.7, 594.4, 0.67) & (736.0, 184.2, 0.33) & 0.008 \\
S1 & (-50, 576) & (700, 300, 0.2), (-700, 300, 0.2) & (-906.0, 659.9, 0.50) &  (733.8, 185.6, 0.33), (-658.2, 127.9, 0.17) & 0.72 \\
S2 & (443, -144) & (-900, 300, 0.2) &(-42.4, 1145.8, 0.77)  & (-851.0, 99.2, 0.23)& 0.082 \\

S3 &  (302, -1706) & (800, 300, 0.2)& (-819.8, 404.8, 0.23) & (818.4, 512.5, 0.76)& 0.009\\

S4 &  (-1225, 181)&(-400, 300, 0.2) & (--, --, --) & (-291.1, 577.3, 1.00)& --\\
\enddata
\end{deluxetable*}

\subsection{Summary of the substructure properties}\label{struct_summary}
Having identified many significant substructures in both A2142 and RXJ1720, 
and characterized them dynamically using the KMM method, we now briefly discuss 
each of them and, in particular, their properties that are most pertinent to 
our study.
\subsubsection{A2142}
\begin{itemize}
\item {{\bf S1:} This substructure is coincident with the second BCG 
$\sim 180$\,kpc to the northwest of the central BCG. There is a local
overdensity in the projected galaxy distribution here, although the 
$\kappa$-test does not reveal strong evidence that the local velocity 
distribution differs significantly from the global one. The substructure
is prominent as an overdensity in the ${\overline v_{\rm pec}}=1500\,{\rm and}\,
2000$\kms\ tomograms, consistent with a grouping of galaxies around the 
velocity of the second BCG ($v_{\rm pec}=1760$\kms). The KMM analysis returns
a mean $v_{\rm pec} = 1732.5$\kms\ and dispersion $\sigma_{\rm vpec}=224.2$\kms\ for
a group of $\sim 7-10$ galaxies surrounding this substructure, although a
bimodal fit is not favored. The lack of evidence for dynamical substructure
in the $\kappa$- and KMM tests is probably due to S1's proximity to the
cluster core where projection effects caused by the main cluster population
are large. This is likely the remnant core of a more massive system which 
has had its less well bound members stripped during its infall.}

\item {{\bf S2:} This is a more loose agglomeration of spirals. There doesn't 
appear to be a dominant early type member within this substructure and there is 
only a mild excess in the galaxy surface density here. The velocity field is
significantly positive in this region and the tomograms reveal local 
overdensities here which are most prominent at ${\overline v_{\rm pec}}=1000\,
{\rm and}\,1500$\kms. The KMM fit reveals a significant bimodality where the 
substructure has mean velocity $v_{\rm pec}= 1681.0$\kms\ and dispersion of 
$\sigma_{\rm vpec}=368.8$\kms. \citet{okabe2008} find a mildly significant peak in 
their projected mass maps at approximately the same position.}

\item {{\bf S3:} This substructure is clearly distinguished by the
$\kappa$-test.  It lies $\sim 2.5$\,Mpc from the cluster center and it is
conspicuous as a local overdensity in the projected galaxy
distribution, as features in the velocity field maps and as an
overdensity in the ${\overline v_{\rm pec}}=-500$\kms\ tomogram. The KMM 
analysis returns significant bimodality and the substructure has mean velocity 
$v_{\rm pec}=-450.8$\kms\ and dispersion $\sigma_{\rm vpec}=442.5$\kms, consistent with a 
large group or small cluster.}

\item {{\bf S4:} Another substructure in the outskirts located $\sim 2.4$\,Mpc
from the center. This substructure manifests itself as a local overdensity in the
projected galaxy distribution and in the ${\overline v_{\rm pec}}=-1000$\kms\ tomogram.
The KMM test reveals significant bimodality and the substructure has mean 
velocity $v_{\rm pec}=-913.4$\kms\ with dispersion $\sigma=178.5$\kms, consistent 
with a group-sized substructure.}

\item {{\bf S5:} This substructure is a significant local overdensity in the
projected galaxy distribution although the $\kappa$-test does not show
conclusive results for local velocity substructure. The substructure is seen as
an overdensity in the ${\overline v_{\rm pec}}=-1500\,{\rm and}\,500$\kms\ tomograms.
The KMM analysis reveals the velocity distribution is trimodal with two low 
dispersion substructures with mean velocity $v_{\rm pec}=-1620.4\, {\rm and}\, 
658.8$\kms\, and dispersions $\sigma_{v_{\rm pec}}=144.1 \, {\rm and}\, 68.9$\kms, 
respectively. The low velocity dispersions indicate that the substructures have low 
mass and their large peculiar velocity separation indicates they are not 
associated. The projection effects caused by these two separate structures
aligned along the line of sight have enhanced the projected galaxy density 
here.}
\item {{\bf S6:} There is only a mild increase in the local galaxy surface
density in this region, however the $\kappa$-test shows that the local kinematics
are significantly different from the global kinematics.
The KMM analysis showed that the reason for this difference was a strong
bimodality with two substructures separated by a peculiar velocity of
$\sim 1860$\kms. The two substructures have mean velocity 
$v_{\rm pec}=-701,1\,{\rm and}\, 1156.5$\kms\, and dispersion 
$\sigma_{\rm vpec}=522.5\,{\rm and}\, 357.8$\kms, respectively. The tomograms show
that the negative peculiar velocity component is fairly dispersed, while the
$v_{\rm pec}\sim1000$\kms\ component does appear to be more concentrated in this 
region, although it is not as prominent as other substructures such as S3. The 
galaxies in this region may be part of coherent substructures which are infalling
from the surrounding large scale structure.}

\item {{\bf S7:} This substructure is prominent as a local overdensity in the
projected galaxy distribution, although the $\kappa$-test does not
reveal any significant velocity substructure here. There are local density peaks 
here in the ${\overline v_{\rm pec}}=500\,{\rm and}\,1000$\kms\ tomograms
(Figure~\ref{a2142_tomo}) and the KMM algorithm fits a substructure with mean velocity 
$v_{\rm pec}=902.6$\kms\ and dispersion $\sigma_{\rm vpec}=172.2$, 
although bimodality is not strongly favored. The low dispersion indicates a low
mass, although given the minimal evidence for velocity substructure, the local
peak in the galaxy density here may simply due to projection effects from 
unrelated galaxies.}

\end{itemize}

\subsubsection{RXJ1720}
\begin{itemize}
\item {{\bf S1:} This substructure is a well defined local peak in the galaxy
surface density distribution although it has unusual kinematic properties.
The $\kappa$-test reveals significant local kinematic substructure in this 
region, while the tomograms reveal local overdensities in this region at 
${\overline v_{\rm pec}}=-500,\,-1000,\,500\,{\rm and}\,1000$\kms\, but not at
 ${\overline v_{\rm pec}}=0$\kms. The KMM analysis revealed a bimodal fit was 
preferred over uni- and trimodal fits with the two components having mean 
velocity $v_{\rm pec}=-849.7\,{\rm and}\, 736.0$\kms\, and dispersions of 
$\sigma_{\rm vpec}=594.4\,{\rm and}\, 184.2$\kms. The second brightest member in 
the core of the cluster has $v_{\rm pec}=933$\kms\ and is within the S1 region.
\citet{okabe2010} find a second peak in their weak lensing 
convergence maps which is coincident with this substructure, providing further 
evidence that this substructure is real. We tentatively point to a mild excess 
in the X-ray residuals in the right panel of Figure~\ref{rxj1720_all} which is 
coincident with this structure.} 

\item {{\bf S2:} This substructure has a mild enhancement in the projected galaxy
distribution and a local velocity distribution which differs from the global
one according to the $\kappa$-test. The KMM analysis reveals that this 
difference is caused by an extremely low velocity dispersion 
($\sigma_{\rm vpec}=99.2$\kms) substructure at a velocity $v_{\rm pec}=-851.0$\kms, although
the improvement in the fit in going from a unimodal to bimodal distribution is 
marginal. Given the small velocity dispersion and mild enhancement in galaxy 
surface density, this substructure is probably dynamically insignificant although 
we note that \citet{okabe2010} find a weak enhancement at roughly this position 
in their lensing maps.}

\item {{\bf S3:} This substructure shows a mild local peak in the projected
galaxy distribution and the velocity field shows a preference for
positive peculiar velocities here. While the tomograms do not reveal
significant local peaks for the maps centered around positive peculiar 
velocities, the KMM analysis reveals that the velocity distribution here is
bimodal with the majority of the galaxies allocated to a component centered
at $v_{\rm pec}=818.4$\kms\, with dispersion $\sigma_{\rm vpec}=512.5$. This large
dispersion, if real, indicates a significant mass in this region. However, 
visual inspection of the velocity distribution in the lower left panel of
Figure~\ref{rxj1720_KMM} reveals that there may be some contamination from
galaxies at $v_{\rm pec} \sim 1500$\kms\, which may be inflating the dispersion.
More complete spectroscopy is required in this region to better disentangle
the velocity components and confirm the large dispersion measured here.}

\item {{\bf S4:} The $\kappa$-test revealed a number of large 
bubbles in this region which indicates evidence for velocity substructure. There
is no strong galaxy surface density enhancement, although the tomogram
centered at $v_{\rm pec}=-500$\kms\, shows a local peak here while the velocity 
dispersion in this region appears to be low. However, the KMM analysis failed 
to fit a second velocity substructure to the data in this region, although the best
fitting single Gaussian does have a low velocity dispersion 
($\sigma_{\rm vpec}=577.3$\kms) and is centered at $v_{\rm pec}=-291.1$\kms\ indicating a 
departure from the global velocity distribution here. More data are required in 
this region in order to confirm the existence of real dynamical substructure.}
\end{itemize}

\section{Discussion}\label{disc}

The principal objective of this study is to find observational evidence
for merger related substructure in the sloshing cold front clusters A2142 and 
RXJ1720. Furthermore, we wish to determine if the cold fronts observed in the 
cores of these clusters can be related to merger activity, therefore confirming
that cold fronts are excellent signatures of merger activity. The results
of our substructure tests reveal that A2142 and RXJ1720 harbor a number of
group-scale substructures which manifest themselves as enhancements in the 
local galaxy surface density (Section~\ref{2d}), and as regions of localized 
velocity substructure (Section~\ref{3D}). The existence of these group-scale
substructures is consistent with a minor merger origin of the cold fronts, as
seen in the simulations of \citet[][see also \citealt{roediger2011} for a more specific case]{tittley2005, ascasibar2006}  where a low mass group perturbs 
the main cluster core and induces cold fronts. The problem now, particularly 
for A2142, is to determine which of the many substructures detected is the most 
likely perturber.
As a first step, we note that substructures with large \rproj\
are unlikely to have made a pericentric passage and, thus, are unlikely to be 
the perturber. We outline the evidence for this statement in 
subsection~\ref{out} and discuss the substructures which we consider to be the 
most likely perturbers in A2142 (S1 and S2; subsection~\ref{a2142_in}) and 
RXJ1720 (S1 and S3; subsection~\ref{rxj1720_in}).

\subsection{Ruling out the substructures least likely to be perturbers}\label{out}

First, we consider the apocentric distance of a subcluster after its initial
pericentric passage. We obtain an estimate for this distance by considering the 
simulations of \citet{tormen2004} where the average apocentric distances for 
an ensemble of simulated clusters with merging subhalos of varying masses is 
shown in their Figure~11. We only consider results for the ``self bound'' satellite 
particles in \citet{tormen2004}, since these results are more likely to reflect 
our observed substructures. Two important features can be noted here: (i) On 
average, the subhalos do not travel farther than the main cluster's virial
radius, $R_V$, after pericenter and (ii) Due to dynamical friction, larger mass 
subhalos have apocentric distances as little as $\sim 0.5R_V$. Thus, it is 
highly unlikely that any substructure observed at radii larger than the 
cluster virial radius has passed pericenter---they are far more likely to be 
groups which are falling onto the cluster for the first time. 
For A2142 and RXJ1720, we assume that the virial radius is
$R_{\rm Vir} \simeq R_{200}=0.17 \sigma_{v} / H(z)$ 
where $R_{200}$ is the radius where the density of the cluster is 200 times the 
critical density at redshift $z$, $\sigma_{v}$ is the cluster velocity 
dispersion and $H(z)=H_0 \sqrt{\Omega_m(1+z)^3 + \Omega_V}$ 
\citep{carlberg1997}. This results in $R_{\rm vir}=2.36\, {\rm and\,} 2.02$\,Mpc 
for A2142 and RXJ1720, respectively, and we set this as an initial constraint
on the \rproj\ within which the perturber must lie. These limits
are plotted as cyan circles in Figures~\ref{a2142_all} and \ref{rxj1720_all}. 
For RXJ1720, these limits do not strongly rule out any of the substructures as 
potential perturbers, although S3 lies close to the virial radius in projection.
For A2142, the substructures S3 and S5 are outside of the $R_V$ and are therefore 
ruled out as potential perturbers, while S4 and S6 lie at approximately the 
virial radius. Those substructures at the virial radius in A2142 (S4, S6) are 
unlikely to be the perturbers because (i) Projection effects mean that the 
observed radius gives the minimum distance from the cluster center---if the 
substructure lies at some distance in front or behind the cluster, then it is 
likely to lie outside the virial radius and (ii) The limit imposed here is an 
upper limit on the radius. More massive systems which have passed close to the 
cluster core are strongly affected by dynamical friction \citep{tormen2004} and 
thus have smaller apocentric distances.

Further evidence that these substructures at large \rproj\ ($\gtrsim R_V$) 
 are unlikely to have passed pericenter comes from 
considering the timescales required to produce the cold front pairs seen in the 
\chan\ images (Figures~\ref{a2142_all} and \ref{rxj1720_all}).
Given that the fluid velocities
involved are small compared to the Kepler speed \citep[e.g.,][]{ascasibar2006}, 
the sloshing motions that give rise to cold fronts
are likely to be reasonably well described as superpositions of
internal gravity waves excited by merging subclusters \citep{churazov2003}.  
Locally, the oscillation frequency of these waves is the
Brunt-V\"ais\"al\"a frequency
\begin{equation}
\omega_{\rm BV}=\sqrt{{g \over r} {3\over 5} {d\ln s \over d\ln r}} = 
\omega_k \sqrt{{3\over 5} {d\ln s \over d\ln r}},
\end{equation}\label{omega_BV}
where $g = GM(r)/r^2$ is the acceleration
due to gravity, $r$ is the radius, $s$ is the entropy index and
$\omega_{\rm K}$ is the Keplerian orbital frequency.  Thus
$\omega_{\rm BV}$ is expected to provide an estimate of sloshing
frequencies.  The radial entropy profiles for both A2142 and 
RXJ1720 have been measured by \citet{cavagnolo2009} and they have logarithmic 
slopes of $\sim 1.22\, {\rm and}\, 1.39$ at the positions of the cold fronts, 
respectively. Therefore, the approximation $\omega_{\rm BV} \simeq \omega_k = 
V_{\rm circ}/r$ is sufficient for our purposes.  This
justifies the estimate of \citet{simionescu2010}, which employs the
radial dependence of the frequency to determine the time required for
an adjacent pair of cold fronts to get out of phase by $\pi$ radians,
i.e., to appear on opposite sides of the cluster center.  Even in the
linear approximation, a more thorough treatment requires determination
of the spectrum and phasing of the internal modes excited by a merging
subhalo, then following the time development of this superposition of
internal gravity waves.  Apart from nonlinear effects, this leaves
considerable uncertainty in our time estimates.

To determine $\omega_k$, we measure the cold front radii
from the central brightest cluster galaxy which is assumed to mark the potential
minimum and assume a singular isothermal sphere (SIS) density distribution 
meaning the circular velocity is simply $V_{\rm circ}=\sqrt{2} \sigma_{v}$.
We also assume that the time required for the cold front pairs
to become $\pi$ radians out of phase in their orbits is equivalent to the time 
since the perturber's pericentric passage. For A2142, the inner and outer cold 
fronts have radii $R_{\rm proj, in} = 87$\,kpc and $R_{\rm proj, out} = 326$\,kpc, 
respectively, and $V_{\rm circ} = 1407$\kms. The time required for the cold 
fronts to become out of phase by $\pi$ radians is $\tau = \pi/(\omega_{k,{\rm in}} - 
\omega_{k,{\rm out}}) \simeq 0.26$\,Gyr. The same calculation for RXJ1720 with 
$R_{\rm proj, in} = 69$\,kpc, $R_{\rm proj, out} = 185$\,kpc and $V_{\rm circ} = 
1247$\kms\ gives $\tau \simeq 0.27$\,Gyr. \citet{roediger2011} found that the 
cold front age calculated by \citet{simionescu2010} was underestimated by a 
factor of 3.5 when compared to their simulations. We have also calculated 
$\tau$ for several of the cold fronts in Figure~7 of 
\citet{ascasibar2006}, finding that this method generally underestimates the 
time since pericentric passage by factors of 3-4, consistent with the findings
of \citeauthor{roediger2011}. Therefore, we estimate that the 
``true'' time since pericentric passage for both A2142 and RXJ1720 is 
$\tau_{\rm true}\simeq 0.8-1.1$\,Gyrs. We note that viewing
angles and uncertainties in the phasing of the waves mean that these estimates
give only a rough guide for the time since pericenter. Furthermore, we note that
the cold fronts that we observe in A2142 and RXJ1720 occur on much larger scales
than the cold fronts seen at similar times in the simulations. This may
reflect differences in the physical properties of A2142 and RXJ1720  
compared to the simulated clusters, e.g., the shape of the underlying potential 
will affect the oscillation frequencies. The spatial and temporal scales of the
fronts are also likely to be affected by properties of the perturber, including 
its mass and impact parameter.  Nevertheless, the timescales derived above 
provide useful estimates for our purposes and provide strong support for the 
radial constraints derived from \citet{tormen2004}. For a
substructure to have reached the virial radius after pericenter on these 
timescales, it would need to have traveled at an average radial velocity since 
pericenter of $\sim 2100-2890\, {\rm and}\, 1790-2460$\kms\ for A2142 and 
RXJ1720. While velocities of this order are expected at pericenter during a 
merger, they do not persist on gigayear timescales since the subcluster velocity
quickly decays after pericenter due to the effects of gravity and dynamical 
friction as it travels outward. This places strong constraints
on whether substructures S4 and S6 in A2142 have had time to have passed pericenter
and traveled to their current \rproj. Therefore, we disregard 
them as being potential perturbers.

Having ruled  out S3, S4, S5 and S6 in A2142, we now consider substructure S7 which
we rule out for the following reasons. First, while it lies within
the virial radius, it has a relatively large radius ($R_{\rm proj}\sim1.7$\,Mpc)
which, as noted above for S4 and S6, gives its minimum distance from the
cluster center.
Furthermore, as noted in Section~\ref{struct_summary}, while there is a local 
excess in the projected galaxy density, there is little evidence for any 
corresponding localized velocity substructure. This means that if there is a 
real substructure in this region, it is probably not very massive. A further 
possibility is that the excess in galaxy surface density 
is due to projection of galaxies which are not physically associated and 
therefore that S7 is not a bona fide substructure. Given its uncertain nature 
and large \rproj, we find it unlikely that S7 has passed 
pericenter and, if it has, it is probably not massive enough to have caused 
the perturbation required to produce the cold fronts with such large radii and 
contrast. Therefore, we also dismiss S7 as a potential perturber.

The radial limits derived above for RXJ1720 placed no strong limits on any
of the substructures shown in Figure~\ref{rxj1720_all}. However, we rule out
the substructures S2 and S4 because they are unlikely to be significant 
substructures, if real associations at all. Starting with S2, this substructure
consists of a grouping of a small number of galaxies with similar 
velocities and a small dispersion of only $\sim100$\kms. There appears to be
a substructure at approximately this location in the lensing maps of 
\citet{okabe2010}, although it is not concentrated and is only detected at the
$3\sigma$ level. The presence of 
substructure detected using our spectroscopically confirmed members and also
in the lensing maps of \citet{okabe2010} provide strong support that S2 is a 
real substructure. However, the low velocity dispersion and low significance in 
the lensing maps imply that S2 has low mass and is 
therefore not likely to have played an important role in perturbing the cluster 
core. Considering the S4 substructure, while there are hints of localized velocity 
substructure in this region according to the $\kappa$-test and the velocity fields,
there is only a weak excess in galaxy surface density here. Furthermore, the KMM
analysis failed to identify any bimodality in this region. Given the uncertainty
in the existence of velocity substructure and the lack of a compact 2D substructure, 
it is unlikely that S4 is a real substructure. Deeper, more complete 
spectroscopic coverage of RXJ1720 would be required to determine if there is a 
real substructure there.

\subsection{Most likely perturbers}\label{in}

\subsubsection{A2142}\label{a2142_in}

We begin our discussion with the S1 substructure surrounding the second BCG at 
$\sim 180$\,kpc to the northwest of the first ranked BCG.  The KMM fit returns 
$v_{\rm pec}=1732.5\, {\rm and}\, \sigma_v=224.2$\kms, although a bimodal fit to 
the local velocity distribution was not favored statistically over a unimodal 
one. However, the surrounding local overdensity in the galaxy surface density, 
along with the compact local overdensity in the 1500 and 2000\kms\ tomograms 
(Figure~\ref{a2142_tomo}) indicate that there is a kinematically compact grouping
of $\sim7-10$ galaxies associated with this BCG. Any kinematic distinction is 
likely diluted by the projection of the main cluster core galaxies onto S1 due 
to its close proximity in projection to the core. BCGs are expected to be found
at the centers of groups and clusters thus the existence of a second BCG 
provides strong evidence that S1 is a merger remnant. Given the relatively high 
(although not extreme) positive peculiar velocity of S1, we speculate that the 
majority of its motion is aligned with our line of sight. In the scenario where S1 is the perturber responsible
for the cold fronts, it must have already passed pericenter. If the estimated 
time since pericenter of $\sim 1$\,Gyr derived above is correct, then S1 is 
unlikely to have passed apocenter and therefore we surmise that it is currently 
located on the far side of the cluster on its way to apocenter. 

\begin{figure*}
{\includegraphics[angle=0,width=0.45\textwidth]{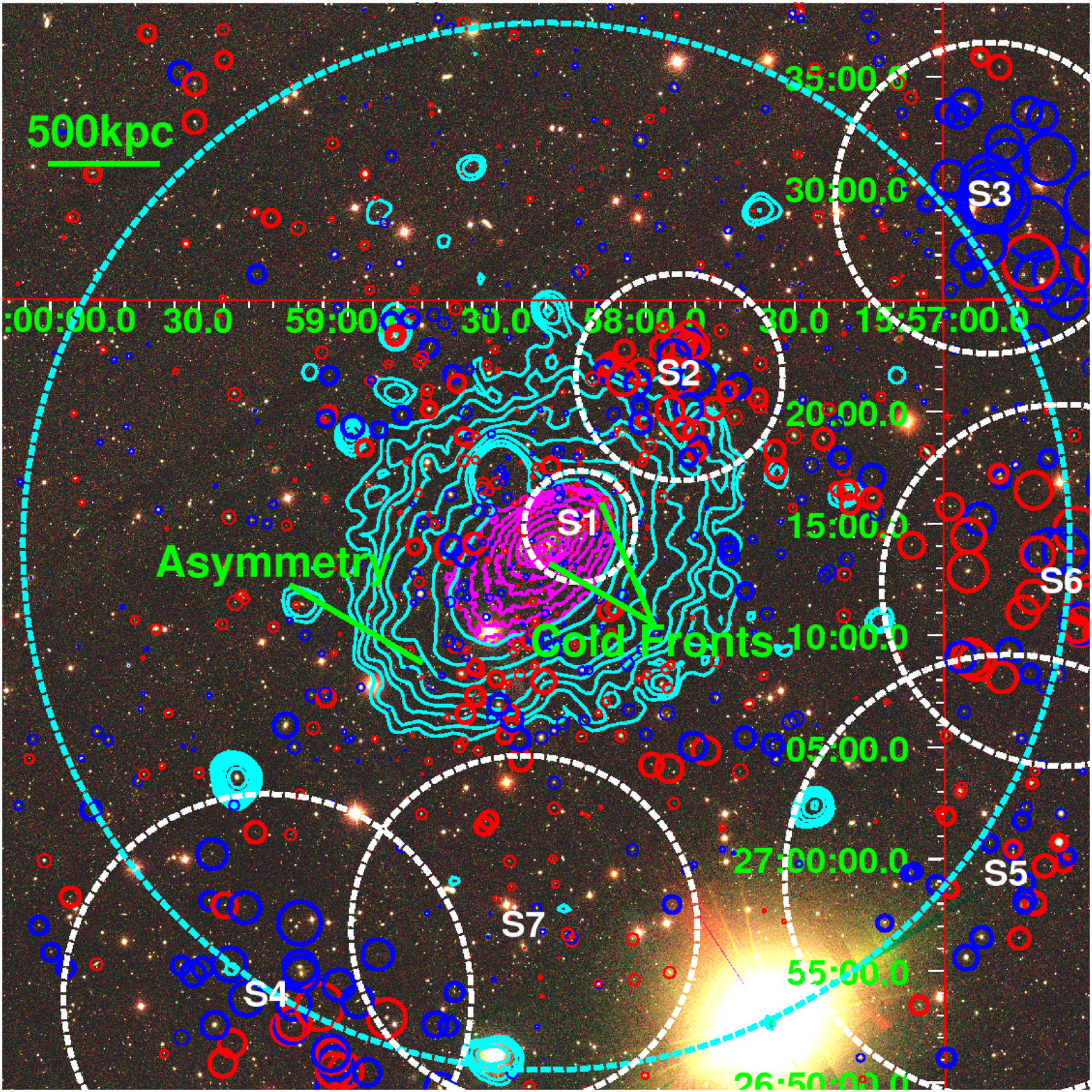}}
{\includegraphics[angle=0,width=0.45\textwidth]{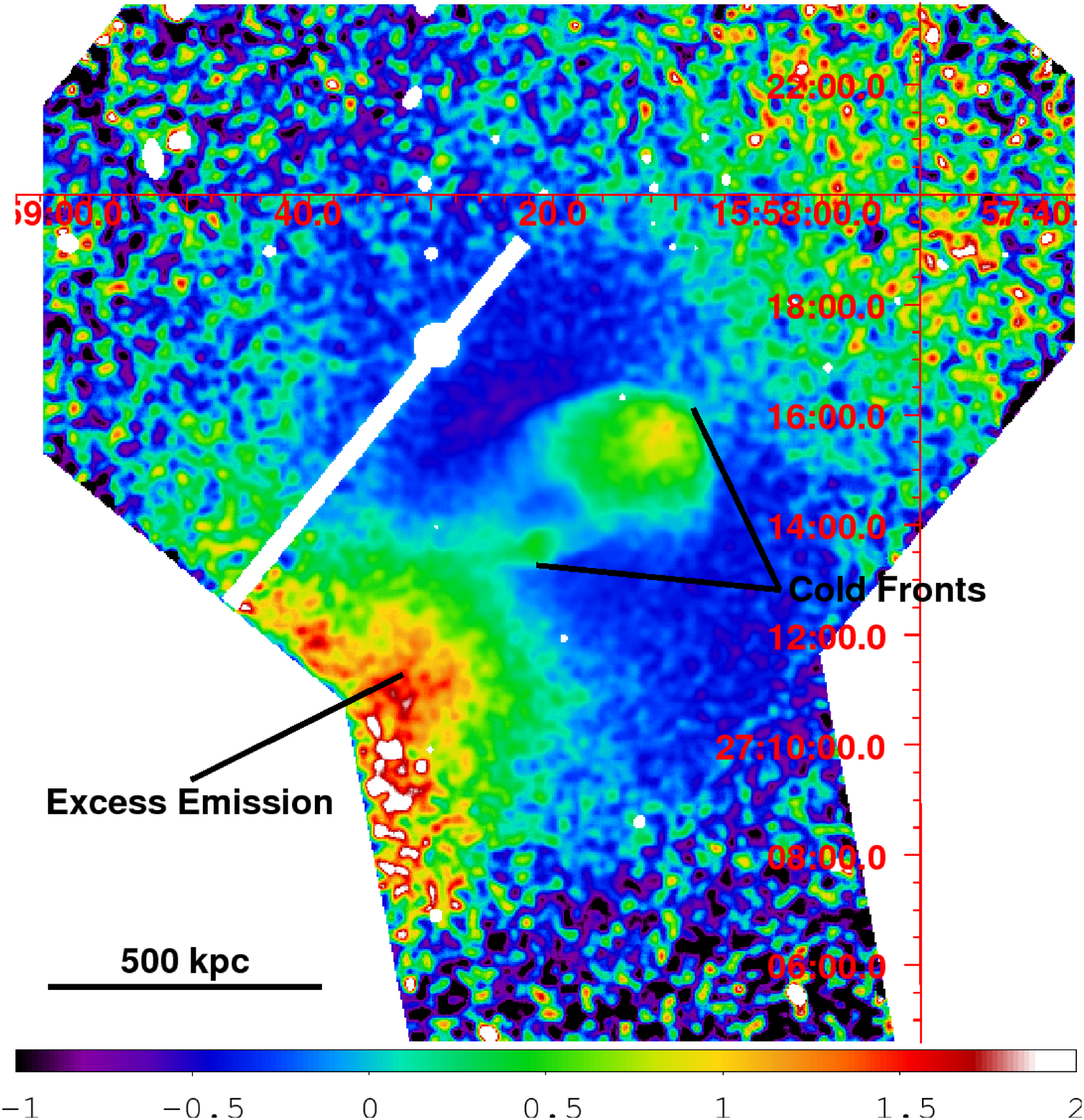}}
\caption{{\it Left panel:} SDSS {\it gri} image of A2142 with X-ray contours 
from ROSAT (cyan) and \chan\, (magenta) overlaid. The white dashed circles show 
the same regions used to define the KMM substructures in Figure~\ref{a2142_KMM}. 
The cyan dashed circle shows the cluster virial radius ($R_V=2.36$\,Mpc) beyond
which we do not expect to find substructure which has passed pericenter. 
{\it Right panel:} Residuals from 
an exposure and background corrected 0.5-7\,keV \chan\ image after subtraction 
of the azimuthal average at each radius. There is no clear spiral structure 
joining the outer cold front to the cluster core in this residual map.}
\label{a2142_all}
\end{figure*}

Further evidence for a perturber which has an orbit which is highly inclined to 
the plane of the sky, such as that proposed for S1, comes from
the morphology of the cold fronts. The \chan\ data for A2142 reveal
two prominent cold fronts \citep{markevitch2000} straddling the cool core 
(Figure~\ref{a2142_all}). There is no clear spiral structure seen 
in the residual map (right panel of Figure~\ref{a2142_all}), nor in the 
temperature maps available in the literature \citep{markevitch2000, owers2009c},
although there are fairly linear structures joining the innermost cold front to 
the outermost one. The morphology of the cold fronts is consistent with a spiral
structure viewed with its axis almost perpendicular to our line of sight, e.g., 
as shown in the middle panel of Figure~19 in \citet{ascasibar2006}. In the 
simulations, the spiral structure is formed when angular momentum is transfered 
to the displaced core gas from a perturber which has undergone an offcenter 
passage \citep{ascasibar2006}. Because the angular momentum responsible for the
spiral structure originates with the offcenter passage of the perturber, the
spiral and perturber orbit must be coplanar. Based on this and the evidence that
we are viewing a spiral structure with its axis almost perpendicular to our 
line of sight, we surmise that the perturber is orbiting in a plane which 
is highly inclined to the plane of the sky. 

We also note here that the 
central BCG has a peculiar velocity of $v_{\rm pec}=212\pm65$\kms\ with respect 
to the cluster velocity distribution. This difference is significant at the 
$2.9\sigma$ level according to the method outlined in \citet{oegerle2001}. 
Assuming that the BCG lies at the bottom of the cluster potential as traced by 
the dark matter, we interpret this as evidence of dark matter core 
oscillations. As pointed out by \citet{ascasibar2006} the observation of  
BCG peculiar velocities is naturally encompassed in their gasless minor merger 
scenario. In that scenario, a purely gravitational perturbation excites 
oscillations in both the dark matter and gas cores leading to cold fronts when 
the gas decouples from the dark matter. However, in the case where a weak shock 
or other purely hydrodynamical disturbance 
\citep{churazov2003, fujita2004a,roediger2011} perturbs the core and
generates the cold fronts, only the gas is expected to oscillate, thus no 
significant BCG peculiar velocity should be observed. We interpret the BCG 
peculiar velocity as evidence that the cold fronts are generated by the 
gravitational disturbance of a minor merger. Furthermore, the 
detection of a line of sight velocity offset lends support to our assertion that
the perturber's orbit is inclined to the line of sight.

As an alternative, we consider the S2 substructure, which is a fairly loose group
of galaxies at a projected distance of $\sim 970$\,kpc from the central BCG, 
has a mean peculiar velocity of $v_{\rm pec}=1680$\kms\ and dispersion 
$\sigma \simeq 370$\kms. Supporting evidence for S2 being a real substructure
comes from the lensing analysis of \citet{okabe2008} who
detect a signal in their projected mass map which is significant at the 
$3\sigma$ level and is approximately at the position of S2.
Similar to S1, the positive peculiar velocity of S2 
implies that it is currently travelling away from us and, assuming that it has 
not passed apocenter, that its initial trajectory began in the foreground to 
the southeast of the main cluster core. It passed pericenter on the SE before 
being deflected behind the cluster core and to the northwest where it traveled 
to its current position. 

\subsubsection{RXJ1720}\label{rxj1720_in}

The most viable perturber candidate for RXJ1720 is the S1 substructure
lying $\sim 550$\,kpc north of the cluster core. This substructure is
conspicuous as an excess in the galaxy surface density maps and shows evidence 
for localized kinematic substructure. A more detailed inspection of the 
velocity distribution surrounding S1 in Section~\ref{kmm} revealed significant 
bimodality with two substructures separated by $\sim 1586$\kms, neither of which 
is centered at the systemic cluster velocity ($\mu_1=-849.7$\kms\ and 
$\mu_2=736.0$\kms). While the velocity structure is complicated here and is hard
to interpret, we are confident that there exists a massive substructure here, 
particularly given that \citet{okabe2010} find significant substructure
in their lensing maps which is spatially coincident with S1.

\begin{figure*}
{\includegraphics[angle=0,width=0.45\textwidth]{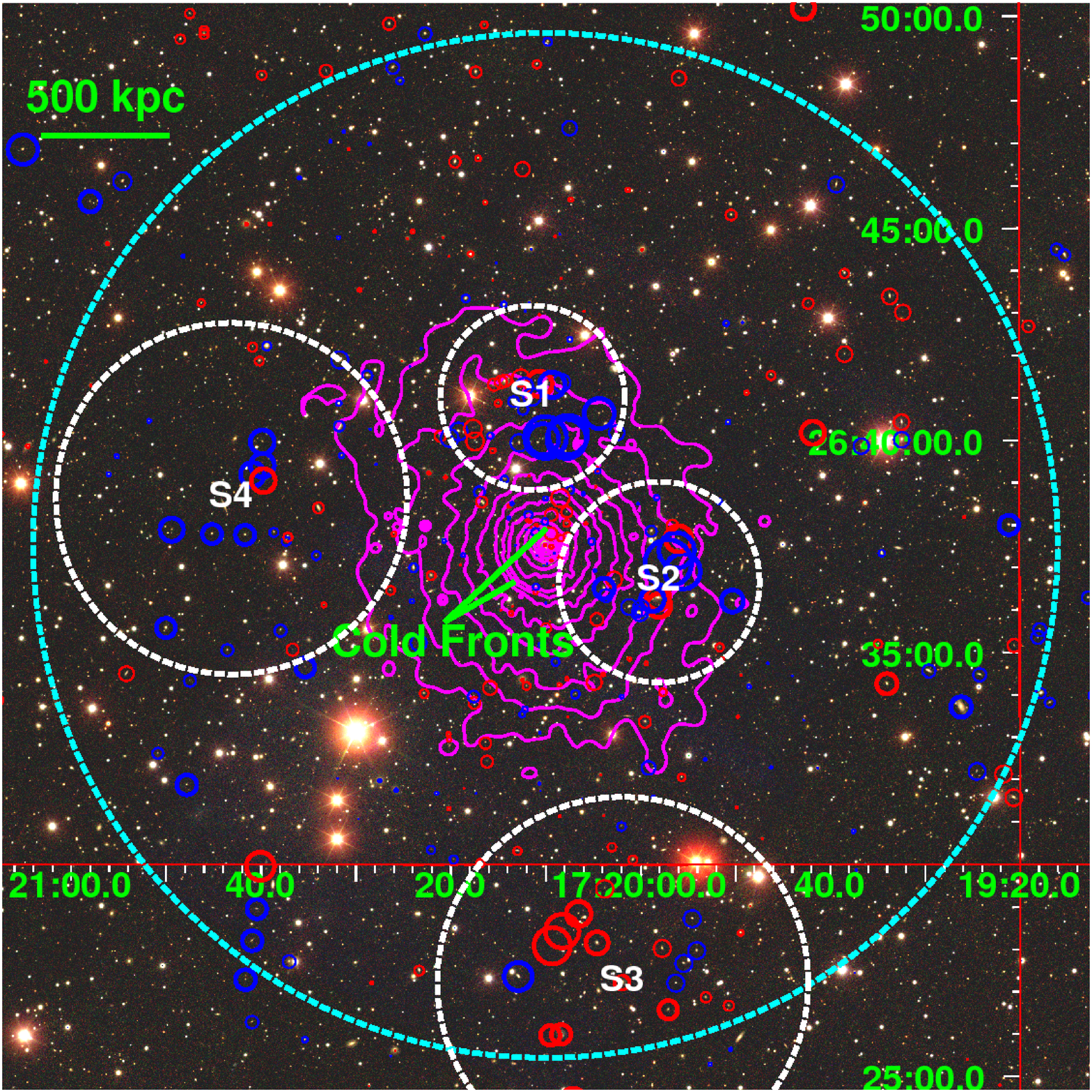}}
{\includegraphics[angle=0,width=0.45\textwidth]{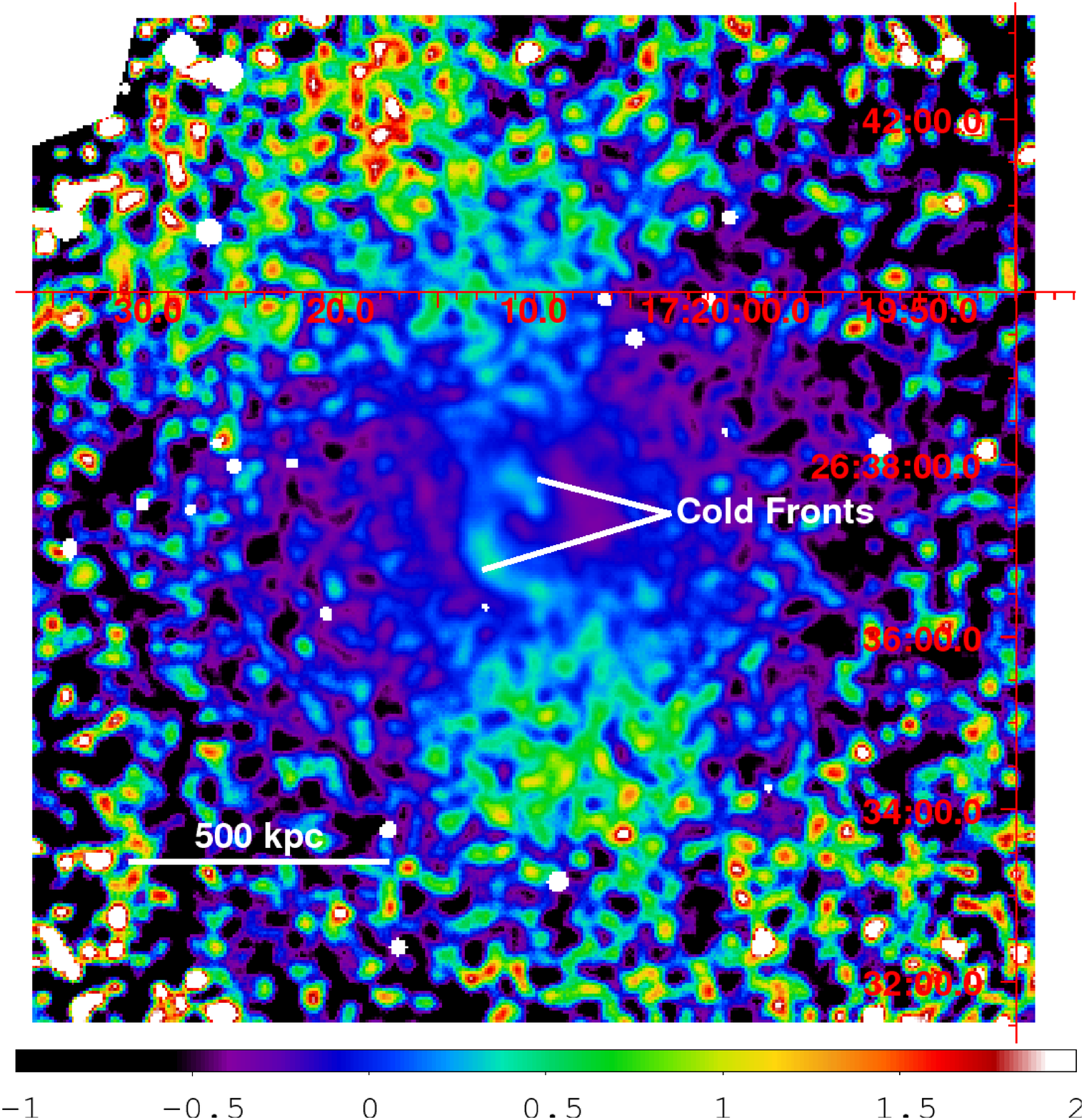}}
\caption{{\it Left and right panels:} The same as Figure~\ref{a2142_all}, but for
RXJ1720. ROSAT contours are not shown in the left panel, and the radius of the cyan dashed 
circles is $R_V=2.02$\,Mpc.}
\label{rxj1720_all}
\end{figure*}

Here, we consider a scenario where S1 has perturbed the core during
its pericenter passage and is currently moving outwards towards the north. 
Considering the \chan\ residual map generated from the \chan\ 
data in the right panel of Figure~\ref{rxj1720_all}, a clear spiral structure 
can be seen which is also seen in the temperature maps of \citet{mazzotta2008} 
and \citet{owers2009c}. Following on from the discussion on the spiral 
morphology in Section~\ref{a2142_in}, this means we must be observing the spiral
structure which has an axis that is aligned roughly with our line of sight. We 
also note that, in contrast to A2142, we see no significant peculiar velocity 
offset in the BCG, consistent with the sloshing occurring mainly in a plane 
perpendicular to our line of sight. By extension, the axis of the perturber's 
orbit is roughly coincident with our line of sight. Furthermore, following the 
spiral from the outermost cold front to the core, the winding direction is 
clockwise. \citet{markevitch2007} noted that the orbit of the perturber and that
of the gas generating the spiral pattern should exhibit prograde rotation, as 
seen in the simulations \citep{ascasibar2006,poole2006,roediger2011}. This 
implies that the S1 substructure follows a clockwise orbit from our perspective.
We propose that S1 has traveled from the south, passed the core region on the 
eastern side providing the necessary perturbation for the generation of the cold
fronts and is now headed to the north. One issue with this interpretation is the
proximity of S1 to the cluster core. If the time since pericenter of $1$\,Gyr 
derived above is correct and the perturber's orbital plane is very close to the 
plane of the sky, then S1 should have had time to travel much further than 
$550$\,kpc. Of course, this assertion relies critically on the velocity of S1 
which is not well constrained. However, even assuming a relatively small radial 
velocity of 1000\kms\ since core passage implies that S1 should be $\sim1$\,Mpc 
from the cluster core. This issue is somewhat relieved if we take the lower 
value of $\tau_{\rm true}=0.8$ derived above and assume that S1's orbit has some 
component parallel with our line of sight. 

Alternatively, if we have underestimated the time since pericenter, then it is 
possible that S1 has passed apocenter and is currently headed towards its second
pericenter. This also relieves the tension between the observed scales of the 
cold fronts in RXJ1720, which occur at much larger radii when compared with those
formed on $\sim 1$\,Gyr post-pericenter timescales seen in the simulations of 
\citet{ascasibar2006}. For example, in the 3.8 Gyr panel of Figure~7 in 
\citet{ascasibar2006}, i.e., $\sim 2.4$ Gyr after pericenter and just prior to 
the second pericenter, the most prominent cold fronts have radii $\sim 80$\,kpc 
and $\sim 160$\,kpc, similar to the cold fronts observed in RXJ1720. However,
\citet{ascasibar2006} note that the contrast in density and temperature at the 
cold fronts decreases with time by a factor of 1.5 at 3.8 Gyrs, compared with 
the factor of $\sim 2$ jump observed in RXJ1720 \citep{owers2009c}. This 
difference in contrast may reflect a difference in the mass, the gas content 
and/or the impact parameter of the perturber between RXJ1720 and the 
simulations presented in Figure~7 of \citeauthor{ascasibar2006}.

The S3 substructure located at $\sim 1.7$\,Mpc south of the cluster core may also 
be considered a potential perturber. The KMM fits revealed that the velocity of 
this substructure is $\sim 820$\kms\ and it has dispersion $\sigma \sim 512$\kms. 
This dispersion is fairly large and comparable to that expected of a reasonably 
large group or poor cluster. However, visual inspection of the velocity 
distribution (lower left panel in Figure~\ref{rxj1720_KMM}) reveals that the 
$820$\kms\ partition has two peaks and thus that this partition may be 
contaminated by galaxies which are not associated with the substructure. This 
contamination may be responsible for inflating the velocity dispersion. More 
data are needed to disentangle any contamination and to confirm this velocity 
dispersion. In any case, there is clearly a dynamical substructure in this 
region. In the scenario where S3 is the perturber, we propose that it has 
traveled from the north and is currently heading to the south. Based on the 
arguments above on the winding direction of the spiral structure seen in the 
right panel of Figure~\ref{rxj1720_all}, we suggest that S3 passed the cluster 
core on the western side and is currently heading towards apocenter for the 
first time.

\section{Summary and Conclusions}\label{sum}

We have presented comprehensive MMT/Hectospec and Keck/DEIMOS multi-object 
spectroscopy of the cold front clusters A2142 and RXJ1720. These 
clusters were selected from the ``relaxed appearing'' subsample of cold front 
clusters from \citet{owers2009c} with the specific aim of testing the 
hypothesis that minor mergers cause the observed cold fronts. To do this,
we used the spatial and kinematic information contained in samples of 956 
and 400 spectroscopically confirmed cluster members for A2142 and RXJ1720, 
respectively, to search for substructure. In both of the clusters we find
group-sized substructure and we identify likely perturbers based on 
comparisons between the observed cold front morphologies and those seen 
in the simulations. Our results are consistent with the cold fronts
being caused by sloshing induced by the gravitational perturbation caused 
by a minor merger. We stress, however, that more detailed merger histories
and better orbital constraints will likely require cluster-specific simulations
aimed at reproducing both the X-ray morphology and the positions and
kinematic properties of the detected substructures. Nonetheless, we conclude 
that our original stated goal of showing cold fronts are related to merger 
activity \citep{owers2009a, owers2009b, owers2009c} has been met.

\acknowledgments

We thank Michael Cooper and Jeffrey Newman for their help with {\sf spec2d}.
MSO and WJC gratefully acknowledge the financial support of an Australian 
Research Council Discovery Project grant throughout the course of this work. 
PEJN was partly supported by NASA grant NAS8-03060. We 
acknowledge the Keck data used in this paper was obtained through the Swinburne 
Time Allocation Committee for Keck (STACK). The authors wish to recognize and 
acknowledge the very significant cultural role and reverence that the summit of 
Mauna Kea has always had within the indigenous Hawaiian community.  We are most 
fortunate to have the opportunity to conduct observations from this mountain. 
Some of the observations reported here were obtained at the MMT Observatory, a 
joint facility of the Smithsonian Institution and the University of Arizona.
We thank the MMT operators and queue-schedule mode scientists for their help 
during observations and the staff at the Harvard-Smithsonian Center for 
Astrophysics Telescope Data Center for reducing the Hectospec data.

{\it Facilities:} \facility{CXO (ACIS)}, \facility{Keck (DEIMOS)}, \facility{MMT (Hectospec)}

\end{document}